\documentclass[aps,twocolumn,prb,longbibliography,showpacs,floatfix,superscriptaddress]{revtex4-2}

\usepackage{graphicx,color}
\usepackage{amsfonts}
\usepackage[figuresright]{rotating}
\usepackage{amssymb}
\usepackage{bm}
\usepackage{natbib}
\usepackage{amsmath}
\usepackage{mathtools}
\usepackage{psfrag}
\usepackage{floatrow}
\usepackage{multirow}
\usepackage{tabularx}
\usepackage{textcomp}
\usepackage{units}
\usepackage{lipsum}
\usepackage{soul}
\usepackage{comment}
\usepackage{titlesec}
\usepackage{times}
\usepackage{hyperref}
\usepackage{enumitem}
\DeclareMathAlphabet\mathbfcal{OMS}{cmsy}{b}{n}

\titlespacing*{\section}
{0pt}{1ex plus .25ex}{1ex plus 1ex}
\titlespacing*{\subsection}
{0pt}{1ex plus .25ex}{1ex plus 1ex}
\titlespacing*{\subsubsection}
{0pt}{1ex plus .25ex}{1ex plus 1ex}

\titleformat{\section}
 {\fontsize{10}{15}\bfseries }
  {\thesection}
  {1em}
  {}

\titleformat{\subsection}
  {\bfseries }
  {\thesubsection}
  {2em}
  {}

\titleformat{\subsubsection}
  {\itshape}
  {\thesubsubsection}
  {2em}
  {}

\def\beq{\begin{eqnarray}}
\def\eeq{\end{eqnarray}}

\newcommand{\gv}[1]{\ensuremath{\mbox{\boldmath$ #1 $}}}

\newcommand{\ket}[1]{\left| #1 \right>} 
\newcommand{\bra}[1]{\left< #1 \right|} 
\newcommand{\grad}[1]{\gv{\nabla} #1} 
\let\baraccent=\= 
\renewcommand{\=}[1]{\stackrel{#1}{=}} 

\makeatletter

\titleclass{\subsubsubsection}{straight}[\subsection]

\newcounter{subsubsubsection}[subsubsection]
\renewcommand\thesubsubsubsection{\thesubsubsection.\arabic{subsubsubsection}}

\titleformat{\subsubsubsection}
   {\normalfont\normalsize\itshape \centering}{\thesubsubsubsection}{1em}{}
\titlespacing*{\subsubsubsection}
{0pt}{1ex plus .3ex}{1ex plus .3ex}

\makeatletter
\renewcommand\paragraph{\@startsection{paragraph}{5}{\z@}%
  {3.25ex \@plus1ex \@minus.2ex}%
  {-1em}%
  {\normalfont\normalsize}}
\renewcommand\subparagraph{\@startsection{subparagraph}{6}{\parindent}%
  {3.25ex \@plus1ex \@minus .2ex}%
  {-1em}%
  {\normalfont\normalsize}}
\def\toclevel@subsubsubsection{4}
\def\toclevel@paragraph{5}
\def\toclevel@paragraph{6}
\def\l@subsubsubsection{\@dottedtocline{4}{7em}{4em}}
\def\l@paragraph{\@dottedtocline{5}{10em}{5em}}
\def\l@subparagraph{\@dottedtocline{6}{14em}{6em}}
\makeatother

\begin{document}
\title{Finite-size topological phases from semimetals}
\author{Adipta Pal}
\affiliation{Max Planck Institute for Chemical Physics of Solids, Nöthnitzer Strasse 40, 01187 Dresden, Germany}
\affiliation{Max Planck Institute for the Physics of Complex Systems, Nöthnitzer Strasse 38, 01187 Dresden, Germany}

\author{Ashley M. Cook}
\affiliation{Max Planck Institute for Chemical Physics of Solids, Nöthnitzer Strasse 40, 01187 Dresden, Germany}
\affiliation{Max Planck Institute for the Physics of Complex Systems, Nöthnitzer Strasse 38, 01187 Dresden, Germany}

\begin{abstract}
Topological semimetals are some of the topological phases of matter most intensely-studied experimentally. The Weyl semimetal phase, in particular, has garned tremendous, sustained interest given fascinating signatures such as the Fermi arc surface states and the chiral anomaly, as well as the minimal requirements to protect this three-dimensional topological phase. Here, we show that thin films of Weyl semimetals (which we call quasi-(3-1)-dimensional, or q(3-1)d) generically realize finite-size topological phases distinct from 3d and 2d topological phases of established classification schemes: response signatures of the 3d bulk topology co-exist with topologically-protected, quasi-(3-2)d Fermi arc states or chiral boundary modes due to a second, previously-unidentified bulk-boundary correspondence. We show these finite-size topological semimetal phases are realized by Hamiltonians capturing the Fermiology of few-layer Van der Waals material  MoTe\textsubscript{2} in experiment. Given the broad experimental interest in few-layer Van der Waals materials and topological semimetals, our work paves the way for extensive future theoretical and experimental characterization of finite-size topological  phases.
\end{abstract}
\maketitle

\section{Introduction}
Topological degeneracies in electronic band structures~\cite{bradlyn2016} are now some of the most widely-studied examples of topology in experiments~\cite{borisenko2014experimental, huang_spectroscopic_2016, jiang_signature_2017, liu2014stable, neupane2013observation, lv2015taas, lv_observation_2015, xu_discovery_2015, suyang2015discovery, liu2014discovery, shekhar_extremely_2015, gyenis_2016}. The most intensely-studied of this vast set of degeneracies, however, are the Weyl nodes of the Weyl semimetal~\cite{yanreview2017, jia_weyl_2016, armitage2018review}, realized generically in three spatial dimensions when either time-reversal symmetry~\cite{burkov_wsmti_2011} or spatial inversion symmetry~\cite{halasz2012time} is broken. The rich physics of the Weyl semimetal Fermi arc surface states~\cite{haldane2014attachment, mathai2017global, hosur2012friedel, potter2014quantum} and transport signatures, such as the chiral anomaly~\cite{nielsen1983anomaly, zyuzin2012topological, son2013chiral, parameswaran2014chiral, huang2015observation, zhang2016signatures}, in combination with the relative ease of its experimental study, make it a workhorse~\cite{huang2015weyl, wan2011topological, wang2012dirac} for experimental study of topological phases of matter.

The recent discovery of finite-size topological phases (FSTs)~\cite{cook2022, calderon2023}, where a quasi-($D$-$\delta$)-dimensional FST is topologically-distinct from strictly $D$- and $\delta$-dimensional topological states---instead exhibiting signatures of both---motivates redoubled interest in topological semimetal physics given tremendous interest in thin film systems~\cite{TI-thin-film-conduc,TI-crossover,TI-thin-phase-trans} and specifically Van der Waals few-layer and heterostructure materials and Moir\'e systems~\cite{vanwaals-hetero,Nature-hetero-antiferr,TI-hetero,robust-2dhetero, TMD-1,TMD-topo}. FSTs are possible when $D$-dimensional topological phases are realized for open-boundary conditions in at least one direction, to yield ($D$-1)-dimensional topologically-protected gapless boundary states. If these gapless boundary states strongly hybridise due to finite system size, FSTs can be stabilised by the hybridization gap: a system thermodynamically large in $\delta < D$ dimensions and finite in size in $D-\delta$ directions exhibits topological response signatures of the $D$-, ($D$-1)-,..., ($\delta$+1)-dimensional bulk topological phase(s) in combination with quasi-($\delta$-1)-dimensional gapless boundary modes. These boundary modes correspond to additional quasi-($D$-$\delta$)-dimensional bulk non-trivial spectral flow in fragments of the topologically non-trivial regions of the underlying $D$-dimensional bulk. Finite-size topological phases are therefore distinct from $D$-dimensional topological states of existing classification schemes~\cite{Ryu_2010, Schnyder2008, kitaev2009} and indicate these schemes must be generalised to classify the full set of topological phases of matter. Instead, finite-size topological phases are consistent with generalisation of the framework of the quantum Hall effect to that of the quantum skyrmion Hall effect~\cite{qskhe2024}. Given established classification schemes, even K-theory~\cite{kitaev2009}, cannot be applied to classify FSTs, it is valuable to first introduce them for canonical examples of topological phases, with one of the most important being the Weyl semimetal, the foundational system for considerable later work on the large family of topological semimetal phases of matter~\cite{armitage2018review}.

In this work, we therefore introduce and characterize the finite-size topological phases for experimentally-relevant thin film, quasi-(3-1)-dimensional (q(3-1)d) geometries, which derive from three-dimensional (3d) topological semimetal phases realized when the system is thermodynamically-large in all three spatial dimensions. We realize these phases in tight-binding models for Weyl semimetals(WSMs), first for (i) open-boundary conditions yielding Fermi arc surface states which hybridise, and then (ii) for open-boundary conditions without Fermi arc surface states. In each case, we characterize bulk-boundary correspondence and topological response signatures distinguishing these finite-size topological phases from previously-known topological phases of matter. Remarkably, case (i) yields topological semimetals with  properties of strictly 2d topological semimetals \textit{co-existing with} properties of a 3d Weyl semimetal, and case (ii) yields a topological state with properties of a 2d topological \textit{insulator}, \textit{co-existing with}  properties of a 3d Weyl semimetal. We finally explore the effects of Weyl node tilting for a model previously-used to interpret experimental results on the Weyl semimetal MoTe\textsubscript{2}~\cite{huang_spectroscopic_2016, jiang_signature_2017, yan2015prediction}. Past work also indicates few-layer 1T'-MoTe\textsubscript{2} is semi-metallic~\cite{Song_2018}, while the monolayer is predicted to be a quantum spin Hall insulator~\cite{qian_2014}, suggesting the few-layer topology derives from the Weyl semimetal phase of the three-dimensional bulk, further motivating our analysis. Given the very intense theoretical and experimental interest in Weyl semimetals, our work is foundational to study of topological semimetals and finite-size topological phases of matter.

\begin{center}
\begin{figure*}[tb]
    \centering
    \includegraphics[width=0.9\textwidth]{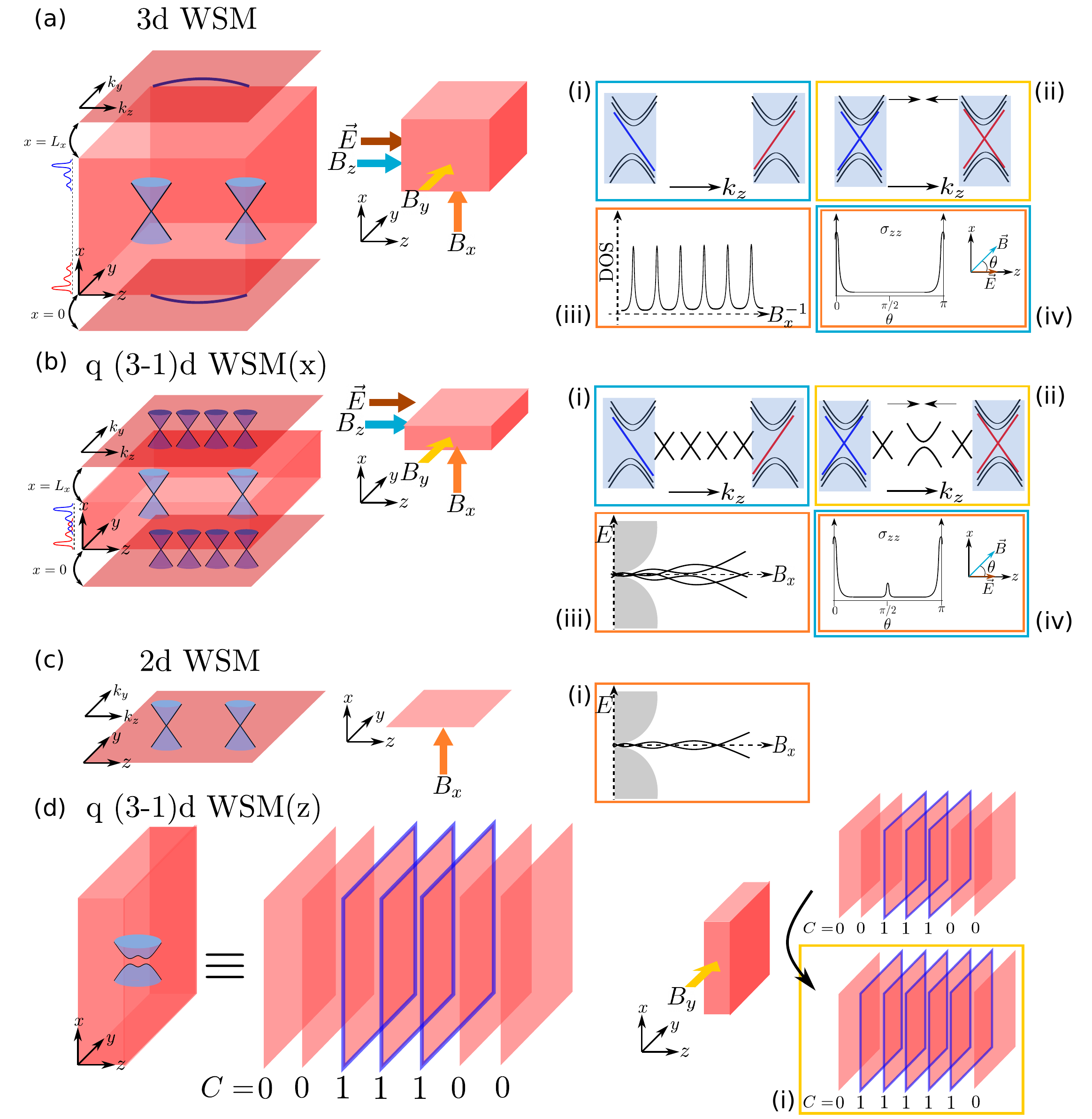}
    \caption{ Schematic diagrams of band structures, bulk-boundary correspondences, and topological response signatures for (a) 3d Weyl semimetal (WSM), (b) finite-size topological semimetal phase for quasi (3-1)d geometry (thin film) along x (normal vector of film in the x direction) from underlying 3d WSM state,  (c) 2d WSM and (d)  finite-size topological insulator phase for quasi (3-1)d geometry (thin film along z  or equivalently normal vector of film in the z direction) from underlying 3d WSM state. For band structure schematics, blue marks bulk spectra and purple represents surface spectra. The quasi (3-1)d finite-size topological phases (FSTs) emerge from strong finite size effects, which can yield (b) the FST semimetal state possessing 2d Weyl nodes on the top and bottom surfaces of the thin film (surface Weyl nodes, or SWNs) in addition to retaining signatures of 3d Weyl nodes in its bulk or (d) the FST insulator state with chiral edge modes ostensibly of a quantum anomalous Hall insulator (purple lines depict the chiral edge mode contributions from 2d submanifolds of the underlying 3d Brillouin zone). We have shown the relevant response to magnetic field for each case on the right, labeled (i) to (iv). The relation between different parts of this schematic figure to different sections in the manuscript has been outlined towards the end of the Introduction. Sub-parts (i), (ii), (iii), (iv) for subfigures (a) and (b) show the relevant responses for magnetic field along z, y, x and zx plane respectively indicated by color coding cyan, yellow, orange and cyan-orange respectively. The same color coding depicts the response for (c) in (i) for magnetic field along x(out-of-plane) and for (d) in (i) for magnetic field along y(in-plane).}
    \label{fig:fWSMschematic}
\end{figure*}
\end{center}

We outline the organization of this paper below. The reader may find it helpful to associate each section with one of the figures in our schematic diagram Fig.~\ref{fig:fWSMschematic} which illustrates qualitatively our main results for each section. In Section~\ref{WSMreview} we provide a short introduction to the 3d time reversal broken Weyl semimetal and discuss the important aspects of its band structure, topology and surface states which make it an interesting system to study from the perspective of finite-size topological phases. We also show the toy model Hamiltonian for the 3d bulk system that we will be considering in the first half of the manuscript. The schematic diagram Fig.~\ref{fig:fWSMschematic}(a) depicts topological features of the WSM when the system is thermodynamically-large in all three spatial directions, such as the bulk band structure, the Fermi arc surface states resulting from bulk-boundary correspondence, and topological response signatures, which will be modified as a result of finite size effects.

In Section~\ref{q2dWSM}, we consider the model Hamiltonian realizing the 3d WSM for open boundary conditions and small system size along one coordinate axis (the number of unit cells in one direction is comparable to the characteristic decay length scale of potential boundary modes). In this geometry, we characterize possible FSTs of the model. This section is subdivided into two subsections based on whether OBCs and small system size are along an axis parallel or perpendicular to the axis in the 3d BZ along which the Weyl nodes are separated, or the Weyl node axis. In subsection~\ref{fWSMthinx}, we consider the system with OBCs and small number of unit cells in the x-direction, and subsection~\ref{fWSMthinz}, where we consider the system with OBCs and small number of unit cells in the z-direction. We refer these two cases as the Finite-size topological Weyl semimetal (FST-WSM) and Finite-size topological Weyl insulator (FST-WI), respectively. We have again shown schematically the exact configuration we are studying in subsection~\ref{fWSMthinx} in Fig.~\ref{fig:fWSMschematic}(b), with a strictly 2d ``Weyl semimetal" depicted schematically in Fig.~\ref{fig:fWSMschematic}(c) for comparison. We similarly show, schematically, the exact configuration we are studying in subsection~\ref{fWSMthinz} in Fig.~\ref{fig:fWSMschematic}(d). In each of Fig.~\ref{fig:fWSMschematic}(b) and (d), the hybridization mechanism (surface or bulk, respectively) yielding the FST is depicted.

For each of Fig.~\ref{fig:fWSMschematic}(a), (b), (c), and (d), we furthermore depict response signatures of the 3d WSM, the FST-WSM, the 2d WSM, and the FST-WI, respectively, which we discuss in detail in Section~\ref{fWSMmagfield}. In correspondence with subsection~\ref{fWSMthinx} and subsection~\ref{fWSMthinz}, this section on response signatures is divided into subsection~\ref{fWSMxmag} for OBCs and small system size in the x-direction, and subsection~\ref{fWSMzmag} for OBCs and small system size in the z-direction. The relevant effects for the first subsection have been schematically illustrated in Fig.~\ref{fig:fWSMschematic}(b)(i), (ii), (iii) and (iv), which consider the magnetic field along each of the z, y, x axes and evidence of the chiral anomaly by probing response signatures while varying the orientation of an external magnetic field relative to external electric field, respectively. These response signatures of the FST-WSM are, schematically, compared directly with those of the 3d WSM in Fig.~\ref{fig:fWSMschematic}(a)(i), (ii), (iii) and (iv) for the same cases of external magnetic field orientation relative to the Weyl axis, and also the response signature of 2d WSMs, shown schematically in Fig.~\ref{fig:fWSMschematic}(c). In the last row, the result in subsection~\ref{fWSMzmag} is schematically reproduced in Fig.~\ref{fig:fWSMschematic}(d).

Furthermore, we have sections not schematically represented in Fig.~\ref{fig:fWSMschematic}. In Sec.~\ref{TRIWSM} and Sec.~\ref{TRWSMresponse}, we study a toy model for time reversal symmetric Weyl semimetals with four Weyl nodes, which is fermiologically similar to the Weyl semimetal phase is materials like MoTe$_2$. In Sec~\ref{trwsmslabsec}, we show the slab spectra for the time-reversal-symmetric WSM in a quasi (3-1)d geometry with thin film normal vector oriented along each of the x-, y-, and z-axes and describe the effects of surface state and bulk node hybridization corresponding to myriad open boundary conditions for a given thin film orientation. Sec.~\ref{trwsmtiltsec} shows the effect of tilting on the topology and slab spectra in the thin film limit. Finally, Sec.~\ref{TRWSMresponse} studies responses to magnetic field for the thin film time reversal symmetric WSM.

Again we clarify some notation before jumping into the main content. FSTs are realized here in quasi (3-1)d geometries which imply thin films derived from a 3d system by thinning the system lattice along one of the coordinate axes. We will use the terms quasi (3-1)d geometry and thin film interchangeably. Whenever we write a phrase like ``thin film along x", we mean a quasi (3-1)d system thinned along x or a thin film thermodynamically large in the yz plane with normal vector to the plane along the x axis along which the system size is small.

\section{Three-dimensional Weyl semimetal}\label{WSMreview}
We first review key features of the three-dimensional Weyl semimetal (WSM) topological phase relevant to the present work. The 3d WSM is characterized by pairs of two-fold, topologically-protected band structure degeneracies, known as Weyl nodes (WNs), and linear energy dispersion in their vicinity in momentum space. For a Weyl node located at zero momentum, a small $\boldsymbol{k}$ expansion of the Bloch Hamiltonian $\mathcal{H}(\boldsymbol{k})$ in the vicinity of the Weyl node may be written as
\begin{equation}
\mathcal{H}(\boldsymbol{k}) = v\boldsymbol{k}\cdot\boldsymbol{\sigma},
\end{equation}
where the sign of $v$ determines the chirality of the Weyl node.  The Weyl node is  a Berry curvature monopole, meaning one interpretation of a Weyl node is that it corresponds to the closing of the minimum direct bulk energy gap when tuning a two-dimensional system between phases characterized by distinct total Chern numbers \cite{burkov_wsmti_2011, wan2011topological, armitage2018review}. A lattice model of a Weyl semimetal respecting inversion symmetry---with \textit{time-reversal symmetry broken (TRB)}---is then obtained from a lattice model for a Chern insulator if additional dependence on a third momentum component, $k_z$, is added to the Chern insulator model, such that there exist $k_z$ values across which the total Chern number changes. In lattice models, Weyl nodes of opposite chiralities therefore occur in pairs. In the following few sections, we will only consider the TRB WSM. There is another variant of the 3d WSM, where the time reversal symmetry is preserved and the inversion symmetry is broken. We will refer to it as the \textit{time-reversal invariant (TRI)} WSM and such a Weyl semimetal has a minimum of two pairs of Weyl nodes. We will probe the TRI WSM for finite size effects in Sec.~\ref{TRIWSM}.

Focusing on the TRB WSM, we therefore construct our lattice model from a canonical model for a Chern insulator, known as the QWZ model~\cite{qi2006topological}, considered previously in study of finite-size topological phases derived from the Chern insulator~\cite{cook2022}. Our Bloch Hamiltonian for the QWZ model is therefore
\begin{equation}
\begin{split}
H_{QWZ}(\boldsymbol{k})=&(M-2t\cos k_x-2t\cos k_y-2t\cos k_z)\sigma^z\\
&+2\Delta\sin k_x\sigma^x+2\Delta\sin k_y\sigma^y.
\end{split}
\label{qwz}
\end{equation}
The topological phase diagram of the QWZ model includes regions with Chern number $C=0$ and $C=\pm1$, which are realized for different values of the mass parameter $M$. We therefore promote $M$ to a $k_z$-dependent function, $m(k_z)$, yielding a minimal model for a TRB WSM of the form,
\begin{equation}
\begin{split}
H_{WSM}(\boldsymbol{k})=&(m(k_z)-2t\cos k_x-2t\cos k_y)\sigma^z\\
&+2\Delta\sin k_x\sigma^x+2\Delta\sin k_y\sigma^y,
\end{split}
\label{bulk3d}
\end{equation}
where we have chosen $m(k_z)=M-2t\cos k_z$. The positions of the Weyl nodes are not subject to any particular parametric dependence besides $M$ and $t$ and are located at $\boldsymbol{k}_0=(0,0,\pm\cos^{-1}(\frac{M}{2t}-2))$. However, while toy models for WSMs often consider the case of $\Delta =t$, the cases in which $\Delta \neq t$ are important in studying finite-size topological phases of matter~\cite{cook2022} and we therefore keep these two parameters independent of one another.

By construction, the WSM in Eqn.~\ref{bulk3d} can be described as a stacking of 2d Chern insulators in the $k_z$ direction. Hence, if the system has open boundary conditions (OBCs) in the $x$-direction, yielding two surfaces, we expect to observe chiral edge states in the slab spectrum along $k_y$ for thermodynamically-large system size in the $x$-direction, which extend between the two Weyl nodes separated in $k_z$, and which decay into the bulk at the Weyl nodes. The projection of the chiral edge states on the remaining 2d BZ is an important signature of the WSM known as Fermi Arc surface states \cite{armitage2018review, wan2011topological}. These states connect Weyl node pairs of opposite chirality on the surface and have important consequences in transport properties of the system \cite{armitage2018review, zyuzin2012topological, hosur_wsmtransport_2013, potter2014quantum}. The Fermi arc surface states decay exponentially to the bulk along the open $x$-direction, such that hybridization between the Fermi arc states on opposite surfaces is negligible for sufficiently-large system size in the $x$-direction well-approximated by the thermodynamic limit. Similarly to the case of the FSTs derived from the Chern insulator~\cite{cook2022}, where FSTs emerged from hybridization of chiral edge modes, the Fermi arc surface states are a natural starting point in efforts to realize FSTs from the Weyl semimetal. We will later show, in section ~\ref{fWSMthinz}, that it is possible to derive other FSTs from WSMs, via hybridization of the bulk Weyl nodes, rather than hybridization of Fermi arc surface states.

\section{The thin-film quasi (3-1)d Weyl semimetal}\label{q2dWSM}
Considering Hamiltonian Eqn.~\ref{bulk3d} for the TRB WSM, with a single pair of Weyl nodes separated in momentum-space along the $k_z$ axis and positioned at $\boldsymbol{k}_0=(0,0,\pm\cos^{-1}(\frac{M}{2t}-2))$, we will examine two possible thin film orientations. In the first case (i), we consider OBCs and finite system size in the $x$ or $y$ direction, of $L_x$ or $L_y$, respectively, which are the directions perpendicular to the Weyl axis. In the second case (ii), we consider OBCs and finite system size in the $z$ direction, of $L_z$, the direction parallel to the Weyl axis.  The latter orientation (ii) has also been previously studied in work demonstrating the quantum Hall effect induced by chiral Landau levels in semimetal thin films \cite{nguyen_chiralqhe_2021}, but did not consider or demonstrate FST realization.

Characterizing Hamiltonian Eqn.~\ref{bulk3d} under these conditions, we will show the system in orientation (i) and (ii), for small finite system size, say, relative to the characteristic decay length scale of any Fermi arc surface states in the former, is actually driven into previously-unidentified FSTs. In these FST phases, subsets of the phenomena are naively associated with strictly 2d topological phases or strictly 3d topological phases, but we show these phenomena co-exist in a topological phase distinct from previously-known 2d and 3d states and are generically richer than counterpart phenomena of strictly 2d or 3d topology.

\subsection{Wilson loop characterization of bulk and surface topology}
We provide a short review of the Wilson loop characterization of bulk topology, which will be used throughout this article. Our explanation is similar to one of our previous work \cite{pal2024mkc}.

One of the most robust methods to characterize bulk topology is by characterizing the polarization or peak of the Wannier state in a unit cell. The bulk is topologically non-trivial if the Wannier peak is centered on one of the orbitals of the unit cell which is equivalent to a polarization of $\pi\mod{2\pi}$. We calculate the position of the Wannier peak, or the Wannier spectra by considering the Wilson loop spectrum. The Wilson loop is a unitary operator defined over a closed path in momentum space,
\begin{equation}
\mathcal{W}=\overline{\exp}\int_{BZ} d\mathbf{k}\cdot\mathbf{A}(\mathbf{k}),
\end{equation}
where, $\mathbf{A}_{mn}(\mathbf{k})=i\bra{u_m(\mathbf{k})}\grad_{\mathbf{k}}\ket{u_n{\mathbf{k}}}$ is the non-abelian Berry connection. Here $\ket{u_n(\mathbf{k})}$ are Bloch states over an occupied subspace and $\mathbf{A}$ is a Hermitian operator. The diagonal elements can be viewed as the position expectation value for a particular band index and since $\mathcal{W}$ is a unitary operator, its eigenvalues are $e^{i\nu_j}$, so that $\nu_j$ are related to the digonalized position expectation and hence to the polarization modulo $2\pi$. We call the list $\{\nu_j\}$ the Wannier spectra. All our characterization will be based on finding this Wannier spectra $\{\nu_j\}$ for different closed loops in the Brillouin zone across different momenta directions. For thin film systems, where we consider the surface topology, we will calculate Wilson loops along the remaining momenta directions over all occupied bands, keeping the Fermi energy at zero. The size of the Wannier spectra should be identical to the number of occupied bands and we consider the system to be topologically non-trivial if at least one of the elements of the Wannier spectra is valued $\pi\text{mod} 2\pi$ which corresponds to spectral flow along that particular closed momenta direction.\\

As an example, let us consider the WSM in Eqn.~\ref{bulk3d} with thin film along the x-axis with system size, $L_x$. The number of occupied bands for Fermi energy zero is $L_x$, so we calculate the Wilson loop for $L_x$ bands around the BZ along $k_y$. This provides $L_x$ elements in the Wannier spectra, and we look for a $\pi\mod{2\pi}$ element to infer the topological character of the system corresponding to a spectral flow around $k_y$. In the rest of the manuscript, for simplicity, we have considered the Wannier spectra with a $2\pi$ denominator, so that the Wannier spectra element for non-trivial topology is given as $0.5\mod{1}$.

\begin{center}
\begin{figure*}[tb]
    \centering
    \includegraphics[width=0.95\textwidth]{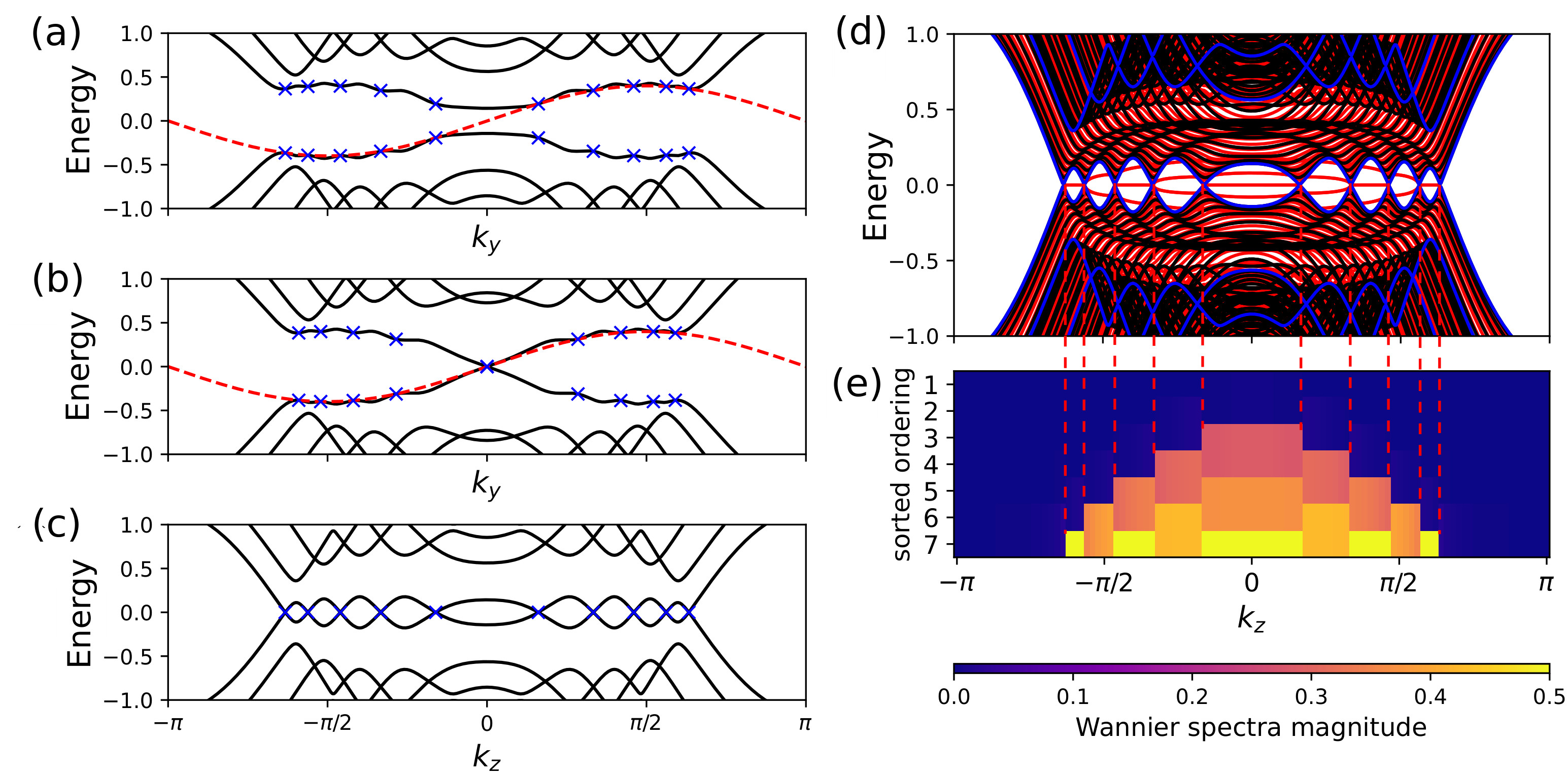}
    \caption{Weyl semimetal with thin film in the x-direction ($L_x=7$) for $M=3$ and $\Delta =0.2t$. The blue crosses indicate the analytically calculated values derived from Eqn. ~\ref{fwsmcrit} and are remnants from the system with large lattice size along x. The red dotted line is the actual expectation in the large lattice limit. (a) show the slab spectra along $k_y$ between two such blue crosses in the slab spectra along $k_z$ shown in (c). In (b) we plot the same slab spectra along $k_y$ at one of the blue crosses in the slab spectra long $k_z$ in (c). In (d) we show the slab spectra along $k_z$ for thin film along x with large lattice size along y for OBC along y denoted by red, PBC along y denoted in black and overlapped with the slab spectra in (c) shown in blue. The plot below in (e) shows Wannier spectra derived from Wilson loop around $k_y$ in a half-filled system with OBC along x. The Wannier spectra indicates that the flat bands along $k_z$ in (d) are topologically non-trivial.}
    \label{fig:fWSMxslabWL}
\end{figure*}
\end{center}

\subsection{Quasi-(3-1)d Weyl semimetal with open boundary conditions in the x-direction}\label{fWSMthinx}
We consider a 3d WSM for OBCs perpendicular to the Weyl axis ($z$), namely along the $x$-direction. By construction, the system should now show Fermi arc surface states derived from the stacking of Chern insulators, as mentioned in the previous section. However, if the system is now restricted to finite system size in the $x$-direction of $L_x$, which is comparable to the characteristic decay length scale of the Fermi arc surface states, finite-size effects become prominent. The ratio of $t$ to $\Delta$ plays an important role in determining the characteristic decay length scale relative to the lattice spacing, as well as whether the Fermi arc wave functions possess an oscillatory profile in real-space or purely exponential decay. We will usually consider $t>\Delta$, with lattice sites in the $x$-direction located from $x=0$ to $x=L_x$ (we take lattice spacing $a$=1).  \\

In the finite-size regime, the Fermi arc states must satisfy boundary conditions, which facilitates FST realization when Fermi arc states form standing waves in the finite-size regime, hence restricting their wavelength. The slab spectrum in the finite-size regime then includes contributions from the 3d bulk spectrum only at a discrete set of momentum values in the direction of OBCs, for a given Weyl node separation. The following equation reveals the points in k-space for which the 3d bulk contributes to the slab,
\begin{equation}
\begin{split}
M-2t\cos k_y-2t\cos k_z = 2\sqrt{t^2-\Delta^2}\cos\frac{n\pi}{L_x+1},\\ (n=1,...,L_x).
\end{split}
\label{fwsmcrit}
\end{equation}
These points lie on the curves $E=\pm 2\Delta\sin k_y$ of the underlying bulk spectrum. (Detailed calculations on derivation of this condition can be found in the Supplementary Materials(SM)~\cite{SuppMat}. We show, numerically, the fragmentation of Fermi arcs in Fig.~\ref{fig:fWSMxslabWL}(a), (b) and (c) in the slab spectra of the thin film and infer that the closing points match the ones calculated from Eqn.~\ref{fwsmcrit} at zero energy. Here, by considering $t<\Delta$, we are transitioning from a critically damped wavefunction $\sim e^{-qx}$ to a damped oscillating wavefunction, $\sim e^{-qx}\sin(\lambda x)$, where the exponential tells us that the peak in probability of this wavefunction is near the left edge. That is, this is the wavefunction for the ``left edge state''. The counterpart wavefunction for the state with probability density peaking near the right edge, or ``right edge state'', takes the same form but with $x\rightarrow L+1-x$.

The left edge state at those specific points reflecting the bulk 3d topology is
\begin{equation}
\begin{split}
\Psi(x,k_y,k_z)\sim e^{ik_yy+ik_zz}\bigg{(}\frac{t-\Delta}{t+\Delta}\bigg{)}^{\frac{x}{2}}\sin\frac{n\pi x}{L_x+1},\\ (n=1,...,L_x).
\end{split}
\label{psistanding}
\end{equation}
Here, $\frac{n\pi}{L_x+1}$ are the wavenumbers for the edge states which fit in the thin film direction between $x=0$ and $x=L_x+1$. The above expression shows that the decay length for the edge state increases with increasing ratio of $\frac{t}{\Delta}$. Henceforth, we consider size along the thin film direction to be of the order of the decay length by carefully choosing the value of $\Delta<t$. \\

From Eqn.~\ref{fwsmcrit}, we observe that each point in k-space at zero energy is a 2d band structure degeneracy characterized by the small $\boldsymbol{k}$ expansion $k_z\sigma^z+k_y\sigma^y$. We henceforth refer to these surface nodes as \textit{2d surface Weyl nodes(SWNs)}.
For a given value of $k_z$, we have a 1d chiral insulator along y, where we expect zero energy edge states for OBCs in a specific mass parameter interval. We therefore can define \textit{2d surface Weyl semimetals(SWSM)} exhibiting multiple pairs of SWNs. For the purposes of discussing serpentine Landau level response signatures of SWSMs, a pair of SWNs is defined here as two SWNs related by inversion symmetry.

Each pair of SWNs is expected to yield doubly degenerate flat bands along $k_z$ when the $y$-direction is opened since the mass is now a function of $k_z$. This flat band should extend between the two nodes of that particular 2d SWSM which has been shown in Fig.~\ref{fig:fWSMxslabWL}(d). However, the WSM thin film  harbors multiple pairs of 2d SWNs, and when multiple such flat band intervals overlap, hybridization occurs. Consequently, the observation of flat bands at zero energy is contingent upon the presence of an odd number of these overlapping intervals, each interval corresponding to a 2d Weyl node pair and a Fermi arc state on each edge. Hence, the intervals between two adjacent SWNs can be classified as  ``odd" and ``even", respectively, based on whether there is an odd or even number of overlapping flat band intervals. This behaviour signals the presence of a non-trivial $\mathbb{Z}_2$ topological invariant. The topological nature of these zero energy flat band states are then validated by the non-trivial Wannier spectra obtained from a Wilson-loop winding around $k_y$, as shown in Fig.~\ref{fig:fWSMxslabWL}(e). We have also shown probability densities for different $k_z$ values in Fig.~\ref{fig:fWSMxslabWL}(d) in SM~\cite{SuppMat}.\\

\begin{center}
\begin{figure}[t!]
    \centering
    \includegraphics[width=0.8\textwidth]{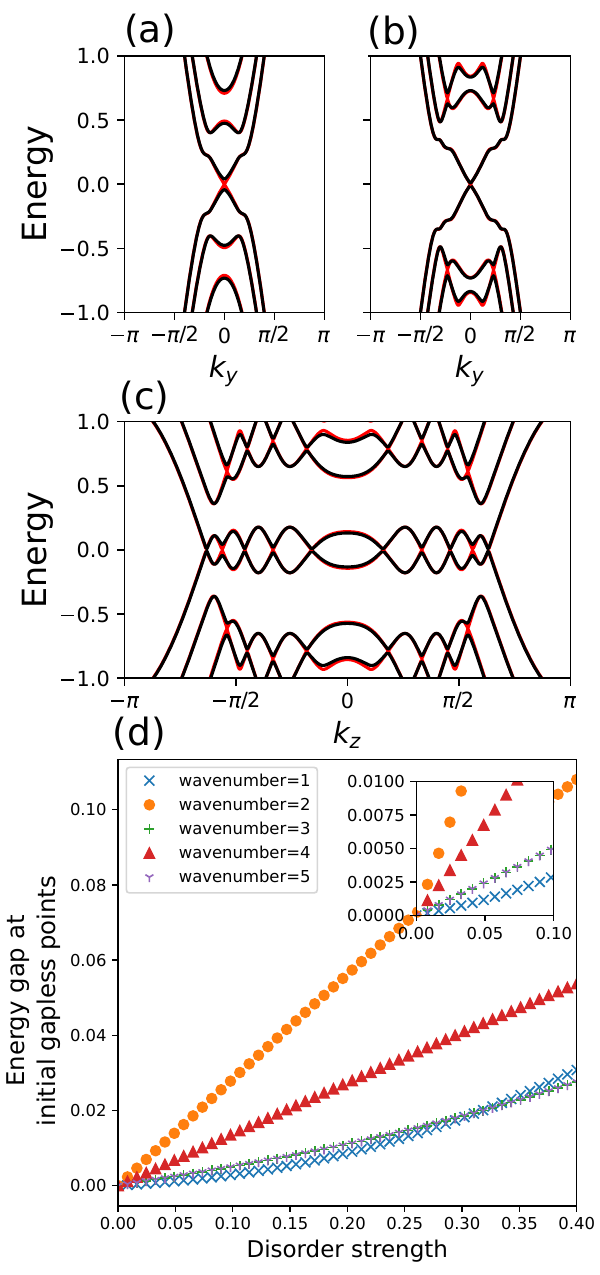}
    \caption{Effect of chirality breaking disorder in a Weyl semimetal thin film with small system size along x $L_x=7$ for $M=3$ and $\Delta =0.2t$. Subfigure (c) shows slab spectra in $k_z$ in the presence of the chirality breaking disorder (black) and in its absence (red). Slab spectra vs. $k_y$ at two gapless points, second from the left in (a) and third from the left in (b) have been compared likewise in the presence and absence of disorder. Subfigure (d) shows that the gapless points have distinctly different robustness in the presence of the aforementioned disorder, such that wavenumbers 1, 3 and 5 appear more robust compared to the others. For each disorder strength, we have averaged over $50$ disorder realizations.}
    \label{fig:fWSMdisorder}
\end{figure}
\end{center}

 Standard dimensional reduction, corresponding to moving upwards and left within the ten-fold way classification table~\cite{Ryu_2010}, implies that the surface effective Hamiltonian is  in class BDI~\cite{Schnyder2008, Ryu_2010}, with particle-hole, time-reversal and chiral symmetry. This implies the effective slab Hamiltonian for these OBCs should gain a chiral symmetry. Therefore, it is pertinent to study the effects of a chiral symmetry-breaking disorder term on the thin film system to test the robustness of the 2d SWNs. We introduce a random-onsite disorder proportional to $\sigma^x$ within certain maximum absolute value, the disorder strength, and consider the average profile across all disorder realizations. Numerical simulations in Fig.~\ref{fig:fWSMdisorder}(a), (b) and (c) reveal that there exists a variation in the gap size among the 2d SWNs for a particular disorder strength. Moreover, we find the critical points associated with the even wavenumber for the standing wave edge mode in Eqn.~\ref{psistanding} are gapped more severely, with these gaps increasing exponentially as opposed to the gaps at odd wavenumbers, which increase more gradually with increasing disorder strength as shown in Fig.~\ref{fig:fWSMdisorder}(d). We observe that only 2d Weyl nodes, which are end points of ``odd" regimes---as defined in the previous paragraph---are more robust than the nodes which are boundary points of even regimes.\\

\subsection{Quasi (3-1)d Weyl semimetal with open boundary conditions in the z-direction}\label{fWSMthinz}
\begin{center}
\begin{figure}[t]
    \centering
    \includegraphics[width=0.8\textwidth]{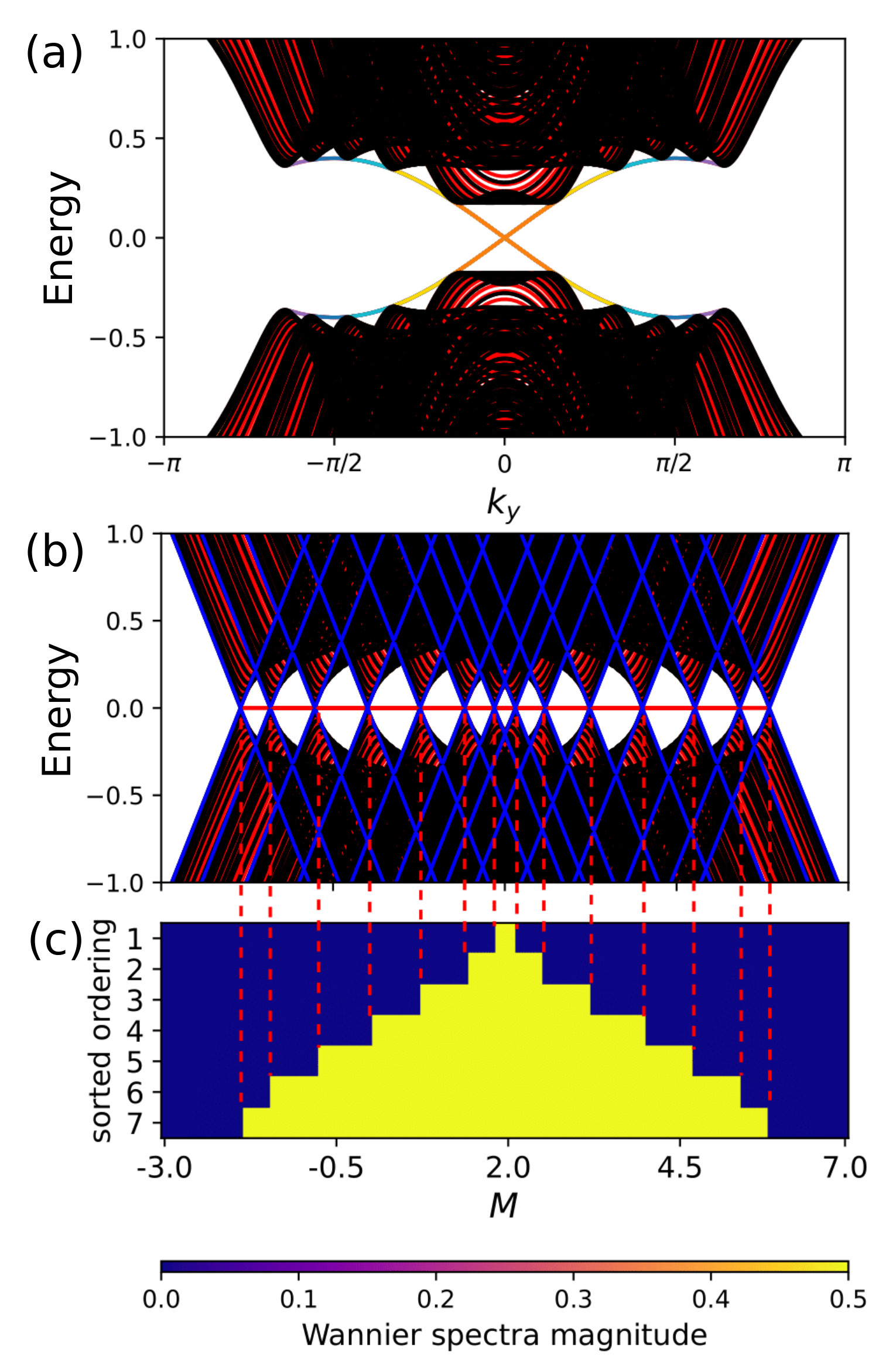}
    \caption{Weyl semimetal thin film along the Weyl axis ($L_z=7$) with x open and large for $M=3$ and $\Delta =0.2t$. Slab spectra vs. $k_y$ in (a) shows there exists multiple chiral edge states labelled by distinct colors (black for pbc along x). The energy vs. M spectra at $k_y=0$ (red) with the same for pbc along x (black) and for pbc along x and y with $k_x,k_y=0,0$ and $k_x,k_y=\pi,0$ (blue). The corresponding Wannier spectra for Wilson loop around $k_x$ with $k_y=0$ shows that indeed the gapless points along M separate phases with different Chern numbers.}
    \label{fig:fWSMzopen}
\end{figure}
\end{center}

For the system we have considered in Eqn. \ref{bulk3d}, the two Weyl nodes are separated in momentum space in $k_z$. The total anomalous Hall conductivity of the Weyl semimetal is determined by the extent of this momentum space separation and the total Chern number of the 2d submanifold of the Brillouin zone for a given value of $k_z$ \cite{burkov_wsmti_2011, hosur_wsmtransport_2013, grushin_lorentz_2012}. However, if the system is restricted to a quasi (3-1)d geometry extending over the xy-plane with normal vector in the z-direction, $k_z$ is no longer a good quantum number. Unlike the surface state hybridization we discussed previously, here we observe hybridization of bulk Weyl nodes of opposite chirality yielding gapped band structure for the thin film. This can be understood by considering a particle in a box analogy along the Weyl axis direction: $k_z$ is restricted to a discrete set of values yielding wavelengths, which satisfy the thin film boundary conditions. Taking the thin film thickness to be $L_z$, we now have the allowed momenta values $k_z = \frac{n_z\pi}{L_z+1}$, for $(n_z=1,...,L_z)$~\cite{nguyen_chiralqhe_2021}, which are the standing wave wavelengths in the $z$-direction. In this standing wave basis, the system can be expanded as follows,
\begin{equation}
\begin{split}
H(k_x,k_y,n_z) =& \sum_{n_z=1}^{L_z}((M_n-2t\cos k_x-2t\cos k_y)\sigma^z\\
&+2\Delta\sin k_x\sigma^x+2\Delta\sin k_y\sigma^y)\ket{n_z}\bra{n_z},
\end{split}
\label{WSMblockCI}
\end{equation}
where, we denote $M_n=M-2t\cos\frac{n_z\pi}{L_z+1}$. This expansion shows that the system is now a collection of gapped insulators, which in our case each have the functional form of the QWZ model~\cite{qi2006topological}. The Hall conductivities of these Chern insulators in the block diagonal representation---and their total value when considered together---determine the total Hall conductivity of the thin film Weyl semimetal. The correlation between the Weyl node separation in momentum space and the anomalous Hall effect still exists, since it is possible to change the topological nature of each individual block Chern insulator by varying the value of $M$, which adjusts the Weyl node separation in Eqn.~\ref{bulk3d}. However, the total Hall conductivity is now quantized, with a maximum value determined by the Chern numbers of the individual Chern insulators and the thickness of the thin film, which determines the number of these effective Chern insulators. We have shown this behaviour numerically in Fig.~\ref{fig:fWSMzopen}(a) where the different colors represent different Chern insulators and the Wannier spectra in Fig.~\ref{fig:fWSMzopen}(c) compared to Fig.\ref{fig:fWSMzopen}(b) which also illustrates the higher Chern numbers from the ordered Wannier spectra eigenvalues.\\
The quasi (3-1)d Weyl semimetal in this specific thin film regime is ostensibly similar to a 2d quantum anomalous Hall insulator. However, we can differentiate between the finite-size topological phase and a quantum anomalous Hall insulator by probing the system with an external magnetic field, as discussed in \ref{fWSMthinz}.

\begin{figure*}[tb!]
    \centering
    \includegraphics[width=0.95\textwidth]{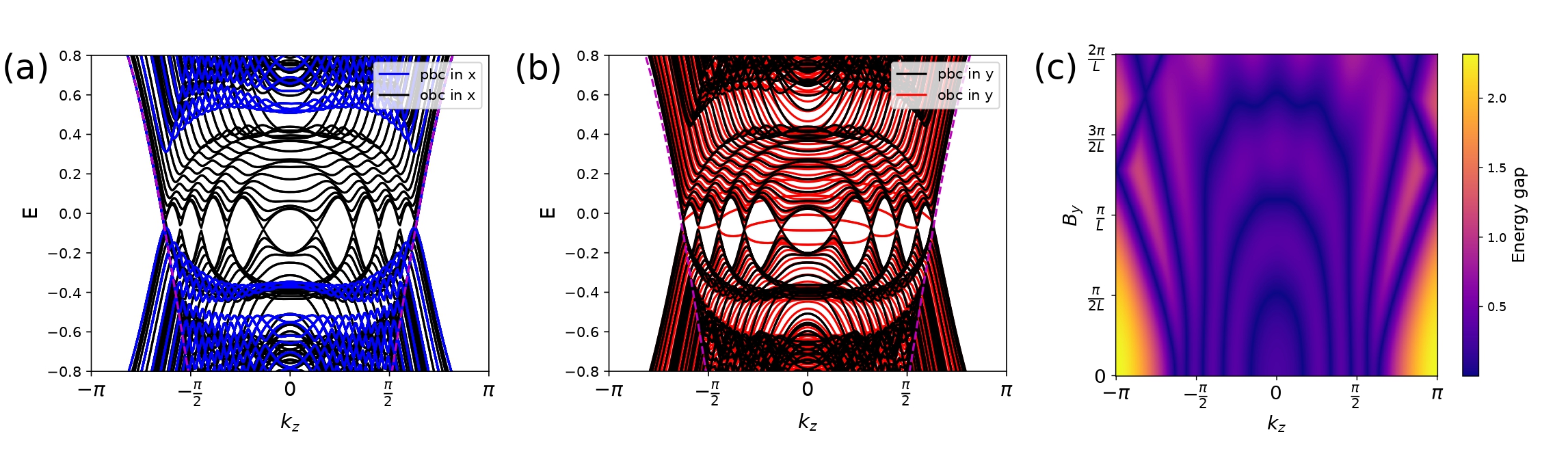}
    \caption{Effect of magnetic field $B_z\hat{z}$ and $B_y\hat{y}$ on FST-WSM with thin film along x ($L_x=7$) for $M=3$ and $\Delta =0.2t$. (a)Spectra vs. $k_z$ for applied magnetic field parallel to the Weyl axis (z) for both periodic(blue) and open(black) boundary conditions along x and periodic boundaries along y. The analytical expression for the chiral Landau level has been marked with dotted magenta line. (b) Spectra vs. $k_z$ for applied magnetic field along z for open boundaries along x and periodic boundaries along y(black) and open boundaries along y(red). The flat bands between inversion symmetric SWNs respond to the magnetic field and are perturbed. (c) Energy gap of lowest lying bands vs. $k_z$ for applied magnetic field perpendicular to the Weyl axis along y, $B_y\hat{y}$, shows variation in separation of pairwise 2d SWNs and subsequent annihilation at zero TRIM point as a function of $B_y$.}
    \label{fig:fWSMBzBy}
\end{figure*}

\section{Effect of external magnetic field on finite-size topological phases from Weyl semimetals}\label{fWSMmagfield}
Numerous studies \cite{huang2015observation,zyuzin2012topological,parameswaran2014chiral,potter2014quantum}, report on distinctive response signatures of 3d Weyl semimetals reflecting their topological nature, with the most notable being the chiral anomaly. The chiral anomaly occurs when external electric and magnetic fields are applied in parallel, which induces pumping of electrons between Weyl nodes of opposite chirality separated in momentum space, thus breaking chiral charge conservation. This axial current makes Weyl semimetals more conductive in magnetic fields oriented parallel to the electric field \cite{nielsen1983anomaly,huang2015observation,zyuzin2012topological,son2013chiral, jia_weyl_2016}. It is possible to measure this response from a peak in the longitudinal magnetoconductivity as a function of the angle between magnetic field and the electric field.

FSTs in thin films of systems, which harbor the 3d WSM for system size thermodynamically large in three spatial directions, exhibit co-existing response signatures naively very similar to both those of 3d WSMs and also 2d topological states, such as 2d WSMs and the QAHI, as we show in this section. We consider two subsections based on the two thin film geometries we discussed in Sec.~\ref{q2dWSM}, and here, we consider three separate cases of external applied magnetic field, in which the field is applied in each of three different directions relative to the Weyl axis and the plane containing the SWNs. We also consider some specific cases for the direction of applied magnetic field which probes aspects of the chiral anomaly or control over SWN-based responses.

\subsection{Effect of magnetic field on FST with thin film along x(perpendicular to Weyl axis)}\label{fWSMxmag}
\subsubsection{Magnetic field parallel to the Weyl axis(along z)}

We must consider separately the effect of external magnetic field applied parallel to the direction in which there is separation between underlying 3d WNs on the bulk 3d WNs and the 2d SWNs, for OBCs and small system size in the x direction (thin film lying in the yz plane). In the 3d bulk, we expect the spectrum to break down into Landau levels with a chiral Landau level extending between the two 3d Weyl nodes as shown in blue in Fig.~\ref{fig:fWSMBzBy}(a) for PBCs in all directions. However, introducing finite size effects by opening boundary conditions in the x-direction and reducing system size in this direction, $L_x$, shows, in addition, the presence of the 2d SWNs. The incident magnetic field in this case is an in-plane field with respect to the 2d SWNs. Hence the SWNs on a 2d plane should not interact with this magnetic field and cannot be gapped out as states emerging from surface modes of the 3d WSM. However, the SWNs still rely, indirectly, on the 3d bulk topology and therefore are affected by external fields altering the underlying 3d bulk electronic structure.

We now consider OBCs in the y direction, so that we can probe the effects of magnetic fields on the flat bands between SWNs. We have shown previously that the flat bands---vs. changing $k_z$---gap out in 'even' regimes, which is not a response one would expect if only the surface 2d topology is considered. Consider now the results in Fig.~\ref{fig:fWSMBzBy}(b) comparing the case for thin film open boundaries along x with open (red) and periodic (black) boundaries along y where system size is large. We observe the initial flat bands, and their gapped out counterparts, have curvature opposite to that for the bulk chiral Landau level (magenta, dotted). Each SWN pair related by inversion harbors one such ``chiral'' level.

We present the still gapless SWNs along with the bulk chiral Landau levels(in blue) in Fig.~\ref{fig:fWSMBzBy}(a). Therefore, the 2d SWNs contribute to the net longitudinal conductivity, $\sigma_{zz}$, which also includes a contribution from the chiral Landau level of the underlying 3d bulk. To demonstrate this extra contribution  in numerics, we first discuss the effect of external magnetic field applied parallel to the y-axis---perpendicular to the 3d Weyl axis but in the plane of the thin film (yz plane).

\subsubsection{Magnetic perpendicular to Weyl axis but in-plane to 2d SWNs(along y)}
For a Weyl semimetal characterized by Hamiltonian Eqn.~\ref{bulk3d}, for OBCs in the x-direction and system size in the x-direction, $L_x$, of the order of a few unit cells, we consider an external magnetic field applied in the y-direction in the gauge modifying the z momentum component as $k_z\rightarrow k_z-eB_yx$, where the magnetic field is given as $B_y\hat{y}$. A recent study \cite{abdulla2024pairwise} has shown that any magnetic field perpendicular to the axis of 3d WNs can reduce the distance between 3d WNs of opposite chirality and annihilate them at TRIM points. We must consider both the intra-BZ and inter-BZ separation in k-space between the 3d WNS, as the smaller separation will go to zero first.

We similarly expect the 2d SWNs to shift in $k_z$ as a part of the 3d system. However, the 2d nature of the SWNs imply that an in-plane magnetic field cannot gap the SWNs immediately at finite field strength. Hence they are only shifted in momentum space, and instead annihilate at specific strengths of $B_y$ when two SWNs of a pair are brought to the same location in the BZ. Therefore, it is possible to modify the number of 2d SWNs via an external magnetic field and hence also control their contribution to surface conductance. We have shown this result numerically in Fig.~\ref{fig:fWSMBzBy}(c). We observe that the SWN pairs successively annihilate with increasing field strength, in order of increasing intra-BZ pairwise separation. Furthermore, the path of the outermost 2d SWN intersects that of another SWN of opposite chirality, but this pair does not annihilate. We observe, irrespective of the separation in k-space of 3d WN pairs, that the outermost SWNs does not annihilate at the BZ edge.

\begin{center}
\begin{figure*}[htb!]
\includegraphics[width=0.97\textwidth]{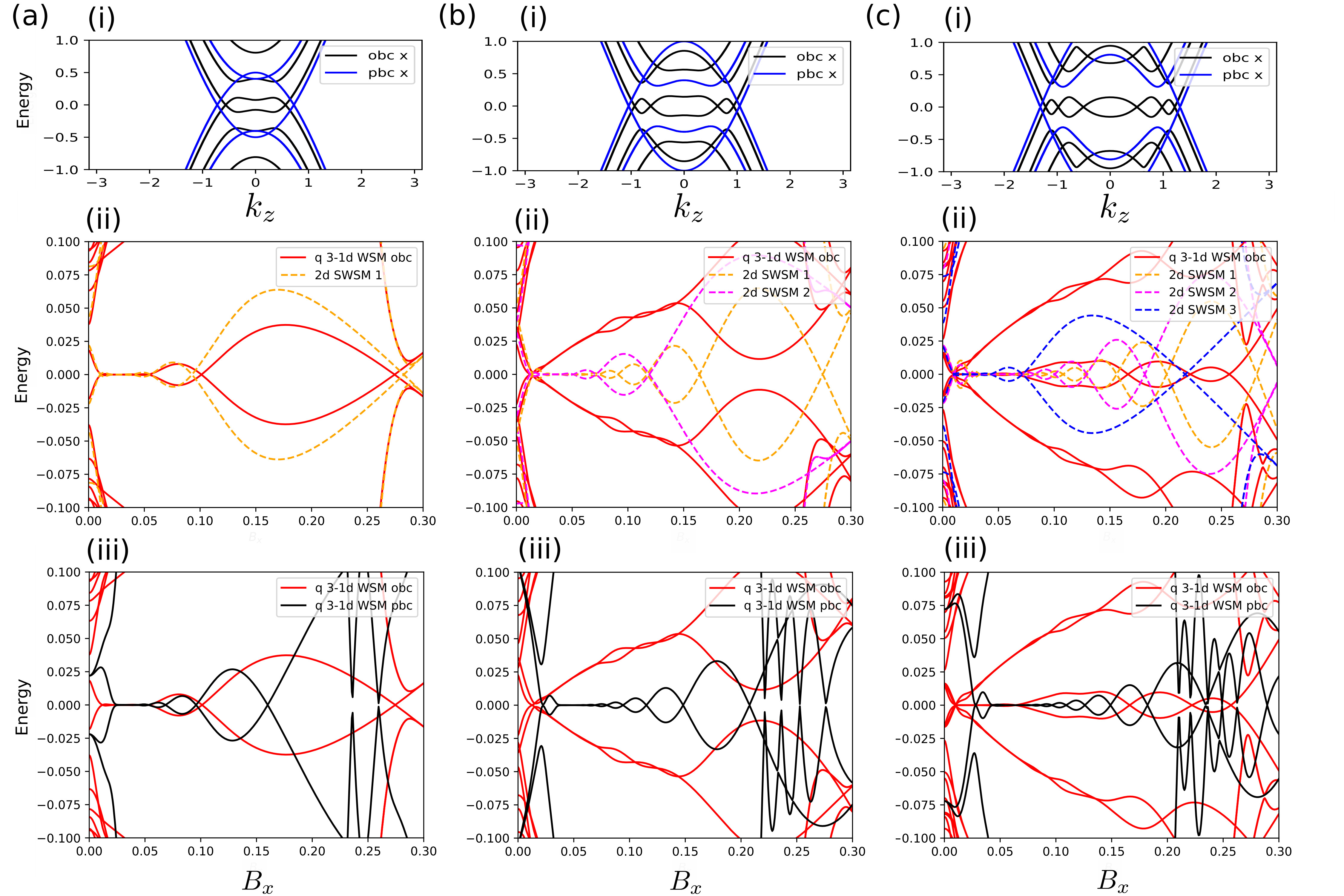}
\caption{Serpentine Landau levels derived from 2d SWNs in the thin film Weyl semimetal with system size $L_x=7$ along x and $\Delta =0.2t$. The three columns (a), (b) and (c) depict mass parameter values, $M=5.5$, $5.0$ and $4.5$ respectively. The upper figures (a)(i), (b)(i) and (c)(i) refer to the slab spectra along $k_z$ at $k_y=0$ for the three mass values respectively with increasing number of pairs of SWNs. The middle figures in (a)(ii), (b)(ii) and (c)(ii) depict the resulting Serpentine Landau levels arising from the 2d SWNs as a function of the out of plane magnetic field $B_x\hat{x}$. As the number of pairs increase, the Landau levels interfere and produce serpentine patterns of increasing complexity. These Landau levels have been compared with standalone 2d SWSMs from \ref{2dWSMs} as dotted lines. The lowest row in (a)(iii), (b)(iii) and (c)(iii) shows the presence of Serpentine LLs in the bulk as shown in Devakul \emph{et al.}\cite{devakul2021serpentine} and the one arising from SWNs in the thin film limit simultaneously for the three mass parameter values respectively. }
\label{fig:serpentineLL}
\end{figure*}
\end{center}

\subsubsection{Magnetic field perpendicular to Weyl axis and out-of-plane to 2d SWNs(along x)}\label{fWSMxBx}
An external magnetic field $B_x\hat{x}$ acts out-of-plane to the 2d SWNs and hence should produce Landau levels by gapping out these degeneracies. In a 3d WSM in the thermodynamic regime, we expect quantum oscillations in the density of states as a function of $B_x^{-1}$ at a particular Fermi energy~\cite{potter_qosc_2014,zhang_qosc_2016}. Essentially, this can be explained semiclassically via the concept of Weyl orbits, which require a continuous Fermi arc along $k_z$. However, finite-size effects reduce the Fermi arc to a discrete set of SWNs, and the original quantum oscillation signature is lost. Instead, it is possible to observe responses arising solely from the 2d SWNs. As shown in a previous study~\cite{devakul2021serpentine}, 2d chiral semimetals exhibit serpentine Landau levels, which are intertwining Landau levels, as the strength of the out-of-plane magnetic field is varied. The closing points of these serpentine Landau levels, referred to as `magic magnetic fields', are responsible for zero Landau level quantum oscillations (ZQOs). The condition proposed in \cite{devakul2021serpentine} to realize ZQOs is equivalent to $t>\Delta$ in our situation, and crossings are expected at certain magnetic field strengths proportional to the 2d SWN separation. The previous study further showed that ZQOs are possible in 3d Weyl semimetals as well for chiral Landau level in the bulk. Here, we show in the thin film limit, there is an added contribution to such serpentine LLs from SWNs.

The serpentine LLs of 2d WSMs are defined for one pair of 2d Weyl nodes, so that there is no ambiguity as to which LLs intertwine. However, in the thin film regime, the system can host multiple pairs of SWNs and there must be two interwining LLs from each pair of SWNs which are part of one 2d SWSM, which are characterized by a distinct SWN separation in the BZ. We provide the Hamiltonian for all the 2d SWSMs with distinct node separation contributing to serpentine Landau levels, derived from Eqn.~\ref{fwsmcrit} with distinct node separations,

\begin{equation}
\begin{split}
H_{2d}(k_y,k_z) = (M_n-2t\cos k_y-2t\cos k_z)\sigma^z+2\Delta\sin k_y\sigma^y,\\
M_n = M-2\sqrt{t^2-\Delta^2}\cos\frac{n\pi}{L_x+1}, \quad (n=1,...,L_x).
\end{split}
\label{2dWSMs}
\end{equation}

Based on \cite{devakul2021serpentine}, the number of `magic magnetic fields' is expected to vary with the separation between the SWNs of a given SWSM. We have shown previously in Sec.~\ref{fWSMthinx} that the SWSMs interfere, as in the case with flat band overlaps. Similarly, a pair of serpentine LLs arising from one such SWSM defined in Eqn.~\ref{2dWSMs}, must interfere with another such pair of serpentine LLs, so that we get multiple pairs of interwining LLs along $B_x$ separated in energy. We show numerically in fig.~\ref{fig:serpentineLL}(a) that one pair of SWNs produces only one pair of serpentine LLs, which is a purely 2d response. As we add another pair of SWNs with a different separation in reciprocal space by tuning $M$, numerics in fig.~\ref{fig:serpentineLL}(b) shows that we get two pairs of interwining LLs separated equally in energy around zero. Similarly, adding another pair of SWNs, with yet another separation in reciprocal space, by tuning $M$ in Fig.~\ref{fig:serpentineLL}(c), yields two pairs of interwining LLs separated equally in energy around zero and one pair of interwining LL at zero energy.

The distribution of the serpentine LLs corresponds to each pair of SWNs producing serpentine LLs near zero energy. Serpentine LLs originating from distinct SWN pairs then hybridize to form bonding-antibonding-nonbonding-type levels. Serpentine LLs for each individual 2d WN pair from Eqn.~\ref{2dWSMs} are shown as dashed lines in Fig.~\ref{fig:serpentineLL}(a)(ii), (b)(ii) and (c)(ii). Comparison between the response due to SWSMs vs. strictly 2d WSMs shows that the responses are identical when only one pair of SWNs exists for the SWSM. In this case of a SWSM with single SWN pair, it is therefore possible to mistake a finite-size topological phase as a strictly 2d topological phase. One can distinguish between these topological states, however, by examining the response---or lack thereof---to other magnetic fields, with which a strictly 2d system should not be able to couple. Of course, this apparent 2d limit is not exact when more SWN pairs are present and the number of intertwining LLs quantify the number of SWN pairs in the system.

There are therefore two sources of serpentine Landau levels -- (i) from oscillations in the lowest Landau level in the bulk, as shown in Devakul~\emph{et al.}~\cite{devakul2021serpentine} and (ii) from SWSMs in the FST-WSM, which we have shown in the last few paragraphs. The former causes oscillations in the transverse conductance originating from the lowest Landau levels, as a function of $B_x$. The latter is responsible for oscillations in the longitudinal conductance through the Fermi arcs, as $B_x$ is varied. We  therefore get two types of magic magnetic fields -- bulk magic fields for closing of the bulk LLs and surface magic fields for closing of surface LLs at zero energy. Transport along the bulk chiral LL can only happen at bulk magic magnetic fields at high field strengths. The electron cannot reach the other surface along the chiral LL elsewhere at high fields, implying there exists a length scale proportional to the gap between the chiral LL restricting the transverse motion along x. However, this implies that, by reducing the system size along x, transverse motion becomes possible when the length scale is of the order of $L_x$. This corresponds to the peaks of transverse conductance broadening around the magic magnetic fields. In general, the surface magic fields and the bulk magic fields are disjoint. However a thin film along x promotes the creation of SWNs and hence surface magic fields and the broadening of the possibility of transverse conductance around magic fields. Combined, this leads to regimes where closed loops created from surface states and chiral LLs become possible.

In the next two subsections, we combine our results for uniaxial magnetic fields and consider the effect of magnetic field on a plane. Investigating these configurations enables us to observe chiral anomaly like phenomena in the large system size WSM or 3d quantum Hall effect due to closed loops of Fermi arcs. In thin films, we show that these particular response signatures of the 3d WSM persist with added contributions from the SWNs. It is furthermore possible to control the 2d response via external fields coupling to 3d bulk which can pairwise annihilate SWNs and in turn reduce their effect. Therefore, it is possible to tune response signatures of the FSTs, between response signatures typically associated with strictly 2d WSMs and 3d WSMs in the thin film limit.

\begin{center}
\begin{figure*}[htb!]
    \centering
    \includegraphics[width=0.97\textwidth]{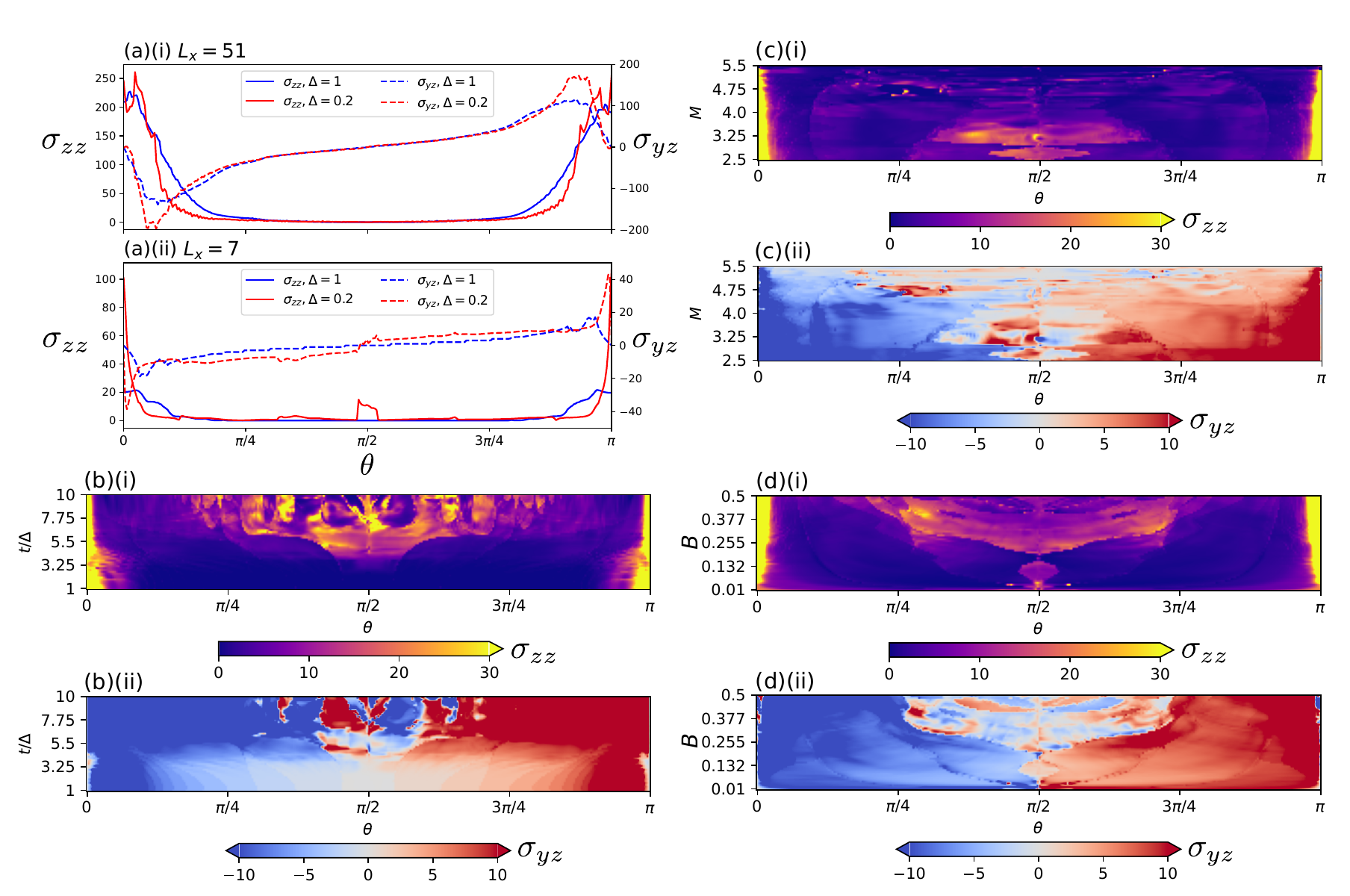}
    \caption{The longitudinal ($\sigma_{zz}$) and transverse ($\sigma_{yz}$) magnetoconductivity (in units of $\frac{e^2}{h}$) for the thin film Weyl semimetal as a function of the angle $\theta$ between the potential difference along z and the applied magnetic field in the zx plane to investigate chiral anomaly based phenomena is studied with respect to various parameters. Figure (a)(i) shows the expected magnetoconductivity for large system size ($L_x=51$) for $\Delta =t$ and $\Delta=0.2 t$ and then for small system size ($L_x=7$) for $\Delta = t$ and $\Delta =0.2t$ in (a)(ii) for magnetic field magnitude $B=0.15$. Subfigure (b)(i) and (ii) shows the variation of $\sigma_{zz}$ and $\sigma_{yz}$ respectively as the parameter $\frac{t}{\Delta}$ is tuned for the FST-WSM ($L_x=7$) at $B=0.15$ and $M=3$. Finite size effects become prominent in the form of SWNs after a threshold value of $\frac{t}{\Delta}$ as shown. Subfigure c(i) and (ii) reveals the the trend of $\sigma_{zz}$ and $\sigma_{yz}$ respectively as a function of the bulk inter-nodal distance between Weyl points at $\Delta=0.2t$ and $B=0.15$. Finally, subfigure (d)(i) and (ii) show the profile of $\sigma_{zz}$ and $\sigma_{yz}$ as a function of magnetic field strength $B$ at $M=3$ and $\Delta=0.2t$ and system size, $L_x=7$. Comparing with (a)(i), significant deviation in the magnetoconductivity profiles take place in the vicinity of $\theta=\frac{\pi}{2}$ which illustrates the importance of SWNs and subsequent serpentine Landau levels derived from it. Subfigures (a), (b), (c) and (d) have been studied at energy $E=0$.}
    \label{fig:chiralanomaly}
\end{figure*}
\end{center}

\subsubsection{Magnetic field in the zx plane -- Chiral anomaly}\label{fWSMchiralanomalysec}
Our discussion about the chiral anomaly in the beginning of Sec.~\ref{fWSMmagfield} discusses the angular dependence of the magnetic field with respect to the potential difference. Basically, in a 3d WSM, for potential difference along the Weyl axis, application of external magnetic field at finite angle $\theta$ with respect to the Weyl axis yields variation in chiral current $\mathbf{j}_5$ between two Weyl nodes in a pair, with change in angle $\theta$, as $\mathbf{j}_5\propto \mathbf{E}\cdot\mathbf{B}$, where $\mathbf{E}$ and $\mathbf{B}$ are the applied electric and magnetic fields, respectively ~\cite{jia_weyl_2016, zhang2016signatures, nielsen1983anomaly, son2013chiral, zyuzin2012topological}. Therefore, we expect peaks in the longitudinal conductivity in the z direction, $\sigma_{zz}$, when the orientation of the external magnetic field is varied in, for instance, the zx plane as $\mathbf{B}=(B\sin\theta,0,B\cos\theta)$. We expect two peaks in $\sigma_{zz}$, at $\theta=0$ and $\pi$. The Hall conductivity $\sigma_{yz}$ also shows a step-like behaviour near $\theta=\frac{\pi}{2}$ as a result of flat Landau levels along $k_z$ which has been studied previously as 3d quantum Hall effect \cite{wang20173dqhe}.

We calculate the magnetoconductivity of the FST-WSM with quasi (3-1)d geometry numerically by constructing the six terminal Hall bar geometry using the KWANT python package for quantum transport calculations\cite{kwantpaper}. Details of the construction have been mentioned in the SM\cite{SuppMat}.

If we choose a Weyl semimetal with small system size in the x direction, corresponding to a thin film over the yz plane, and the Weyl axis is along z, we should expect some modification of the chiral anomaly as realized when the system is thermodynamically-large in the x-direction. Tuning the magnetic field in the zx plane, we showed in Sec.~\ref{fWSMxBx} that the SWNs gap out due to magnetic field along x, and we try to analyze the effect of gap closing in serpentine LLs on the chiral anomaly. We show the usual profile for the longitudinal and transverse conductivity in the large system size limit ($L_x=51$, $L_y=51$, $L_z=51$) as a function of the angle $\theta$ of the magnetic field with respect to the applied electric field along z in Fig.~\ref{fig:chiralanomaly}(a)(i). We observe the expected peaks of longitudinal conductivity at $\theta=0$ and $\pi$ and signatures of the 3d quantum Hall effect~\cite{wang20173dqhe} in the profile for the transverse conductivity. We show both cases for $\Delta=t$ and $t\Delta \neq t$ at $B=0.15$ and $M=3$ where the qualitative picture does not change in the large system size limit. We next show a similar picture for identical parameters but small system size along x, $L_x=7$, while keeping the size along other axes unchanged in Fig.~\ref{fig:chiralanomaly}(a)(ii. For $\Delta=0.2t$, we observe a deviation at $\theta=\frac{\pi}{2}$ from the usual profile. This proves that the deviation is only a result of finite size effects resulting from hybridization of surface states, or alternatively the presence of SWNs in the FST-WSM. Furthermore, the longitudinal conductivity for $\Delta=0.2 t$ at $\theta=0$ and $\pi$ appears much larger than the case with $\Delta =t$ which can also be ascried to the presence of SWNs which provide additional conducting channels.

Next, Fig.~\ref{fig:chiralanomaly}(b)(i) and (ii) shows that the magnetoconductivity ($\sigma_{zz}$ and $\sigma_{yz}$ respectively) for a thin film system ($L_x=7$), with given $M=3$ and $B=0.15$, deviates from the profile in Fig.~\ref{fig:chiralanomaly}(a)(i) after a threshold value of $\frac{t}{\Delta}$. We observe the deviation at $\theta=\frac{\pi}{2}$ appears only when the hybridization between the surface states is large enough. Further, in Fig.~\ref{fig:chiralanomaly}(c)(i) and (ii), we show the magnetoconductivity profile as a function of $M$, which quantifies the distance between Weyl nodes along the Weyl axis in momentum space. The value of $M$ alternatively quantifies the number of SWNs in the system. However, we expect the trend we observe to result from the SWN based serpentine LL which has a gap closing near or at the given magnetic field $(B=0.15)$. The profile shows at least three regions of different sizes which can be attributed to the motion of gapless points in serpentine LLs as $M$ is varied.
Last but not the least, in Fig.~\ref{fig:chiralanomaly}(d)(i) and (ii), we study the effect of magnetic field magnitude on the magnetoconductivity near $\theta=\frac{\pi}{2}$ for given $M=3$ and $\Delta=0.2t$ for system size $L_x=7$. The constriction in the profile near $B=0.17$ directly correspond to the intersection in the serpentine LLs derived from the SWNs. One can refer to this as the ZQO in the conductivity derived from SWNs. The deviation increases at non-magic magnetic fields. We provide a short explanation for this behaviour in the SM~\cite{SuppMat}.
We have thus shown a qualitative picture of the effect of SWNs in modifying the chiral anomaly behaviour in the FST-WSM. From a first impression, one should be able to find intersection points in the serpentine LL by transport measurements on the thin film system. We will consider at a more quantitative picture for the modified chiral anomaly in a future study.

\begin{center}
\begin{figure}[htb!]
    \centering
    \includegraphics[width=0.95\textwidth]{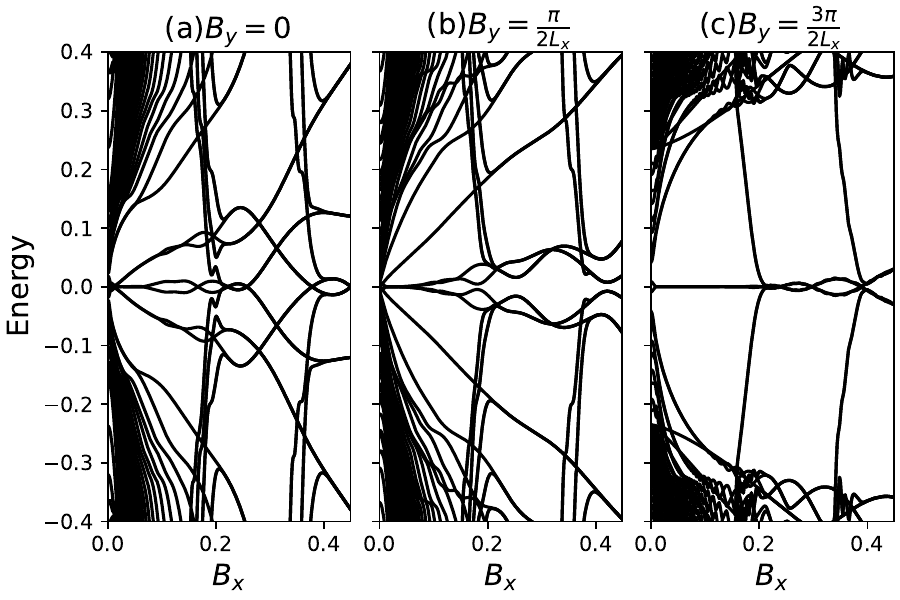}
    \caption{We show the serpentine Landau levels as a function of magnetic field $B_x\hat{x}$ for three different values of $B_y\hat{y}$ in a FST-WSM ($L_x=7$) with $\Delta =0.2t$. We start with a system at $M=4.5$ with three pairs of SWNs and end up eventually with one pair of SWN by increasing the value of magnetic field along y axis. The profiles of serpentine Landau levels in (a) for $B_y=0$, (b) for $B_y=\frac{\pi}{2L_x}$ and (c) for $B_y=\frac{3\pi}{2L_x}$ are directly comparable to Fig.~\ref{fig:serpentineLL} (a), (b) and (c) respectively. }
    \label{fig:zqomanipulation}
\end{figure}
\end{center}

\subsubsection{Magnetic field in the xy plane -- control of ZQOs}
Out of plane magnetic field along the x-axis yields complex dependence of the spectrum on magnetic field strength, which can be understood in terms of contributions from multiple serpentine Landau levels --each associated with a pair of SWNs. In this section, we probe how the profile of spectrum vs. magnetic field strength depends on the number of SWNs. We have shown, in previous sections, that magnetic field along the y-axis manipulates the number of SWNs in the finite size Weyl semimetal. Hence, just by considering an external magnetic field in the xy plane, it is possible to reduce the number of SWNs emerging from finite size effects in a 3d bulk system, down to a single pair. We demonstrate this control of the LL structure in Fig.~\ref{fig:zqomanipulation}. Here, we show LL structure vs. increasing magnetic field component in the x-direction, $B_x$, for three different values of magnetic field component in the y-direction, $B_y$, of (a) $B_y=0$, (b) for $B_y=\frac{\pi}{2L_x}$ and (c) for $B_y=\frac{3\pi}{2L_x}$. The number of SWNs is reduced with increasing $B_y$, reducing the number of serpentine Landau levels and complexity of the overall LL structure vs. $B_x$. For a single pair of SWNs at large $B_y$, the LL structure is similar to that of a 2d WSM. Comparing Fig.~\ref{fig:zqomanipulation} with Fig.~\ref{fig:serpentineLL}, where we consider different mass values for a given system in quasi (3-1)d geometry, we see that tuning $B_y$ achieves the same effect as changing the mass, and serves as a useful probe distinguishing strictly 2d WSM topology from topology of SWNs in a FST state.

\subsection{Effect of magnetic field on FST with thin film along z(parallel to Weyl axis)}\label{fWSMzmag}

We now consider effect of magnetic field on the FST without SWNs, corresponding to the normal vector of the thin film of 3d WSM material parallel to the 3d WN axis (OBC in z, system size in z, $L_z$, of the order of a few unit cells). In this case, the thin film is insulating in the quasi-(3-1)d bulk, and we further characterize the potential FST states in terms of 2d spectral flow, which can correspond to non-trivial Chern number. There is therefore the potential for this type of FST descending from the 3d WSM to be mistaken for a QAHI.

\begin{center}
\begin{figure}[htb!]
    \centering
    \includegraphics[width=0.95\textwidth]{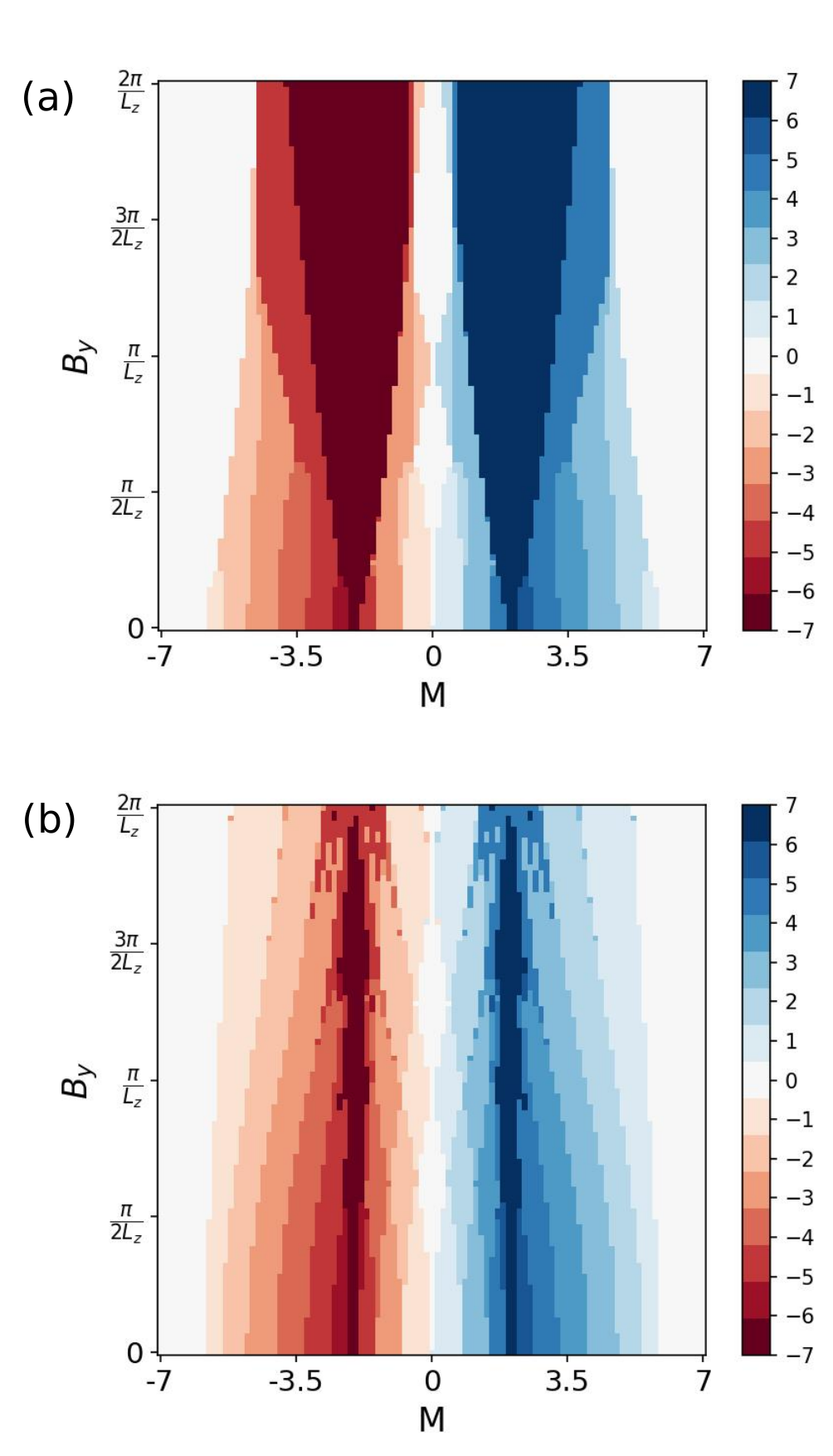}
    \caption{Total Chern number of the  thin film of 3d WSM material, for OBCs in direction of Weyl axis ($L_z=7$) vs. mass term $M$ and the magnitude of the external magnetic field oriented in the y-direction, $B_y\hat{y}$, which is perpendicular to the Weyl axis. We show the $\Delta =t$ case in (a) and for $\Delta =0.2t$ in (b).}
    \label{fig:ChernH7vsMandBy}
\end{figure}
\end{center}

We find external magnetic fields to be useful probes distinguishing these FSTs of the WSM from a QAHI. We first note that, when we open BCs in the direction of the Weyl axis, the Hamiltonian of the thin film system can be block-diagonalized, with some of the blocks corresponding to 2d quantum anomalous Hall insulators in the topological phase as shown in Sec.~\ref{fWSMthinz}. The number of blocks with non-trivial Chern number is determined by separation in k-space of the underlying 3d WNs and system size in the z-direction, $L_z$. Rather than probe the system with an out-of-plane external magnetic field, as might be used to probe a QAHI, then, we probe the system with an external magnetic field that changes separation of the 3d Weyl nodes. We are more interested in the effect of magnetic field perpendicular to the Weyl axis. Results for this case are shown in Fig.~\ref{fig:ChernH7vsMandBy}.

As we know \cite{abdulla2024pairwise} and have shown in the previous subsection, it is possible to modify the distance between the bulk Weyl nodes using, say $B_y\hat{y}$. This further implies that the quantum anomalous Hall effect signature is affected as the number of block diagonalized 2d quantum anomalous Hall insulators in the topologically non-trivial phase is either increased or decreased depending on the initial Weyl node separation, which in our setup depends on the parameter, $M$. We have shown in Fig. \ref{fig:ChernH7vsMandBy}, the magnitude of the quantum anomalous Hall effect varied as a function of $M$ along the x-axis and external magnetic field $B_y\hat{y}$ and have compared the thin film regime for $\Delta =t$ and $\Delta \neq t$ regimes. We notice a marked difference in the trend along the y-axis. Normally we would expect a transition to a quantum anomalous Hall insulator with maximal Chern number determined by the number of layers in the thin film or a topologically trivial insulator depending on the initial Weyl node separation or, equivalently, the value of $M$. However, although there exists such a trend when $\Delta =t$, the total Chern number in the $\Delta <t$ case decreases in magnitude with increasing magnetic field strength for most fixed values of $M$, over the same interval in field strength as considered for the $\Delta =t$ case. For the $\Delta <t$ regime, we expect serpentine Landau levels in the bulk if the system is subjected to external magnetic fields perpendicular to the Weyl axis. We expect this feature to affect the total Chern number and possibly yield the trend observed in Fig.~\ref{fig:ChernH7vsMandBy}(b), which we intend to investigate in future work.

\begin{figure}[htb!]
    \centering
    \includegraphics[width=0.98\textwidth]{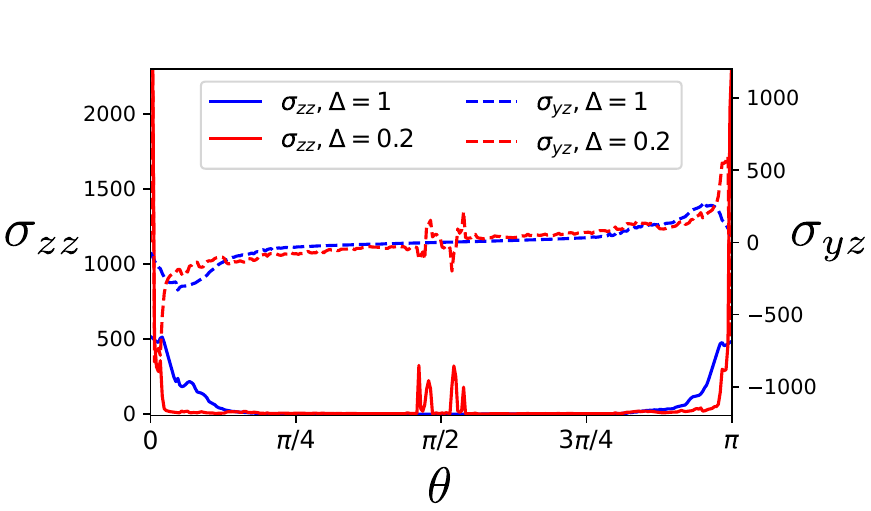}
    \caption{Magnetoconductivity (in units of $\frac{e^2}{h}$) signatures in the WSM with $M=3$ in quasi (3-1)d geometry thinned along Weyl axis($L_z=7$) as a function of the relative angle $\theta$ between the magnetic and electric field. Two cases with $\Delta=t$ (blue) and $\Delta=0.2t$ ($t=1$) are considered. Longitudinal conductivity ($\sigma_{zz}$, solid lines) and transverse conductivity ($\sigma_{yz}$, dotted lines) are compared for the given two cases and show a deviation around $\theta=\frac{\pi}{2}$.}
    \label{fig:WSMz7chiralanomaly}
\end{figure}

For the sake of completion, we also comment on the chiral anomaly behaviour in such an apparent ``2d insulator''. Inspite of the thin film constraint, chiral Landau levels still exist in this system as shown in a previous study~\cite{nguyen_chiralqhe_2021}. We show a small calculation in SM~\cite{SuppMat} which shows that there exist states which still show linear dependence with a magnetic field along z which one should not expect in a normal 2d QAHE insulator. Even with $t>\Delta$, the lowest LL in the system does not undergo any change. Therefore one can still look for chiral anomaly behaviour in terms of longitudinal and transverse conductivity in a six terminal Hall bar geometry as a function of the relative angle between the magnetic field and the electric field. We show the trend in magnetoconductivity as a function of this relative angle $\theta$ in the zx plane (similar setup to Sec.\ref{fWSMchiralanomalysec}) in Fig.~\ref{fig:WSMz7chiralanomaly} for $M=3$ and system size along z, $L_z=7$. Not only does the system show the usual chiral anomaly behaviour, but also one can notice the deviation at $\theta=\frac{\pi}{2}$ between $\Delta=t$ and $\Delta=0.2t$ instances. The second part shows that bulk serpentine Landau levels might indeed play a role even in a quasi (3-1)d system thinned along the Weyl axis direction. These results further cements the importance of looking at insulating heterostructures which may be an ``FST Weyl insulator" (FST-WI) in disguise.

\begin{figure*}[tb!]
    \centering
    \includegraphics[width=0.98\textwidth]{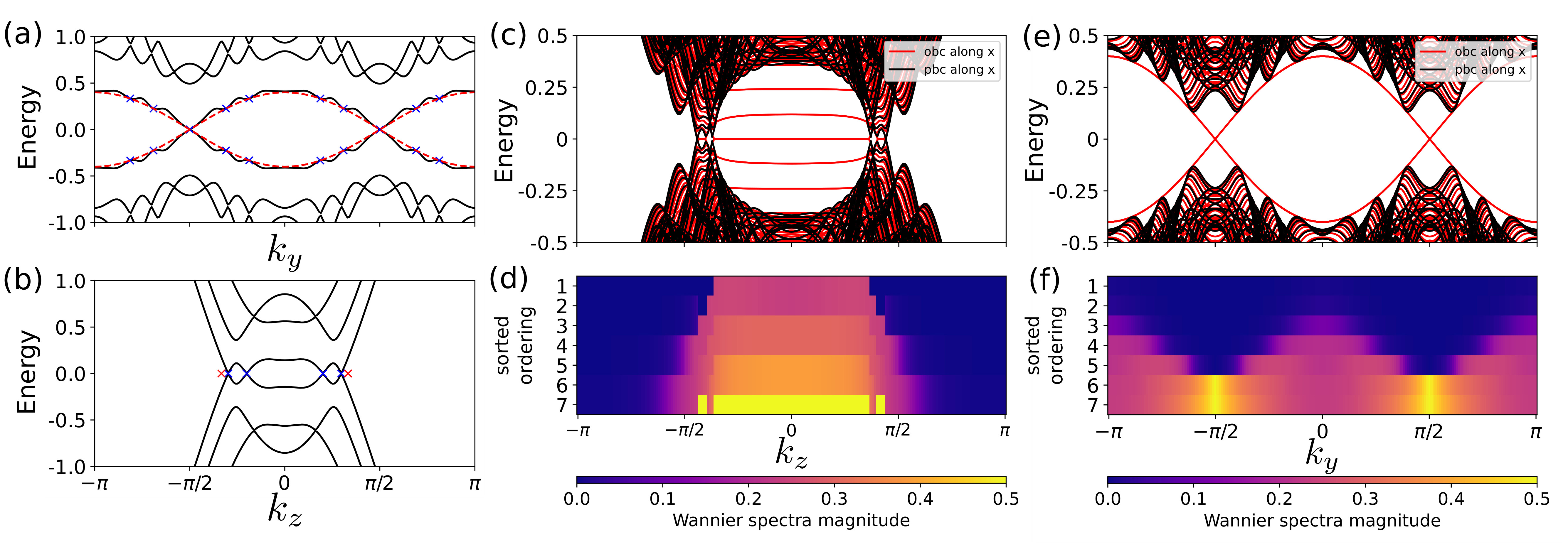}
    \caption{FST for a TRI WSM with quasi (3-1)d geometry by thinning along the x, y and z axes and their corresponding slab spectra and spectral flow for $k_0=\frac{\pi}{3}$ and $\Delta =0.2t$. (a) Slab spectra vs. $k_y$ for thin film along x($L_x=7$) at the $k_z$ describing the inner SWN, blue crosses show the analytically calculated energies from Eqn.~\ref{trWSMcrit}  and the dotted red line shows the slab spectra in large lattice limit. (b)Slab spectra vs. $k_z$ for thin film along x($L_x=7$) for $k_y=\frac{\pi}{2}$, blue crosses show the analytically calculated positions for SWNs along the $k_z$ axis and the two red crosses show the position of the bulk 3d Weyl nodes. There exists an identical spectra at the other time reversal partner Weyl nodes. (c) Slab spectra vs. $k_z$ for thin film along y $(L_y=7)$ with large lattice limit and open boundary along x($L_x=51$)(red) and periodic boundary along x(black) with Wannier spectra for Wilson loop about $k_x$ in (d). Flat bands are denoted by $\pi$ phase in Wannier spectra. (e) Slab spectra vs. $k_y$ for thin film along z($L_z=7$) and large lattice with open boundary(red) and periodic boundary(black) along x($L_x=51$). (f) Wannier spectra for spectral flow around $k_x$ shows $\pi$ phase only at $k_y=\pm\frac{\pi}{2}$ where the bulk system becomes chiral and shows multiple overlapping 1d embedded edge modes.}
    \label{fig:TRWSMslab}
\end{figure*}

\section{Finite size topological phases in quasi (3-1)d geometry of 3d WSM material with time reversal symmetry}\label{TRIWSM}
A Weyl semimetal can either preserve $\mathcal{I}$ symmetry or $\mathcal{T}$ symmetry but not both. In the previous few sections, we have analyzed how finite size effects modify the bands in the former case. In this section, we instead consider time reversal invariant (TRI) WSMs, which must have at least two pairs of Weyl nodes, as each pair of Weyl nodes of opposite chirality must itself have a time reversed partner pair of Weyl nodes. This section is motivated by the experimental significance of TRI 3d WSMs, including, for instance, pump-probe ARPES signatures of minimal 3d WN configuration in materials like TaIrTe$_4$~\cite{belopolski2017signature}, as well as experimental study of particular TRI 3d WSM materials in thin films, we investigate the potential of TRI 3d WSM materials for realizing FST phases in such quasi (3-1)d geometries. Given Van der Waals materials of the form MM$^\prime$Te$_4$, which are TRI WSMs in the 3d bulk, are also experimentally-studied in thin film (quasi (3-1)d) geometries ~\cite{liu_2017_vdW}, such FSTs may be important in understanding experimental works on these materials, in particular. Ab initio results furthermore indicate that these materials may transition between harboring Type-I and Type-II 3d WNs by varying elastic strain in the system, which affects the interlayer coupling strength.

We therefore consider a minimal model for a TRI 3d WSM and investigate whether finite-size effects yield FST phases for this model in quasi (3-1)d geometries. Our Hamiltonian is similar to that proposed previously for a TRI WSM with four Weyl nodes \cite{huang_spectroscopic_2016},
\begin{equation}\label{trwsm}
\begin{split}
\mathcal{H}(\boldsymbol{k}) =& (2-2t\cos(k_x)-2t\cos^2(k_y)\\
&-2t(\cos(k_z)-\cos(k_0)))\sigma^z+2\Delta\sin(k_x)\sigma^x\\
&+2\Delta\cos(k_y)\sigma^y
\end{split}
\end{equation}
The above system harbors Weyl nodes at the momenta, $(0,\pm\frac{\pi}{2},\pm k_0)$ and is equivalent to the general continuum model for TRI WSM proposed in Polatkan~\emph{et al.}~\cite{polatkan2020magopt}. The Fermi arcs connect Weyl nodes at the same $k_z$ along $k_y$. Comparing with the TRB WSM of previous sections, finite size effects should become prominent when $t>\Delta$. However, unlike the previously-considered Hamiltonian for a TRB WSM in Eqn. \ref{bulk3d}, the Hamiltonian Eqn.~\ref{trwsm} is anisotropic and therefore finite-size effects in quasi (3-1)d geometries vary depending on whether the thin film lies in the xy, yz, or zx plane.

We first investigate FST realization from hybridization of Fermi arc surface states. Like the TRB WSM, there exists a stacking argument for the TRI WSM for Hamiltonian Eqn.~\ref{trwsm}. In this particular case, the insulator which undergoes stacking is 2d with $\mathcal{T}^2=+1$ but carries an embedded 1d chiral topologically nontrivial insulator~\cite{xie2021trsymm}. A non-trivial topology is then only observed at $k_y=\frac{\pi}{2}$ and hence we are essentially considering the 2d BZ along $k_x$ and $k_z$.

Therefore, following the analysis in \cite{xie2021trsymm}, we can characterize topology of the 3d bulk in terms of Wannier spectra computed for 2d embedded chiral metal along $k_x$ and $k_z$ for specific values $k_y=\pm\frac{\pi}{2}$. \\

\subsection{Slab spectra in thin film TRI WSM}\label{trwsmslabsec}
We consider finite size effects in quasi (3-1)d geometries, first we consider a thin film along the x-direction. We focus on the regime $t<\Delta$ to realize regions in momentum space with finite edge state hybridization gap, separated by a discrete set of points in momentum space, where the spectrum is gapless, corresponding to negligible hybridization of Fermi arc surface states. The slab spectra along $k_y$ and $k_z$ is similar to the $\mathcal{I}$-symmetric case, except there exists double the number of SWNs as a result of time reversal symmetry. We have shown the slab spectra in Fig.~\ref{fig:TRWSMslab}(a) and (b). A similar condition as Eqn.~\ref{fwsmcrit} for the existence of SWNs can be written for the model in Eqn.~\ref{trwsm} as,
\begin{equation}\label{trWSMcrit}
\begin{split}
&2-2t\cos^2(k_y)-2t(\cos(k_z)-\cos(k_0))\\
&-2\sqrt{t^2-\Delta^2}\cos\frac{n\pi}{L_x+1}=0,\quad (n=1,...,L_x),
\end{split}
\end{equation}
where the condition implies that the state must have energy $E=2\Delta\cos(k_y)$. We have compared results derived from this formula with numerical slab spectra in Fig~\ref{fig:TRWSMslab}(a) and (b), as pointed out by blue crosses. However, OBCs in the y-direction do not lead to flat bands along $k_z$, since flat bands pairwise annihilate due to the presence of time reversal symmetry and gap out. This leads to a trivial Wannier spectra, computed from Wilson loops computed along $k_y$ for each $k_z$, for the thin film.

We next consider a thin film of 3d TRI WSM material corresponding to small system size, $L_y$ in the y-direction. OBCs in the y-direction do not yield edge states, as discussed previously, therefore we only consider FST realization via hybridization of the 3d WNs for this case. We observe a discrete set of gapless points in the spectrum vs. $k_z$ for PBCs in the x-direction, shown in black in Fig.~\ref{fig:TRWSMslab}(c), bounding otherwise generically gapped spectra. For OBCs in the x-direction and small system size in the y-direction $L_y$, the spectrum vs. $k_z$, also shown in Fig.~\ref{fig:TRWSMslab}, exhibits zero-energy modes bounded by adjacent pairs of gapless points, in correspondence with non-trivial Wannier charge center values in the corresponding quasi-(3-1)d bulk Wannier spectra shown in Fig.~\ref{fig:TRWSMslab}(d).

For PBCs in the x-direction and large system size $L_x$ in the x-direction, but small system size $L_z$ in the z-direction, we also show the spectrum vs. $k_y$ in Fig.~\ref{fig:TRWSMslab}(e) in black, which is generically gapped. OBCs in the x-direction for this geometry, however, yield counterpart spectra shown in Fig.~\ref{fig:TRWSMslab}(e) in red with in-gap modes traversing the bulk gap. Crossings of the in-gap bands are in correspondence with non-trivial Wannier charge center values as shown in  Fig.~\ref{fig:TRWSMslab}(f).

\begin{figure*}[htb!]
    \centering
    \includegraphics[width=0.98\textwidth]{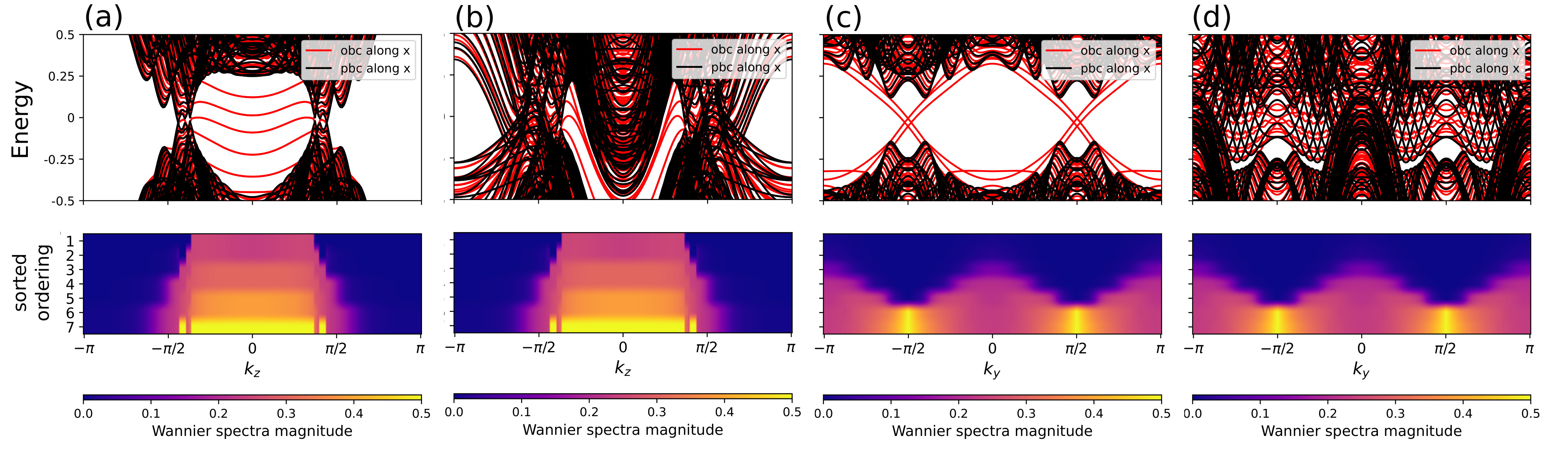}
    \caption{Slab spectra for thin film TRI WSM with finite tilt and corresponding spectral flow for $k_0=\frac{\pi}{3}$ and $\Delta =0.2t$. Slab spectra for TRI WSM thin film along y with tilt magnitude (a) $\gamma=0.2$ and (b) $\gamma=1.2$ and respective Wannier spectra vs. $k_z$.  Slab spectra for TRI WSM thin film along z with tilt magnitude (c) $\gamma=0.2$ and (d) $\gamma=1.2$ and corresponding Wannier spectra vs. $k_y$. Comparing (a) and (b) with Fig.~\ref{fig:TRWSMslab}(c) and (d) and (c) and (d) with Fig.~\ref{fig:TRWSMslab}(e) and (f), we find although the slab spectra is heavily changed, one should still get non-trivial topology in the same regions as the non-tilted case.}
    \label{fig:trwsmtilt}
\end{figure*}

\subsection{Effect of tilting on FSTs in TRI WSM}\label{trwsmtiltsec}
Most TRI WSM materials possess a finite tilting which can be affected via elastic strain. We show that the Wannier spectra indicates non-trivial topology irrespective of the tilting magnitude in thin film TRI WSM systems. We have shown in Fig.~\ref{fig:trwsmtilt}(a) and (b) the slab spectra and corresponding Wannier spectra as a function of $k_z$ in a thin film along y for low and relatively higher value of tilting. Further in Fig.~\ref{fig:trwsmtilt}(c) and (d) we show slab spectra and Wannier spectra for a TRI WSM with thin film along z at low and relatively higher magnitide of tilting. In both cases, the non-trivial topology regime is not affected irrespective of the tilting.

\begin{figure*}[htb!]
    \centering
    \includegraphics[width=0.9\textwidth]{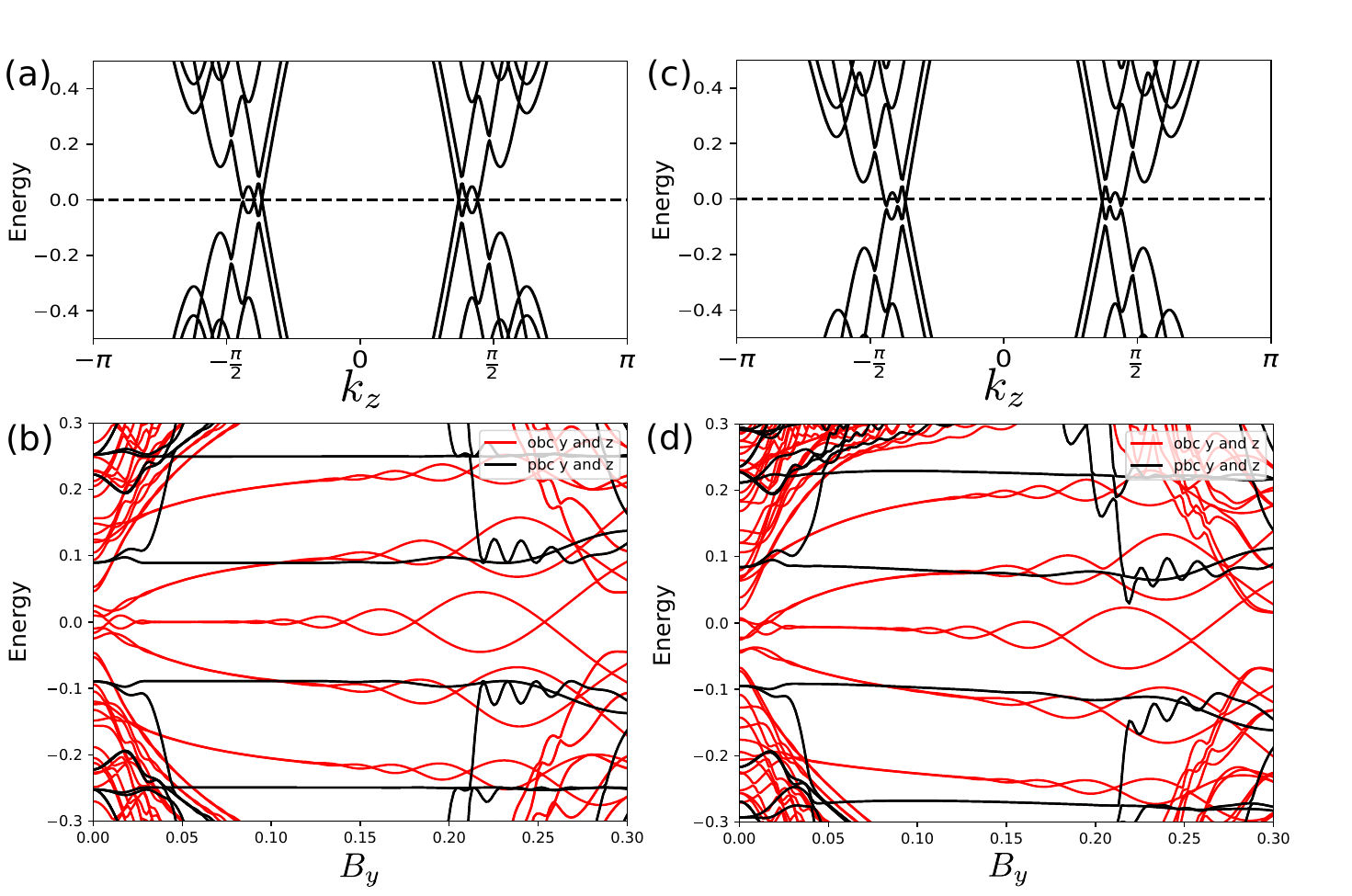}
    \caption{Serpentine Landau levels as a function of magnetic field $B_y$ along the y axis in a TRI WSM thin film along y with system size $L_y=7$. Other parameters include $k_0=\frac{\pi}{3}$ and $\Delta =0.2t$ The slab spectra along $k_z$ for OBC along y is given in (a) for zero tilt, $\gamma$ and (c) for non-zero tilt $\gamma=0.2$. The respective Energy vs. $B_y$ plots showing serpentine Landau levels are shown in (b) for zero tilt and (d) for non-zero tilt.}
    \label{fig:trwsmserpentine}
\end{figure*}

\section{Effect of magnetic field on FSTs in TRI WSM}\label{TRWSMresponse}
Having demonstrated bulk-boundary correspondence in Sec.~\ref{trwsmslabsec} potentially associated with FSTs descending from the 3d TRI WSM, we now employ external magnetic fields to show that signatures of SWNs exist in the form of serpentine LLs similar to the TRB WSM system.

\begin{figure}[htb!]
    \centering
    \includegraphics[width=0.95\textwidth]{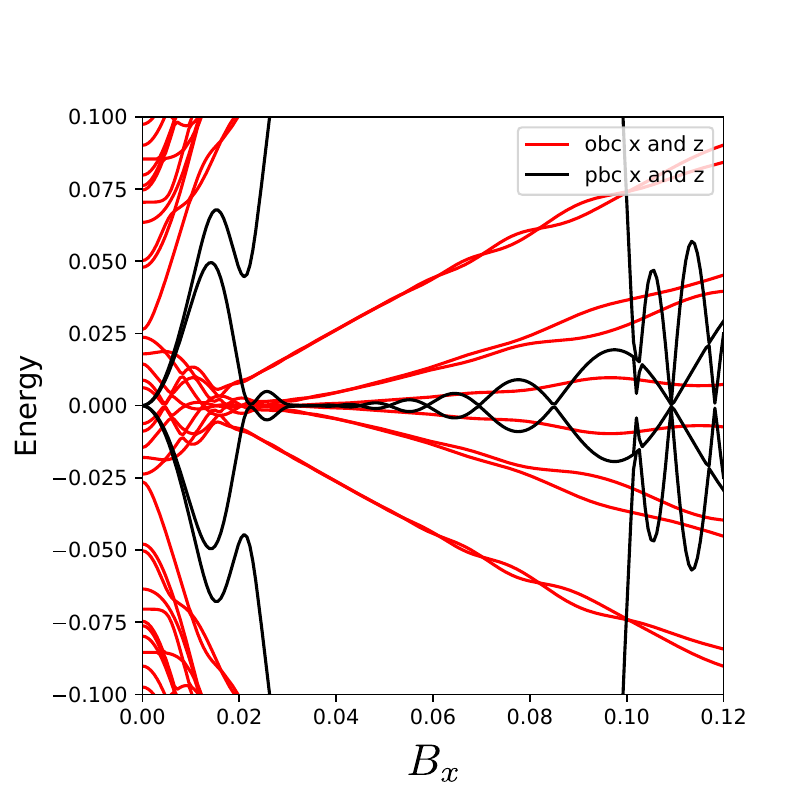}
    \caption{Landau levels for a TRI WSM with system size small along x($L_x=7$) and large along z($L_z=51$) in presence of magnetic field along x. Other parameters include $k_0=\frac{\pi}{3}$ and $\Delta =0.2t$. Black lines represent PBC along both x and z while red lines OBC along both axes. We calculate for $k_y=\frac{\pi}{2}$ and consider $k_0=\frac{2\pi}{3}$ which produces 5 pairs of SWNs along $k_z$ at $k_y=\frac{\pi}{2}$. We find serpentine LLs from the SWNs gap out and there exists no closings at zero energy. However the serpentine LLs from the bulk remain.}
    \label{fig:trwsmxopenBx}
\end{figure}

The TRI WSM with small system size along x shows 2d SWNs at $k_y=\pm\frac{\pi}{2}$ in Fig.~\ref{fig:TRWSMslab}(a) and (b) for $\Delta = 0.2t$. Subjecting the thin film TRI WSM to external magnetic field along x, we find serpentine Landau levels from SWNs at $k_y=\frac{\pi}{2}$ hybridize with its time reversal symmetric partner so that there are no gap closings in serpentine Landau levels at zero energy. However, there exists still serpentine Landau level contribution from the bulk since the bulk is a 2d semimetal at $k_y=\pm\frac{\pi}{2}$. We show this feature in Fig.~\ref{fig:trwsmxopenBx}.

From Fig.~\ref{fig:TRWSMslab}(c), we have found 2d semimetal nodes which behave similar to the 2d SWNs in the TRB WSM. However, these nodes are a result of hybridization of the Weyl nodes along $k_y$. We expect these nodes to be 2d semimetals in the zx plane and investigate the effect of magnetic field applied along the y axis. We find serpentine Landau levels arising from these 2d nodes as shown in Fig.~\ref{fig:trwsmserpentine}. We show that these Landau levels are not present in the bulk as evident from the black colored lines in the plot denoting periodic boundary conditions and are instead a surface effect (red line for OBC). We show the slab spectra and corresponding serpentine Landau levels in Fig.~\ref{fig:trwsmserpentine}(a) and (b) respectively for zero tilt, $\gamma=0$. We also consider a case where the tilting is nonzero, i.e., $\gamma=0.2$ which is reflected in the slab spectra and configuration of the corresponding serpentine Landau levels in Fig.~\ref{fig:trwsmserpentine}(c) and (d).

Therefore the responses we obtain in the thin film along y with magnetic field along y are somewhat opposite to what we get in a thin film along x for magnetic field along x. In both cases we get serpentine Landau levels, but in the former, it arises from the SWNs, while in the latter it is bulk contribution. This can be attributed to the specific structure of the model, Eqn.~\ref{trwsm} we have studied, where depending on which axis we subject to open boundaries, the resulting SWNs may or may not have TR symmetric partners which alternatively results in the 2d semimetal from the bulk to show a complementary effect. Given that Eqn. \ref{trwsm} is a minimal model for a four node TRI WSM, we expect equivalent models to show similar behaviour along some pair of orthogonal axes.

\begin{figure}[htb!]
    \centering
    \includegraphics[width=0.95\textwidth]{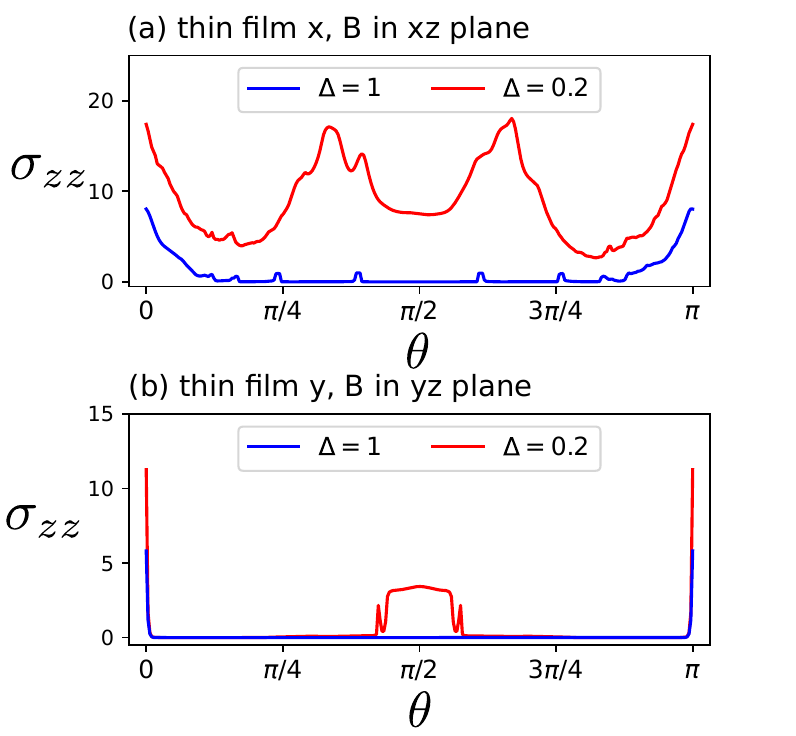}
    \caption{Longitudinal magnetoconductivity($\sigma_{zz}$) in the TRI WSM for thin film along x in (a) and along y in (b) at energy $E=0.01t$. The magnetic field is varied accordingly, in the zx plane for the former and in the yz plane for the latter. The magnetic field strength is $B=0.15$ and the two cases for $\Delta=t$ (blue) and $\Delta=0.2t$ ($t=1$) (red) are compared for $L_x=7$ and $L_y=7$ respectively with the other two axes having 51 lattice sites.}
    \label{fig:trwsmchiralanomaly}
\end{figure}

Lastly we consider the chiral anomaly behaviour in the TRI WSM by probing the longitudinal magnetoconductivity along z. The choice of the plane in which the magnetic field lies, with its orientation varied in relative angle with respect to the electric field along z, depends on the axes over which the thin film extends. Earlier results show that serpentine LLs are contributed from the bulk for the normal vector of the thin film along the x-axis and external magnetic field applied along the x-axis. In contrast, serpentine LLs are contributed from the SWNs for a thin film with normal axis oriented in the y direction and external magnetic field applied in the y direction.Hence, we consider an external magnetic field oriented in the zx plane for a thin film with normal vector oriented in the x-axis direction, i.e. $(B\sin\theta,0,B\cos\theta )$, and an external magnetic field oriented in the yz plane for a thin film with normal vector along the y-axis, i.e. $(0,B\sin\theta,B\cos\theta)$.  Here $\theta$ is the relative angle between the applied electric and magnetic fields. We show the longitudinal magnetoconductivity calculated for both the cases in Fig.~\ref{fig:trwsmchiralanomaly}.

For a thin film with normal vector oriented in the x direction, we observe in Fig.~\ref{fig:trwsmchiralanomaly}(a) there are two separate peaks along the $\theta$ axis for $\Delta=0.2t$(in red) which is not observed for $\Delta=t$(blue). For a thin film with normal vector in the y direction, Fig.~\ref{fig:trwsmchiralanomaly}(b) shows a deviation near $\theta=\frac{\pi}{2}$ in the $\Delta=0.2t$(red) system compared to one with $\Delta=t$. This difference in behaviour can be attributed to the difference between these scenarios in the source---bulk or SWNs---of the serpentine LLs. More generally, serpentine LL contributions come from some combination of the bulk and SWNs. This is the case for the TRB WSM thin films, in which case both bulk and SWNs contribute to serpentine LLs. Analysis of serpentine LL origin and careful breakdown of the nature of myriad contributions is therefore a rich topic of future work.

\section{Discussion \& Conclusion}

In this work, we consider three-dimensional topological semimetals---those characterized by topologically-protected band structure degeneracies in a three-dimensional Brillouin zone--- in quasi (3-1)d or thin film geometries given the recent discovery of finite-size topological (FST) phases and their relation to the broader phenomena of the quantum skyrmion Hall effect~\cite{qskhe2024}. This is the first study of finite-size topology realized by compactifying topological semimetals, with past works on FSTs instead focused on compactification of topological insulators. In the quantum skyrmion Hall effect, a generalization of the framework of the quantum Hall effect, the notion of an incompressible state is generalized due to generalization of the notion of a (quasi)particle, to an object carrying charge but necessarily encoded in terms of (pseudo)spin angular momentum. As the quantum skyrmion Hall effect necessitates generalization of ostensibly ($d$+1)-dimensional quantum field theories to ($d$+$D$+1)-dimensional quantum field theories, with $D>0$, \textit{subjected to generalized compactification procedures over the $D$ extra dimensions}, evidence of finite-size topological phases and the broader quantum skyrmion Hall effect in experiments have profound implications for condensed matter, high-energy, and quantum information physics.

Given tremendous theoretical and experimental interest in Van der Waal heterostructure and Moiré materials, we focus on thin film geometries relevant to these systems, where, more specifically, the system is thermodynamically large in two spatial directions while system size is finite yet far from the thermodynamic regime and boundary conditions are open in the third spatial direction, studying lattice tight-binding models for canonical examples of Weyl semimetals (WSMs) as well as examples with significant experimental and ab initio support~\cite{huang_spectroscopic_2016}. We find these systems generically realize previously unidentified finite-size topological phases of matter, distinguished from strictly (2+1)d and (3+1)d topological states by their topological response signatures and bulk-boundary correspondence. That is, in quasi-(3-1)d geometries, FST semimetals are realized from strong hybridization of 3d WSM Fermi arc states, and a FST insulator is realized in the absence of underlying Fermi arc surface states from strong hybridization of the 3d Weyl nodes themselves.

FST semimetals realize quasi-(3-1)d bulk chiral or helical topologically-protected degeneracies and Fermi arc-like states on the quasi-(3-2)d boundary, co-existing with topological response signatures of the underlying 3d WSM phase and the chiral anomaly. These boundary modes, furthermore, are distinct from those of strictly (2+1)d semimetals, distinguished, for instance, by their localization and response to external fields.

The FST insulator, in contrast, appears at first glance to be similar in phenomenology to the quantum anomalous Hall insulator. One can characterize the quasi-(3-1)d bulk topology in terms of a total Chern number, but the FST insulator topology still depends on the topology of an underlying 3d WSM state, even though 3d Weyl nodes are strongly gapped out by finite-size hybridization. As a result of this dependence on an underlying 3d WSM state, the FST insulator serves as a mechanism for realizing arbitrarily large total Chern number and corresponding quasi-(3-2)d chiral boundary modes limited only by the number of layers in the stacking direction. Studies have already found evidence that van der Waals heterostructure materials can host high Chern numbers, with the maximum magnitude of the total Chern number bounded in part by the number of layers in the stack~\cite{bosnar2023chernvdw, koshino2009graphene, zhang2011layergraphene, wang2024chern}. These results may actually indicate realization of FST insulator states and should be re-examined using methods of the present work.

Strikingly, the total Chern number of the FST insulator can be changed through application of an \textit{in-plane} external magnetic field via coupling between this field and the underlying 3d Weyl nodes, unlike the case of the strictly (2+1)d quantum anomalous Hall insulator. In addition, the FST insulator exhibits topological response signatures of the chiral anomaly and underlying 3d WSM phase, in stark contrast to the quantum anomalous Hall insulator. These results may have profound implications for the considerable body of work on non-trivial topology in Van der Waals heterostructure and Moiré materials in particular, which focuses on topology thus far unambiguously associated with quantum Hall and quantum anomalous Hall states, but which may instead be due to finite-size topological states in some cases. This possibility is especially interesting given how generically 3d WSMs are realized in materials ~\cite{liu_2017_vdW, Song_2018, belopolski2017signature}.

Given the significance of these previously-unidentified finite-size topological states associated with discovery of the quantum skyrmion Hall effect, we confirm this physics for both canonical toy models describing WSM phases and tight-binding models capturing the WSM phase and Fermiology of MoTe\textsubscript{2}  previously used to analyze experimental results~\cite{huang_spectroscopic_2016}. We find bulk-boundary correspondences and topological response signatures of FSTs persist with tilting in type-I WSMs: quasi-(3-1)d bulk spectral flow of the FST state persists, for instance, although bulk-boundary correspondence is obscured for type-II WSM tilting. Finite-size topological states from 3d WSMs therefore yield robust signatures desirable for experimental realization, with the present work serving to motivate theoretical and experimental efforts to experimentally confirm finite-size topological phases of matter associated with the quantum skyrmion Hall effect.

We have already elaborated on some possible material candidates for the realization of such FST WSMs. Here, we briefly comment on broader aspects of experimental realization. To observe evidence of the finite-size topological states introduced in this work, the FST-WSM and the FST-WI, , the magnetic field needed is of the order, $\frac{2\pi}{L}$ where $L$ is the system size in the stacking direction of the thin film, or effectively the number of layers in the thin film. Requisite magnetic field strengths might be so large as to be experimentally inaccessible if $L$ is too small. As well, current fabrication methods, in the case of some materials, might only yield thin films of much larger $L$ than generally discussed in this work. However, it is possible to keep the system size along the thin film comparatively larger while still  observing finite-size topological response signatures. The crucial point is that finite-size topological signatures become more prominent when the characteristic decay length scales of boundary modes are comparable to system size $L$ . Theoretically, we tuned the system to this regime by enforcing that parameters $t$ and $\Delta$, of the minimal Hamiltonians considered here, satisfy $t>\Delta$ and then adjusting the ratio of $t$ to $\Delta$ to adjust the characteristic decay length scale. We specifically focused on reducing parameter $\Delta$ while keeping the hopping coefficient $t$ constant. A similar situation can be obtained if one instead increases the hopping parameter $t$ while keeping $\Delta$ constantm which in many VdW materials can be readily obtained by high pressure or tuning the elastic strain~\cite{liu_2017_vdW}. This not only increases the decay length but also provides some leeway in fabricating systems of desired thickness for observing FST effects. Increasing the system size in the thin film stacking direction, or system thickness, further decreases the magnetic field strength needed to observe such interesting FST responses. Therefore one might be able to observe similar results as detailed in our manuscript by considering VdW materials in the presence of high external pressure or strain.

By investigating the potential of the canonical Weyl semimetal for realizing finite-size topological phases of matter, and confirming the physics in models previously-used to interpret experiments, we show finite-size topological semimetal phases are widespread and relevant to experiments. Our results are especially significant given the tremendous interest in experimental realization and study of topological semimetal phases, quantum Hall states and quantum anomalous Hall states in Van der Waals layered materials and heterostructures.

\section{Acknowledgements}
AP would like to thank J. H. Winter and M. J. Pacholski for many useful discussions. This research was supported in part by the National Science Foundation under Grants No.NSF PHY-1748958 and PHY-2309135, and undertaken in part at Aspen Center for Physics, which is supported by National Science Foundation grant PHY-2210452.

\newpage
\bibliography{ref1.bib}

\begin{thebibliography}{69}%
\makeatletter
\providecommand \@ifxundefined [1]{%
 \@ifx{#1\undefined}
}%
\providecommand \@ifnum [1]{%
 \ifnum #1\expandafter \@firstoftwo
 \else \expandafter \@secondoftwo
 \fi
}%
\providecommand \@ifx [1]{%
 \ifx #1\expandafter \@firstoftwo
 \else \expandafter \@secondoftwo
 \fi
}%
\providecommand \natexlab [1]{#1}%
\providecommand \enquote  [1]{``#1''}%
\providecommand \bibnamefont  [1]{#1}%
\providecommand \bibfnamefont [1]{#1}%
\providecommand \citenamefont [1]{#1}%
\providecommand \href@noop [0]{\@secondoftwo}%
\providecommand \href [0]{\begingroup \@sanitize@url \@href}%
\providecommand \@href[1]{\@@startlink{#1}\@@href}%
\providecommand \@@href[1]{\endgroup#1\@@endlink}%
\providecommand \@sanitize@url [0]{\catcode `\\12\catcode `\$12\catcode
  `\&12\catcode `\#12\catcode `\^12\catcode `\_12\catcode `\%12\relax}%
\providecommand \@@startlink[1]{}%
\providecommand \@@endlink[0]{}%
\providecommand \url  [0]{\begingroup\@sanitize@url \@url }%
\providecommand \@url [1]{\endgroup\@href {#1}{\urlprefix }}%
\providecommand \urlprefix  [0]{URL }%
\providecommand \Eprint [0]{\href }%
\providecommand \doibase [0]{https://doi.org/}%
\providecommand \selectlanguage [0]{\@gobble}%
\providecommand \bibinfo  [0]{\@secondoftwo}%
\providecommand \bibfield  [0]{\@secondoftwo}%
\providecommand \translation [1]{[#1]}%
\providecommand \BibitemOpen [0]{}%
\providecommand \bibitemStop [0]{}%
\providecommand \bibitemNoStop [0]{.\EOS\space}%
\providecommand \EOS [0]{\spacefactor3000\relax}%
\providecommand \BibitemShut  [1]{\csname bibitem#1\endcsname}%
\let\auto@bib@innerbib\@empty
\bibitem [{\citenamefont {Bradlyn}\ \emph {et~al.}(2016)\citenamefont
  {Bradlyn}, \citenamefont {Cano}, \citenamefont {Wang}, \citenamefont
  {Vergniory}, \citenamefont {Felser}, \citenamefont {Cava},\ and\
  \citenamefont {Bernevig}}]{bradlyn2016}%
  \BibitemOpen
  \bibfield  {author} {\bibinfo {author} {\bibfnamefont {B.}~\bibnamefont
  {Bradlyn}}, \bibinfo {author} {\bibfnamefont {J.}~\bibnamefont {Cano}},
  \bibinfo {author} {\bibfnamefont {Z.}~\bibnamefont {Wang}}, \bibinfo {author}
  {\bibfnamefont {M.~G.}\ \bibnamefont {Vergniory}}, \bibinfo {author}
  {\bibfnamefont {C.}~\bibnamefont {Felser}}, \bibinfo {author} {\bibfnamefont
  {R.~J.}\ \bibnamefont {Cava}},\ and\ \bibinfo {author} {\bibfnamefont
  {B.~A.}\ \bibnamefont {Bernevig}},\ }\href
  {https://doi.org/10.1126/science.aaf5037} {\bibfield  {journal} {\bibinfo
  {journal} {Science}\ }\textbf {\bibinfo {volume} {353}},\ \bibinfo {pages}
  {aaf5037} (\bibinfo {year} {2016})},\ \Eprint
  {https://arxiv.org/abs/https://www.science.org/doi/pdf/10.1126/science.aaf5037}
  {https://www.science.org/doi/pdf/10.1126/science.aaf5037} \BibitemShut
  {NoStop}%
\bibitem [{\citenamefont {Borisenko}\ \emph {et~al.}(2014)\citenamefont
  {Borisenko}, \citenamefont {Gibson}, \citenamefont {Evtushinsky},
  \citenamefont {Zabolotnyy}, \citenamefont {B\"uchner},\ and\ \citenamefont
  {Cava}}]{borisenko2014experimental}%
  \BibitemOpen
  \bibfield  {author} {\bibinfo {author} {\bibfnamefont {S.}~\bibnamefont
  {Borisenko}}, \bibinfo {author} {\bibfnamefont {Q.}~\bibnamefont {Gibson}},
  \bibinfo {author} {\bibfnamefont {D.}~\bibnamefont {Evtushinsky}}, \bibinfo
  {author} {\bibfnamefont {V.}~\bibnamefont {Zabolotnyy}}, \bibinfo {author}
  {\bibfnamefont {B.}~\bibnamefont {B\"uchner}},\ and\ \bibinfo {author}
  {\bibfnamefont {R.~J.}\ \bibnamefont {Cava}},\ }\href
  {https://doi.org/10.1103/PhysRevLett.113.027603} {\bibfield  {journal}
  {\bibinfo  {journal} {Phys. Rev. Lett.}\ }\textbf {\bibinfo {volume} {113}},\
  \bibinfo {pages} {027603} (\bibinfo {year} {2014})}\BibitemShut {NoStop}%
\bibitem [{\citenamefont {Huang}\ \emph {et~al.}(2016)\citenamefont {Huang},
  \citenamefont {McCormick}, \citenamefont {Ochi}, \citenamefont {Zhao},
  \citenamefont {Suzuki}, \citenamefont {Arita}, \citenamefont {Wu},
  \citenamefont {Mou}, \citenamefont {Cao}, \citenamefont {Yan}, \citenamefont
  {Trivedi},\ and\ \citenamefont {Kaminski}}]{huang_spectroscopic_2016}%
  \BibitemOpen
  \bibfield  {author} {\bibinfo {author} {\bibfnamefont {L.}~\bibnamefont
  {Huang}}, \bibinfo {author} {\bibfnamefont {T.~M.}\ \bibnamefont
  {McCormick}}, \bibinfo {author} {\bibfnamefont {M.}~\bibnamefont {Ochi}},
  \bibinfo {author} {\bibfnamefont {Z.}~\bibnamefont {Zhao}}, \bibinfo {author}
  {\bibfnamefont {M.-T.}\ \bibnamefont {Suzuki}}, \bibinfo {author}
  {\bibfnamefont {R.}~\bibnamefont {Arita}}, \bibinfo {author} {\bibfnamefont
  {Y.}~\bibnamefont {Wu}}, \bibinfo {author} {\bibfnamefont {D.}~\bibnamefont
  {Mou}}, \bibinfo {author} {\bibfnamefont {H.}~\bibnamefont {Cao}}, \bibinfo
  {author} {\bibfnamefont {J.}~\bibnamefont {Yan}}, \bibinfo {author}
  {\bibfnamefont {N.}~\bibnamefont {Trivedi}},\ and\ \bibinfo {author}
  {\bibfnamefont {A.}~\bibnamefont {Kaminski}},\ }\href
  {https://doi.org/10.1038/nmat4685} {\bibfield  {journal} {\bibinfo  {journal}
  {Nature Materials}\ }\textbf {\bibinfo {volume} {15}},\ \bibinfo {pages}
  {1155} (\bibinfo {year} {2016})}\BibitemShut {NoStop}%
\bibitem [{\citenamefont {Jiang}\ \emph {et~al.}(2017)\citenamefont {Jiang},
  \citenamefont {Liu}, \citenamefont {Sun}, \citenamefont {Yang}, \citenamefont
  {Rajamathi}, \citenamefont {Qi}, \citenamefont {Yang}, \citenamefont {Chen},
  \citenamefont {Peng}, \citenamefont {Hwang}, \citenamefont {Sun},
  \citenamefont {Mo}, \citenamefont {Vobornik}, \citenamefont {Fujii},
  \citenamefont {Parkin}, \citenamefont {Felser}, \citenamefont {Yan},\ and\
  \citenamefont {Chen}}]{jiang_signature_2017}%
  \BibitemOpen
  \bibfield  {author} {\bibinfo {author} {\bibfnamefont {J.}~\bibnamefont
  {Jiang}}, \bibinfo {author} {\bibfnamefont {Z.}~\bibnamefont {Liu}}, \bibinfo
  {author} {\bibfnamefont {Y.}~\bibnamefont {Sun}}, \bibinfo {author}
  {\bibfnamefont {H.}~\bibnamefont {Yang}}, \bibinfo {author} {\bibfnamefont
  {C.}~\bibnamefont {Rajamathi}}, \bibinfo {author} {\bibfnamefont
  {Y.}~\bibnamefont {Qi}}, \bibinfo {author} {\bibfnamefont {L.}~\bibnamefont
  {Yang}}, \bibinfo {author} {\bibfnamefont {C.}~\bibnamefont {Chen}}, \bibinfo
  {author} {\bibfnamefont {H.}~\bibnamefont {Peng}}, \bibinfo {author}
  {\bibfnamefont {C.-C.}\ \bibnamefont {Hwang}}, \bibinfo {author}
  {\bibfnamefont {S.}~\bibnamefont {Sun}}, \bibinfo {author} {\bibfnamefont
  {S.-K.}\ \bibnamefont {Mo}}, \bibinfo {author} {\bibfnamefont
  {I.}~\bibnamefont {Vobornik}}, \bibinfo {author} {\bibfnamefont
  {J.}~\bibnamefont {Fujii}}, \bibinfo {author} {\bibfnamefont
  {S.}~\bibnamefont {Parkin}}, \bibinfo {author} {\bibfnamefont
  {C.}~\bibnamefont {Felser}}, \bibinfo {author} {\bibfnamefont
  {B.}~\bibnamefont {Yan}},\ and\ \bibinfo {author} {\bibfnamefont
  {Y.}~\bibnamefont {Chen}},\ }\href {https://doi.org/10.1038/ncomms13973}
  {\bibfield  {journal} {\bibinfo  {journal} {Nature Communications}\ }\textbf
  {\bibinfo {volume} {8}},\ \bibinfo {pages} {13973} (\bibinfo {year}
  {2017})}\BibitemShut {NoStop}%
\bibitem [{\citenamefont {Liu}\ \emph {et~al.}(2014{\natexlab{a}})\citenamefont
  {Liu}, \citenamefont {Jiang}, \citenamefont {Zhou}, \citenamefont {Wang},
  \citenamefont {Zhang}, \citenamefont {Weng}, \citenamefont {Prabhakaran},
  \citenamefont {Mo}, \citenamefont {Peng}, \citenamefont {Dudin} \emph
  {et~al.}}]{liu2014stable}%
  \BibitemOpen
  \bibfield  {author} {\bibinfo {author} {\bibfnamefont {Z.}~\bibnamefont
  {Liu}}, \bibinfo {author} {\bibfnamefont {J.}~\bibnamefont {Jiang}}, \bibinfo
  {author} {\bibfnamefont {B.}~\bibnamefont {Zhou}}, \bibinfo {author}
  {\bibfnamefont {Z.}~\bibnamefont {Wang}}, \bibinfo {author} {\bibfnamefont
  {Y.}~\bibnamefont {Zhang}}, \bibinfo {author} {\bibfnamefont
  {H.}~\bibnamefont {Weng}}, \bibinfo {author} {\bibfnamefont {D.}~\bibnamefont
  {Prabhakaran}}, \bibinfo {author} {\bibfnamefont {S.~K.}\ \bibnamefont {Mo}},
  \bibinfo {author} {\bibfnamefont {H.}~\bibnamefont {Peng}}, \bibinfo {author}
  {\bibfnamefont {P.}~\bibnamefont {Dudin}}, \emph {et~al.},\ }\href
  {https://doi.org/10.1038/nmat3990} {\bibfield  {journal} {\bibinfo  {journal}
  {Nature Mater}\ }\textbf {\bibinfo {volume} {13}},\ \bibinfo {pages} {677}
  (\bibinfo {year} {2014}{\natexlab{a}})}\BibitemShut {NoStop}%
\bibitem [{\citenamefont {Neupane}\ \emph {et~al.}(2014)\citenamefont
  {Neupane}, \citenamefont {Xu}, \citenamefont {Sankar}, \citenamefont
  {Alidoust}, \citenamefont {Bian}, \citenamefont {Liu}, \citenamefont
  {Belopolski}, \citenamefont {Chang}, \citenamefont {Jeng}, \citenamefont
  {Lin} \emph {et~al.}}]{neupane2013observation}%
  \BibitemOpen
  \bibfield  {author} {\bibinfo {author} {\bibfnamefont {M.}~\bibnamefont
  {Neupane}}, \bibinfo {author} {\bibfnamefont {S.}~\bibnamefont {Xu}},
  \bibinfo {author} {\bibfnamefont {R.}~\bibnamefont {Sankar}}, \bibinfo
  {author} {\bibfnamefont {N.}~\bibnamefont {Alidoust}}, \bibinfo {author}
  {\bibfnamefont {G.}~\bibnamefont {Bian}}, \bibinfo {author} {\bibfnamefont
  {C.}~\bibnamefont {Liu}}, \bibinfo {author} {\bibfnamefont {I.}~\bibnamefont
  {Belopolski}}, \bibinfo {author} {\bibfnamefont {T.-R.}\ \bibnamefont
  {Chang}}, \bibinfo {author} {\bibfnamefont {H.-T.}\ \bibnamefont {Jeng}},
  \bibinfo {author} {\bibfnamefont {H.}~\bibnamefont {Lin}}, \emph {et~al.},\
  }\href {https://doi.org/10.1038/ncomms4786} {\bibfield  {journal} {\bibinfo
  {journal} {Nat Commun}\ }\textbf {\bibinfo {volume} {5}} (\bibinfo {year}
  {2014})}\BibitemShut {NoStop}%
\bibitem [{\citenamefont {Lv}\ \emph {et~al.}(2015{\natexlab{a}})\citenamefont
  {Lv}, \citenamefont {Weng}, \citenamefont {Fu}, \citenamefont {Wang},
  \citenamefont {Miao}, \citenamefont {Ma}, \citenamefont {Richard},
  \citenamefont {Huang}, \citenamefont {Zhao}, \citenamefont {Chen},
  \citenamefont {Fang}, \citenamefont {Dai}, \citenamefont {Qian},\ and\
  \citenamefont {Ding}}]{lv2015taas}%
  \BibitemOpen
  \bibfield  {author} {\bibinfo {author} {\bibfnamefont {B.~Q.}\ \bibnamefont
  {Lv}}, \bibinfo {author} {\bibfnamefont {H.~M.}\ \bibnamefont {Weng}},
  \bibinfo {author} {\bibfnamefont {B.~B.}\ \bibnamefont {Fu}}, \bibinfo
  {author} {\bibfnamefont {X.~P.}\ \bibnamefont {Wang}}, \bibinfo {author}
  {\bibfnamefont {H.}~\bibnamefont {Miao}}, \bibinfo {author} {\bibfnamefont
  {J.}~\bibnamefont {Ma}}, \bibinfo {author} {\bibfnamefont {P.}~\bibnamefont
  {Richard}}, \bibinfo {author} {\bibfnamefont {X.~C.}\ \bibnamefont {Huang}},
  \bibinfo {author} {\bibfnamefont {L.~X.}\ \bibnamefont {Zhao}}, \bibinfo
  {author} {\bibfnamefont {G.~F.}\ \bibnamefont {Chen}}, \bibinfo {author}
  {\bibfnamefont {Z.}~\bibnamefont {Fang}}, \bibinfo {author} {\bibfnamefont
  {X.}~\bibnamefont {Dai}}, \bibinfo {author} {\bibfnamefont {T.}~\bibnamefont
  {Qian}},\ and\ \bibinfo {author} {\bibfnamefont {H.}~\bibnamefont {Ding}},\
  }\href {https://doi.org/10.1103/PhysRevX.5.031013} {\bibfield  {journal}
  {\bibinfo  {journal} {Phys. Rev. X}\ }\textbf {\bibinfo {volume} {5}},\
  \bibinfo {pages} {031013} (\bibinfo {year} {2015}{\natexlab{a}})}\BibitemShut
  {NoStop}%
\bibitem [{\citenamefont {Lv}\ \emph {et~al.}(2015{\natexlab{b}})\citenamefont
  {Lv}, \citenamefont {Xu}, \citenamefont {Weng}, \citenamefont {Ma},
  \citenamefont {Richard}, \citenamefont {Huang}, \citenamefont {Zhao},
  \citenamefont {Chen}, \citenamefont {Matt}, \citenamefont {Bisti},
  \citenamefont {Strocov}, \citenamefont {Mesot}, \citenamefont {Fang},
  \citenamefont {Dai}, \citenamefont {Qian}, \citenamefont {Shi},\ and\
  \citenamefont {Ding}}]{lv_observation_2015}%
  \BibitemOpen
  \bibfield  {author} {\bibinfo {author} {\bibfnamefont {B.~Q.}\ \bibnamefont
  {Lv}}, \bibinfo {author} {\bibfnamefont {N.}~\bibnamefont {Xu}}, \bibinfo
  {author} {\bibfnamefont {H.~M.}\ \bibnamefont {Weng}}, \bibinfo {author}
  {\bibfnamefont {J.~Z.}\ \bibnamefont {Ma}}, \bibinfo {author} {\bibfnamefont
  {P.}~\bibnamefont {Richard}}, \bibinfo {author} {\bibfnamefont {X.~C.}\
  \bibnamefont {Huang}}, \bibinfo {author} {\bibfnamefont {L.~X.}\ \bibnamefont
  {Zhao}}, \bibinfo {author} {\bibfnamefont {G.~F.}\ \bibnamefont {Chen}},
  \bibinfo {author} {\bibfnamefont {C.~E.}\ \bibnamefont {Matt}}, \bibinfo
  {author} {\bibfnamefont {F.}~\bibnamefont {Bisti}}, \bibinfo {author}
  {\bibfnamefont {V.~N.}\ \bibnamefont {Strocov}}, \bibinfo {author}
  {\bibfnamefont {J.}~\bibnamefont {Mesot}}, \bibinfo {author} {\bibfnamefont
  {Z.}~\bibnamefont {Fang}}, \bibinfo {author} {\bibfnamefont {X.}~\bibnamefont
  {Dai}}, \bibinfo {author} {\bibfnamefont {T.}~\bibnamefont {Qian}}, \bibinfo
  {author} {\bibfnamefont {M.}~\bibnamefont {Shi}},\ and\ \bibinfo {author}
  {\bibfnamefont {H.}~\bibnamefont {Ding}},\ }\href
  {https://doi.org/10.1038/nphys3426} {\bibfield  {journal} {\bibinfo
  {journal} {Nature Phys}\ }\textbf {\bibinfo {volume} {11}},\ \bibinfo {pages}
  {724} (\bibinfo {year} {2015}{\natexlab{b}})}\BibitemShut {NoStop}%
\bibitem [{\citenamefont {Xu}\ \emph {et~al.}(2015{\natexlab{a}})\citenamefont
  {Xu}, \citenamefont {Alidoust}, \citenamefont {Belopolski}, \citenamefont
  {Yuan}, \citenamefont {Bian}, \citenamefont {Chang}, \citenamefont {Zheng},
  \citenamefont {Strocov}, \citenamefont {Sanchez}, \citenamefont {Chang},
  \citenamefont {Zhang}, \citenamefont {Mou}, \citenamefont {Wu}, \citenamefont
  {Huang}, \citenamefont {Lee}, \citenamefont {Huang}, \citenamefont {Wang},
  \citenamefont {Bansil}, \citenamefont {Jeng}, \citenamefont {Neupert},
  \citenamefont {Kaminski}, \citenamefont {Lin}, \citenamefont {Jia},\ and\
  \citenamefont {Zahid~Hasan}}]{xu_discovery_2015}%
  \BibitemOpen
  \bibfield  {author} {\bibinfo {author} {\bibfnamefont {S.-Y.}\ \bibnamefont
  {Xu}}, \bibinfo {author} {\bibfnamefont {N.}~\bibnamefont {Alidoust}},
  \bibinfo {author} {\bibfnamefont {I.}~\bibnamefont {Belopolski}}, \bibinfo
  {author} {\bibfnamefont {Z.}~\bibnamefont {Yuan}}, \bibinfo {author}
  {\bibfnamefont {G.}~\bibnamefont {Bian}}, \bibinfo {author} {\bibfnamefont
  {T.-R.}\ \bibnamefont {Chang}}, \bibinfo {author} {\bibfnamefont
  {H.}~\bibnamefont {Zheng}}, \bibinfo {author} {\bibfnamefont {V.~N.}\
  \bibnamefont {Strocov}}, \bibinfo {author} {\bibfnamefont {D.~S.}\
  \bibnamefont {Sanchez}}, \bibinfo {author} {\bibfnamefont {G.}~\bibnamefont
  {Chang}}, \bibinfo {author} {\bibfnamefont {C.}~\bibnamefont {Zhang}},
  \bibinfo {author} {\bibfnamefont {D.}~\bibnamefont {Mou}}, \bibinfo {author}
  {\bibfnamefont {Y.}~\bibnamefont {Wu}}, \bibinfo {author} {\bibfnamefont
  {L.}~\bibnamefont {Huang}}, \bibinfo {author} {\bibfnamefont {C.-C.}\
  \bibnamefont {Lee}}, \bibinfo {author} {\bibfnamefont {S.-M.}\ \bibnamefont
  {Huang}}, \bibinfo {author} {\bibfnamefont {B.}~\bibnamefont {Wang}},
  \bibinfo {author} {\bibfnamefont {A.}~\bibnamefont {Bansil}}, \bibinfo
  {author} {\bibfnamefont {H.-T.}\ \bibnamefont {Jeng}}, \bibinfo {author}
  {\bibfnamefont {T.}~\bibnamefont {Neupert}}, \bibinfo {author} {\bibfnamefont
  {A.}~\bibnamefont {Kaminski}}, \bibinfo {author} {\bibfnamefont
  {H.}~\bibnamefont {Lin}}, \bibinfo {author} {\bibfnamefont {S.}~\bibnamefont
  {Jia}},\ and\ \bibinfo {author} {\bibfnamefont {M.}~\bibnamefont
  {Zahid~Hasan}},\ }\href {https://doi.org/10.1038/nphys3437} {\bibfield
  {journal} {\bibinfo  {journal} {Nature Phys}\ }\textbf {\bibinfo {volume}
  {11}},\ \bibinfo {pages} {748} (\bibinfo {year}
  {2015}{\natexlab{a}})}\BibitemShut {NoStop}%
\bibitem [{\citenamefont {Xu}\ \emph {et~al.}(2015{\natexlab{b}})\citenamefont
  {Xu}, \citenamefont {Belopolski}, \citenamefont {Alidoust}, \citenamefont
  {Neupane}, \citenamefont {Bian}, \citenamefont {Zhang}, \citenamefont
  {Sankar}, \citenamefont {Chang}, \citenamefont {Yuan}, \citenamefont {Lee},
  \citenamefont {Huang}, \citenamefont {Zheng}, \citenamefont {Ma},
  \citenamefont {Sanchez}, \citenamefont {Wang}, \citenamefont {Bansil},
  \citenamefont {Chou}, \citenamefont {Shibayev}, \citenamefont {Lin},
  \citenamefont {Jia},\ and\ \citenamefont {Hasan}}]{suyang2015discovery}%
  \BibitemOpen
  \bibfield  {author} {\bibinfo {author} {\bibfnamefont {S.-Y.}\ \bibnamefont
  {Xu}}, \bibinfo {author} {\bibfnamefont {I.}~\bibnamefont {Belopolski}},
  \bibinfo {author} {\bibfnamefont {N.}~\bibnamefont {Alidoust}}, \bibinfo
  {author} {\bibfnamefont {M.}~\bibnamefont {Neupane}}, \bibinfo {author}
  {\bibfnamefont {G.}~\bibnamefont {Bian}}, \bibinfo {author} {\bibfnamefont
  {C.}~\bibnamefont {Zhang}}, \bibinfo {author} {\bibfnamefont
  {R.}~\bibnamefont {Sankar}}, \bibinfo {author} {\bibfnamefont
  {G.}~\bibnamefont {Chang}}, \bibinfo {author} {\bibfnamefont
  {Z.}~\bibnamefont {Yuan}}, \bibinfo {author} {\bibfnamefont {C.-C.}\
  \bibnamefont {Lee}}, \bibinfo {author} {\bibfnamefont {S.-M.}\ \bibnamefont
  {Huang}}, \bibinfo {author} {\bibfnamefont {H.}~\bibnamefont {Zheng}},
  \bibinfo {author} {\bibfnamefont {J.}~\bibnamefont {Ma}}, \bibinfo {author}
  {\bibfnamefont {D.~S.}\ \bibnamefont {Sanchez}}, \bibinfo {author}
  {\bibfnamefont {B.}~\bibnamefont {Wang}}, \bibinfo {author} {\bibfnamefont
  {A.}~\bibnamefont {Bansil}}, \bibinfo {author} {\bibfnamefont
  {F.}~\bibnamefont {Chou}}, \bibinfo {author} {\bibfnamefont {P.~P.}\
  \bibnamefont {Shibayev}}, \bibinfo {author} {\bibfnamefont {H.}~\bibnamefont
  {Lin}}, \bibinfo {author} {\bibfnamefont {S.}~\bibnamefont {Jia}},\ and\
  \bibinfo {author} {\bibfnamefont {M.~Z.}\ \bibnamefont {Hasan}},\ }\href
  {https://doi.org/10.1126/science.aaa9297} {\bibfield  {journal} {\bibinfo
  {journal} {Science}\ }\textbf {\bibinfo {volume} {349}},\ \bibinfo {pages}
  {613} (\bibinfo {year} {2015}{\natexlab{b}})},\ \Eprint
  {https://arxiv.org/abs/https://www.science.org/doi/pdf/10.1126/science.aaa9297}
  {https://www.science.org/doi/pdf/10.1126/science.aaa9297} \BibitemShut
  {NoStop}%
\bibitem [{\citenamefont {Liu}\ \emph {et~al.}(2014{\natexlab{b}})\citenamefont
  {Liu}, \citenamefont {Zhou}, \citenamefont {Zhang}, \citenamefont {Wang},
  \citenamefont {Weng}, \citenamefont {Prabhakaran}, \citenamefont {Mo},
  \citenamefont {Shen}, \citenamefont {Fang}, \citenamefont {Dai},
  \citenamefont {Hussain},\ and\ \citenamefont {Chen}}]{liu2014discovery}%
  \BibitemOpen
  \bibfield  {author} {\bibinfo {author} {\bibfnamefont {Z.~K.}\ \bibnamefont
  {Liu}}, \bibinfo {author} {\bibfnamefont {B.}~\bibnamefont {Zhou}}, \bibinfo
  {author} {\bibfnamefont {Y.}~\bibnamefont {Zhang}}, \bibinfo {author}
  {\bibfnamefont {Z.~J.}\ \bibnamefont {Wang}}, \bibinfo {author}
  {\bibfnamefont {H.~M.}\ \bibnamefont {Weng}}, \bibinfo {author}
  {\bibfnamefont {D.}~\bibnamefont {Prabhakaran}}, \bibinfo {author}
  {\bibfnamefont {S.-K.}\ \bibnamefont {Mo}}, \bibinfo {author} {\bibfnamefont
  {Z.~X.}\ \bibnamefont {Shen}}, \bibinfo {author} {\bibfnamefont
  {Z.}~\bibnamefont {Fang}}, \bibinfo {author} {\bibfnamefont {X.}~\bibnamefont
  {Dai}}, \bibinfo {author} {\bibfnamefont {Z.}~\bibnamefont {Hussain}},\ and\
  \bibinfo {author} {\bibfnamefont {Y.~L.}\ \bibnamefont {Chen}},\ }\href
  {https://doi.org/10.1126/science.1245085} {\bibfield  {journal} {\bibinfo
  {journal} {Science}\ }\textbf {\bibinfo {volume} {343}},\ \bibinfo {pages}
  {864} (\bibinfo {year} {2014}{\natexlab{b}})},\ \Eprint
  {https://arxiv.org/abs/https://www.science.org/doi/pdf/10.1126/science.1245085}
  {https://www.science.org/doi/pdf/10.1126/science.1245085} \BibitemShut
  {NoStop}%
\bibitem [{\citenamefont {Shekhar}\ \emph {et~al.}(2015)\citenamefont
  {Shekhar}, \citenamefont {Nayak}, \citenamefont {Sun}, \citenamefont
  {Schmidt}, \citenamefont {Nicklas}, \citenamefont {Leermakers}, \citenamefont
  {Zeitler}, \citenamefont {Skourski}, \citenamefont {Wosnitza}, \citenamefont
  {Liu}, \citenamefont {Chen}, \citenamefont {Schnelle}, \citenamefont
  {Borrmann}, \citenamefont {Grin}, \citenamefont {Felser},\ and\ \citenamefont
  {Yan}}]{shekhar_extremely_2015}%
  \BibitemOpen
  \bibfield  {author} {\bibinfo {author} {\bibfnamefont {C.}~\bibnamefont
  {Shekhar}}, \bibinfo {author} {\bibfnamefont {A.~K.}\ \bibnamefont {Nayak}},
  \bibinfo {author} {\bibfnamefont {Y.}~\bibnamefont {Sun}}, \bibinfo {author}
  {\bibfnamefont {M.}~\bibnamefont {Schmidt}}, \bibinfo {author} {\bibfnamefont
  {M.}~\bibnamefont {Nicklas}}, \bibinfo {author} {\bibfnamefont
  {I.}~\bibnamefont {Leermakers}}, \bibinfo {author} {\bibfnamefont
  {U.}~\bibnamefont {Zeitler}}, \bibinfo {author} {\bibfnamefont
  {Y.}~\bibnamefont {Skourski}}, \bibinfo {author} {\bibfnamefont
  {J.}~\bibnamefont {Wosnitza}}, \bibinfo {author} {\bibfnamefont
  {Z.}~\bibnamefont {Liu}}, \bibinfo {author} {\bibfnamefont {Y.}~\bibnamefont
  {Chen}}, \bibinfo {author} {\bibfnamefont {W.}~\bibnamefont {Schnelle}},
  \bibinfo {author} {\bibfnamefont {H.}~\bibnamefont {Borrmann}}, \bibinfo
  {author} {\bibfnamefont {Y.}~\bibnamefont {Grin}}, \bibinfo {author}
  {\bibfnamefont {C.}~\bibnamefont {Felser}},\ and\ \bibinfo {author}
  {\bibfnamefont {B.}~\bibnamefont {Yan}},\ }\href
  {https://doi.org/10.1038/nphys3372} {\bibfield  {journal} {\bibinfo
  {journal} {Nature Phys}\ }\textbf {\bibinfo {volume} {11}},\ \bibinfo {pages}
  {645} (\bibinfo {year} {2015})}\BibitemShut {NoStop}%
\bibitem [{\citenamefont {Gyenis}\ \emph {et~al.}(2016)\citenamefont {Gyenis},
  \citenamefont {Inoue}, \citenamefont {Jeon}, \citenamefont {Zhou},
  \citenamefont {Feldman}, \citenamefont {Wang}, \citenamefont {Li},
  \citenamefont {Jiang}, \citenamefont {Gibson}, \citenamefont {Kushwaha},
  \citenamefont {Krizan}, \citenamefont {Ni}, \citenamefont {Cava},
  \citenamefont {Bernevig},\ and\ \citenamefont {Yazdani}}]{gyenis_2016}%
  \BibitemOpen
  \bibfield  {author} {\bibinfo {author} {\bibfnamefont {A.}~\bibnamefont
  {Gyenis}}, \bibinfo {author} {\bibfnamefont {H.}~\bibnamefont {Inoue}},
  \bibinfo {author} {\bibfnamefont {S.}~\bibnamefont {Jeon}}, \bibinfo {author}
  {\bibfnamefont {B.~B.}\ \bibnamefont {Zhou}}, \bibinfo {author}
  {\bibfnamefont {B.~E.}\ \bibnamefont {Feldman}}, \bibinfo {author}
  {\bibfnamefont {Z.}~\bibnamefont {Wang}}, \bibinfo {author} {\bibfnamefont
  {J.}~\bibnamefont {Li}}, \bibinfo {author} {\bibfnamefont {S.}~\bibnamefont
  {Jiang}}, \bibinfo {author} {\bibfnamefont {Q.~D.}\ \bibnamefont {Gibson}},
  \bibinfo {author} {\bibfnamefont {S.~K.}\ \bibnamefont {Kushwaha}}, \bibinfo
  {author} {\bibfnamefont {J.~W.}\ \bibnamefont {Krizan}}, \bibinfo {author}
  {\bibfnamefont {N.}~\bibnamefont {Ni}}, \bibinfo {author} {\bibfnamefont
  {R.~J.}\ \bibnamefont {Cava}}, \bibinfo {author} {\bibfnamefont {B.~A.}\
  \bibnamefont {Bernevig}},\ and\ \bibinfo {author} {\bibfnamefont
  {A.}~\bibnamefont {Yazdani}},\ }\href
  {https://doi.org/10.1088/1367-2630/18/10/105003} {\bibfield  {journal}
  {\bibinfo  {journal} {New J. Phys.}\ }\textbf {\bibinfo {volume} {18}},\
  \bibinfo {pages} {105003} (\bibinfo {year} {2016})}\BibitemShut {NoStop}%
\bibitem [{\citenamefont {Yan}\ and\ \citenamefont
  {Felser}(2017)}]{yanreview2017}%
  \BibitemOpen
  \bibfield  {author} {\bibinfo {author} {\bibfnamefont {B.}~\bibnamefont
  {Yan}}\ and\ \bibinfo {author} {\bibfnamefont {C.}~\bibnamefont {Felser}},\
  }\href {https://doi.org/10.1146/annurev-conmatphys-031016-025458} {\bibfield
  {journal} {\bibinfo  {journal} {Annual Review of Condensed Matter Physics}\
  }\textbf {\bibinfo {volume} {8}},\ \bibinfo {pages} {337} (\bibinfo {year}
  {2017})},\ \Eprint
  {https://arxiv.org/abs/https://doi.org/10.1146/annurev-conmatphys-031016-025458}
  {https://doi.org/10.1146/annurev-conmatphys-031016-025458} \BibitemShut
  {NoStop}%
\bibitem [{\citenamefont {Jia}\ \emph {et~al.}(2016)\citenamefont {Jia},
  \citenamefont {Xu},\ and\ \citenamefont {Hasan}}]{jia_weyl_2016}%
  \BibitemOpen
  \bibfield  {author} {\bibinfo {author} {\bibfnamefont {S.}~\bibnamefont
  {Jia}}, \bibinfo {author} {\bibfnamefont {S.-Y.}\ \bibnamefont {Xu}},\ and\
  \bibinfo {author} {\bibfnamefont {M.~Z.}\ \bibnamefont {Hasan}},\ }\href
  {https://doi.org/10.1038/nmat4787} {\bibfield  {journal} {\bibinfo  {journal}
  {Nature Mater}\ }\textbf {\bibinfo {volume} {15}},\ \bibinfo {pages} {1140}
  (\bibinfo {year} {2016})}\BibitemShut {NoStop}%
\bibitem [{\citenamefont {Armitage}\ \emph {et~al.}(2018)\citenamefont
  {Armitage}, \citenamefont {Mele},\ and\ \citenamefont
  {Vishwanath}}]{armitage2018review}%
  \BibitemOpen
  \bibfield  {author} {\bibinfo {author} {\bibfnamefont {N.~P.}\ \bibnamefont
  {Armitage}}, \bibinfo {author} {\bibfnamefont {E.~J.}\ \bibnamefont {Mele}},\
  and\ \bibinfo {author} {\bibfnamefont {A.}~\bibnamefont {Vishwanath}},\
  }\href {https://doi.org/10.1103/RevModPhys.90.015001} {\bibfield  {journal}
  {\bibinfo  {journal} {Rev. Mod. Phys.}\ }\textbf {\bibinfo {volume} {90}},\
  \bibinfo {pages} {015001} (\bibinfo {year} {2018})}\BibitemShut {NoStop}%
\bibitem [{\citenamefont {Burkov}\ and\ \citenamefont
  {Balents}(2011)}]{burkov_wsmti_2011}%
  \BibitemOpen
  \bibfield  {author} {\bibinfo {author} {\bibfnamefont {A.~A.}\ \bibnamefont
  {Burkov}}\ and\ \bibinfo {author} {\bibfnamefont {L.}~\bibnamefont
  {Balents}},\ }\href {https://doi.org/10.1103/PhysRevLett.107.127205}
  {\bibfield  {journal} {\bibinfo  {journal} {Phys. Rev. Lett.}\ }\textbf
  {\bibinfo {volume} {107}},\ \bibinfo {pages} {127205} (\bibinfo {year}
  {2011})}\BibitemShut {NoStop}%
\bibitem [{\citenamefont {Hal\'asz}\ and\ \citenamefont
  {Balents}(2012)}]{halasz2012time}%
  \BibitemOpen
  \bibfield  {author} {\bibinfo {author} {\bibfnamefont {G.~B.}\ \bibnamefont
  {Hal\'asz}}\ and\ \bibinfo {author} {\bibfnamefont {L.}~\bibnamefont
  {Balents}},\ }\href {https://doi.org/10.1103/PhysRevB.85.035103} {\bibfield
  {journal} {\bibinfo  {journal} {Phys. Rev. B}\ }\textbf {\bibinfo {volume}
  {85}},\ \bibinfo {pages} {035103} (\bibinfo {year} {2012})}\BibitemShut
  {NoStop}%
\bibitem [{\citenamefont {Haldane}(2014)}]{haldane2014attachment}%
  \BibitemOpen
  \bibfield  {author} {\bibinfo {author} {\bibfnamefont {F.}~\bibnamefont
  {Haldane}},\ }\href@noop {} {\bibfield  {journal} {\bibinfo  {journal} {arXiv
  preprint arXiv:1401.0529}\ } (\bibinfo {year} {2014})}\BibitemShut {NoStop}%
\bibitem [{\citenamefont {Mathai}\ and\ \citenamefont
  {Thiang}(2017)}]{mathai2017global}%
  \BibitemOpen
  \bibfield  {author} {\bibinfo {author} {\bibfnamefont {V.}~\bibnamefont
  {Mathai}}\ and\ \bibinfo {author} {\bibfnamefont {G.~C.}\ \bibnamefont
  {Thiang}},\ }\href
  {https://iopscience.iop.org/article/10.1088/1751-8121/aa59b2} {\bibfield
  {journal} {\bibinfo  {journal} {J. Phys. A: Math. Theor.}\ }\textbf {\bibinfo
  {volume} {50}},\ \bibinfo {pages} {11LT01} (\bibinfo {year}
  {2017})}\BibitemShut {NoStop}%
\bibitem [{\citenamefont {Hosur}(2012)}]{hosur2012friedel}%
  \BibitemOpen
  \bibfield  {author} {\bibinfo {author} {\bibfnamefont {P.}~\bibnamefont
  {Hosur}},\ }\href {https://doi.org/10.1103/PhysRevB.86.195102} {\bibfield
  {journal} {\bibinfo  {journal} {Phys. Rev. B}\ }\textbf {\bibinfo {volume}
  {86}},\ \bibinfo {pages} {195102} (\bibinfo {year} {2012})}\BibitemShut
  {NoStop}%
\bibitem [{\citenamefont {Potter}\ \emph
  {et~al.}(2014{\natexlab{a}})\citenamefont {Potter}, \citenamefont {Kimchi},\
  and\ \citenamefont {Vishwanath}}]{potter2014quantum}%
  \BibitemOpen
  \bibfield  {author} {\bibinfo {author} {\bibfnamefont {A.~C.}\ \bibnamefont
  {Potter}}, \bibinfo {author} {\bibfnamefont {I.}~\bibnamefont {Kimchi}},\
  and\ \bibinfo {author} {\bibfnamefont {A.}~\bibnamefont {Vishwanath}},\
  }\href {https://www.nature.com/articles/ncomms6161} {\bibfield  {journal}
  {\bibinfo  {journal} {Nature Commun}\ }\textbf {\bibinfo {volume} {5}},\
  \bibinfo {pages} {1} (\bibinfo {year} {2014}{\natexlab{a}})}\BibitemShut
  {NoStop}%
\bibitem [{\citenamefont {Nielsen}\ and\ \citenamefont
  {Ninomiya}(1983)}]{nielsen1983anomaly}%
  \BibitemOpen
  \bibfield  {author} {\bibinfo {author} {\bibfnamefont {H.}~\bibnamefont
  {Nielsen}}\ and\ \bibinfo {author} {\bibfnamefont {M.}~\bibnamefont
  {Ninomiya}},\ }\href
  {https://doi.org/https://doi.org/10.1016/0370-2693(83)91529-0} {\bibfield
  {journal} {\bibinfo  {journal} {Physics Letters B}\ }\textbf {\bibinfo
  {volume} {130}},\ \bibinfo {pages} {389} (\bibinfo {year}
  {1983})}\BibitemShut {NoStop}%
\bibitem [{\citenamefont {Zyuzin}\ and\ \citenamefont
  {Burkov}(2012)}]{zyuzin2012topological}%
  \BibitemOpen
  \bibfield  {author} {\bibinfo {author} {\bibfnamefont {A.~A.}\ \bibnamefont
  {Zyuzin}}\ and\ \bibinfo {author} {\bibfnamefont {A.~A.}\ \bibnamefont
  {Burkov}},\ }\href {https://doi.org/10.1103/PhysRevB.86.115133} {\bibfield
  {journal} {\bibinfo  {journal} {Phys. Rev. B}\ }\textbf {\bibinfo {volume}
  {86}},\ \bibinfo {pages} {115133} (\bibinfo {year} {2012})}\BibitemShut
  {NoStop}%
\bibitem [{\citenamefont {Son}\ and\ \citenamefont
  {Spivak}(2013)}]{son2013chiral}%
  \BibitemOpen
  \bibfield  {author} {\bibinfo {author} {\bibfnamefont {D.~T.}\ \bibnamefont
  {Son}}\ and\ \bibinfo {author} {\bibfnamefont {B.~Z.}\ \bibnamefont
  {Spivak}},\ }\href {https://doi.org/10.1103/PhysRevB.88.104412} {\bibfield
  {journal} {\bibinfo  {journal} {Phys. Rev. B}\ }\textbf {\bibinfo {volume}
  {88}},\ \bibinfo {pages} {104412} (\bibinfo {year} {2013})}\BibitemShut
  {NoStop}%
\bibitem [{\citenamefont {Parameswaran}\ \emph {et~al.}(2014)\citenamefont
  {Parameswaran}, \citenamefont {Grover}, \citenamefont {Abanin}, \citenamefont
  {Pesin},\ and\ \citenamefont {Vishwanath}}]{parameswaran2014chiral}%
  \BibitemOpen
  \bibfield  {author} {\bibinfo {author} {\bibfnamefont {S.~A.}\ \bibnamefont
  {Parameswaran}}, \bibinfo {author} {\bibfnamefont {T.}~\bibnamefont
  {Grover}}, \bibinfo {author} {\bibfnamefont {D.~A.}\ \bibnamefont {Abanin}},
  \bibinfo {author} {\bibfnamefont {D.~A.}\ \bibnamefont {Pesin}},\ and\
  \bibinfo {author} {\bibfnamefont {A.}~\bibnamefont {Vishwanath}},\ }\href
  {https://doi.org/10.1103/PhysRevX.4.031035} {\bibfield  {journal} {\bibinfo
  {journal} {Phys. Rev. X}\ }\textbf {\bibinfo {volume} {4}},\ \bibinfo {pages}
  {031035} (\bibinfo {year} {2014})}\BibitemShut {NoStop}%
\bibitem [{\citenamefont {Huang}\ \emph
  {et~al.}(2015{\natexlab{a}})\citenamefont {Huang}, \citenamefont {Zhao},
  \citenamefont {Long}, \citenamefont {Wang}, \citenamefont {Chen},
  \citenamefont {Yang}, \citenamefont {Liang}, \citenamefont {Xue},
  \citenamefont {Weng}, \citenamefont {Fang}, \citenamefont {Dai},\ and\
  \citenamefont {Chen}}]{huang2015observation}%
  \BibitemOpen
  \bibfield  {author} {\bibinfo {author} {\bibfnamefont {X.}~\bibnamefont
  {Huang}}, \bibinfo {author} {\bibfnamefont {L.}~\bibnamefont {Zhao}},
  \bibinfo {author} {\bibfnamefont {Y.}~\bibnamefont {Long}}, \bibinfo {author}
  {\bibfnamefont {P.}~\bibnamefont {Wang}}, \bibinfo {author} {\bibfnamefont
  {D.}~\bibnamefont {Chen}}, \bibinfo {author} {\bibfnamefont {Z.}~\bibnamefont
  {Yang}}, \bibinfo {author} {\bibfnamefont {H.}~\bibnamefont {Liang}},
  \bibinfo {author} {\bibfnamefont {M.}~\bibnamefont {Xue}}, \bibinfo {author}
  {\bibfnamefont {H.}~\bibnamefont {Weng}}, \bibinfo {author} {\bibfnamefont
  {Z.}~\bibnamefont {Fang}}, \bibinfo {author} {\bibfnamefont {X.}~\bibnamefont
  {Dai}},\ and\ \bibinfo {author} {\bibfnamefont {G.}~\bibnamefont {Chen}},\
  }\href {https://doi.org/10.1103/PhysRevX.5.031023} {\bibfield  {journal}
  {\bibinfo  {journal} {Phys. Rev. X}\ }\textbf {\bibinfo {volume} {5}},\
  \bibinfo {pages} {031023} (\bibinfo {year} {2015}{\natexlab{a}})}\BibitemShut
  {NoStop}%
\bibitem [{\citenamefont {Zhang}\ \emph
  {et~al.}(2016{\natexlab{a}})\citenamefont {Zhang}, \citenamefont {Xu},
  \citenamefont {Belopolski}, \citenamefont {Yuan}, \citenamefont {Lin},
  \citenamefont {Tong}, \citenamefont {Bian}, \citenamefont {Alidoust},
  \citenamefont {Lee}, \citenamefont {Huang}, \citenamefont {Chang},
  \citenamefont {Chang}, \citenamefont {Hsu}, \citenamefont {Jeng},
  \citenamefont {Neupane}, \citenamefont {Sanchez}, \citenamefont {Zheng},
  \citenamefont {Wang}, \citenamefont {Lin}, \citenamefont {Zhang},
  \citenamefont {Lu}, \citenamefont {Shen}, \citenamefont {Neupert},
  \citenamefont {Zahid~Hasan},\ and\ \citenamefont
  {Jia}}]{zhang2016signatures}%
  \BibitemOpen
  \bibfield  {author} {\bibinfo {author} {\bibfnamefont {C.-L.}\ \bibnamefont
  {Zhang}}, \bibinfo {author} {\bibfnamefont {S.-Y.}\ \bibnamefont {Xu}},
  \bibinfo {author} {\bibfnamefont {I.}~\bibnamefont {Belopolski}}, \bibinfo
  {author} {\bibfnamefont {Z.}~\bibnamefont {Yuan}}, \bibinfo {author}
  {\bibfnamefont {Z.}~\bibnamefont {Lin}}, \bibinfo {author} {\bibfnamefont
  {B.}~\bibnamefont {Tong}}, \bibinfo {author} {\bibfnamefont {G.}~\bibnamefont
  {Bian}}, \bibinfo {author} {\bibfnamefont {N.}~\bibnamefont {Alidoust}},
  \bibinfo {author} {\bibfnamefont {C.-C.}\ \bibnamefont {Lee}}, \bibinfo
  {author} {\bibfnamefont {S.-M.}\ \bibnamefont {Huang}}, \bibinfo {author}
  {\bibfnamefont {T.-R.}\ \bibnamefont {Chang}}, \bibinfo {author}
  {\bibfnamefont {G.}~\bibnamefont {Chang}}, \bibinfo {author} {\bibfnamefont
  {C.-H.}\ \bibnamefont {Hsu}}, \bibinfo {author} {\bibfnamefont {H.-T.}\
  \bibnamefont {Jeng}}, \bibinfo {author} {\bibfnamefont {M.}~\bibnamefont
  {Neupane}}, \bibinfo {author} {\bibfnamefont {D.~S.}\ \bibnamefont
  {Sanchez}}, \bibinfo {author} {\bibfnamefont {H.}~\bibnamefont {Zheng}},
  \bibinfo {author} {\bibfnamefont {J.}~\bibnamefont {Wang}}, \bibinfo {author}
  {\bibfnamefont {H.}~\bibnamefont {Lin}}, \bibinfo {author} {\bibfnamefont
  {C.}~\bibnamefont {Zhang}}, \bibinfo {author} {\bibfnamefont {H.-Z.}\
  \bibnamefont {Lu}}, \bibinfo {author} {\bibfnamefont {S.-Q.}\ \bibnamefont
  {Shen}}, \bibinfo {author} {\bibfnamefont {T.}~\bibnamefont {Neupert}},
  \bibinfo {author} {\bibfnamefont {M.}~\bibnamefont {Zahid~Hasan}},\ and\
  \bibinfo {author} {\bibfnamefont {S.}~\bibnamefont {Jia}},\ }\href
  {https://doi.org/10.1038/ncomms10735} {\bibfield  {journal} {\bibinfo
  {journal} {Nature Communications}\ }\textbf {\bibinfo {volume} {7}},\
  \bibinfo {pages} {10735} (\bibinfo {year} {2016}{\natexlab{a}})}\BibitemShut
  {NoStop}%
\bibitem [{\citenamefont {Huang}\ \emph
  {et~al.}(2015{\natexlab{b}})\citenamefont {Huang}, \citenamefont {Xu},
  \citenamefont {Belopolski}, \citenamefont {Lee}, \citenamefont {Chang},
  \citenamefont {Wang}, \citenamefont {Alidoust}, \citenamefont {Bian},
  \citenamefont {Neupane}, \citenamefont {Zhang}, \citenamefont {Jia},
  \citenamefont {Bansil}, \citenamefont {Lin},\ and\ \citenamefont
  {Hasan}}]{huang2015weyl}%
  \BibitemOpen
  \bibfield  {author} {\bibinfo {author} {\bibfnamefont {S.-M.}\ \bibnamefont
  {Huang}}, \bibinfo {author} {\bibfnamefont {S.-Y.}\ \bibnamefont {Xu}},
  \bibinfo {author} {\bibfnamefont {I.}~\bibnamefont {Belopolski}}, \bibinfo
  {author} {\bibfnamefont {C.-C.}\ \bibnamefont {Lee}}, \bibinfo {author}
  {\bibfnamefont {G.}~\bibnamefont {Chang}}, \bibinfo {author} {\bibfnamefont
  {B.}~\bibnamefont {Wang}}, \bibinfo {author} {\bibfnamefont {N.}~\bibnamefont
  {Alidoust}}, \bibinfo {author} {\bibfnamefont {G.}~\bibnamefont {Bian}},
  \bibinfo {author} {\bibfnamefont {M.}~\bibnamefont {Neupane}}, \bibinfo
  {author} {\bibfnamefont {C.}~\bibnamefont {Zhang}}, \bibinfo {author}
  {\bibfnamefont {S.}~\bibnamefont {Jia}}, \bibinfo {author} {\bibfnamefont
  {A.}~\bibnamefont {Bansil}}, \bibinfo {author} {\bibfnamefont
  {H.}~\bibnamefont {Lin}},\ and\ \bibinfo {author} {\bibfnamefont {M.~Z.}\
  \bibnamefont {Hasan}},\ }\href {https://doi.org/10.1038/ncomms8373}
  {\bibfield  {journal} {\bibinfo  {journal} {Nature Communications}\ }\textbf
  {\bibinfo {volume} {6}},\ \bibinfo {pages} {7373} (\bibinfo {year}
  {2015}{\natexlab{b}})}\BibitemShut {NoStop}%
\bibitem [{\citenamefont {Wan}\ \emph {et~al.}(2011)\citenamefont {Wan},
  \citenamefont {Turner}, \citenamefont {Vishwanath},\ and\ \citenamefont
  {Savrasov}}]{wan2011topological}%
  \BibitemOpen
  \bibfield  {author} {\bibinfo {author} {\bibfnamefont {X.}~\bibnamefont
  {Wan}}, \bibinfo {author} {\bibfnamefont {A.~M.}\ \bibnamefont {Turner}},
  \bibinfo {author} {\bibfnamefont {A.}~\bibnamefont {Vishwanath}},\ and\
  \bibinfo {author} {\bibfnamefont {S.~Y.}\ \bibnamefont {Savrasov}},\ }\href
  {https://doi.org/10.1103/PhysRevB.83.205101} {\bibfield  {journal} {\bibinfo
  {journal} {Phys. Rev. B}\ }\textbf {\bibinfo {volume} {83}},\ \bibinfo
  {pages} {205101} (\bibinfo {year} {2011})}\BibitemShut {NoStop}%
\bibitem [{\citenamefont {Wang}\ \emph {et~al.}(2012)\citenamefont {Wang},
  \citenamefont {Sun}, \citenamefont {Chen}, \citenamefont {Franchini},
  \citenamefont {Xu}, \citenamefont {Weng}, \citenamefont {Dai},\ and\
  \citenamefont {Fang}}]{wang2012dirac}%
  \BibitemOpen
  \bibfield  {author} {\bibinfo {author} {\bibfnamefont {Z.}~\bibnamefont
  {Wang}}, \bibinfo {author} {\bibfnamefont {Y.}~\bibnamefont {Sun}}, \bibinfo
  {author} {\bibfnamefont {X.-Q.}\ \bibnamefont {Chen}}, \bibinfo {author}
  {\bibfnamefont {C.}~\bibnamefont {Franchini}}, \bibinfo {author}
  {\bibfnamefont {G.}~\bibnamefont {Xu}}, \bibinfo {author} {\bibfnamefont
  {H.}~\bibnamefont {Weng}}, \bibinfo {author} {\bibfnamefont {X.}~\bibnamefont
  {Dai}},\ and\ \bibinfo {author} {\bibfnamefont {Z.}~\bibnamefont {Fang}},\
  }\href {https://doi.org/10.1103/PhysRevB.85.195320} {\bibfield  {journal}
  {\bibinfo  {journal} {Phys. Rev. B}\ }\textbf {\bibinfo {volume} {85}},\
  \bibinfo {pages} {195320} (\bibinfo {year} {2012})}\BibitemShut {NoStop}%
\bibitem [{\citenamefont {Cook}\ and\ \citenamefont
  {Nielsen}(2023)}]{cook2022}%
  \BibitemOpen
  \bibfield  {author} {\bibinfo {author} {\bibfnamefont {A.~M.}\ \bibnamefont
  {Cook}}\ and\ \bibinfo {author} {\bibfnamefont {A.~E.~B.}\ \bibnamefont
  {Nielsen}},\ }\href {https://doi.org/10.1103/PhysRevB.108.045144} {\bibfield
  {journal} {\bibinfo  {journal} {Phys. Rev. B}\ }\textbf {\bibinfo {volume}
  {108}},\ \bibinfo {pages} {045144} (\bibinfo {year} {2023})}\BibitemShut
  {NoStop}%
\bibitem [{\citenamefont {Flores-Calderon}\ \emph {et~al.}(2023)\citenamefont
  {Flores-Calderon}, \citenamefont {Moessner},\ and\ \citenamefont
  {Cook}}]{calderon2023}%
  \BibitemOpen
  \bibfield  {author} {\bibinfo {author} {\bibfnamefont {R.}~\bibnamefont
  {Flores-Calderon}}, \bibinfo {author} {\bibfnamefont {R.}~\bibnamefont
  {Moessner}},\ and\ \bibinfo {author} {\bibfnamefont {A.~M.}\ \bibnamefont
  {Cook}},\ }\href {https://doi.org/10.1103/PhysRevB.108.125410} {\bibfield
  {journal} {\bibinfo  {journal} {Phys. Rev. B}\ }\textbf {\bibinfo {volume}
  {108}},\ \bibinfo {pages} {125410} (\bibinfo {year} {2023})}\BibitemShut
  {NoStop}%
\bibitem [{\citenamefont {Leis}\ \emph {et~al.}(2021)\citenamefont {Leis},
  \citenamefont {Schleenvoigt}, \citenamefont {Cherepanov}, \citenamefont
  {Lüpke}, \citenamefont {Schüffelgen}, \citenamefont {Mussler},
  \citenamefont {Grützmacher}, \citenamefont {Voigtländer},\ and\
  \citenamefont {Tautz}}]{TI-thin-film-conduc}%
  \BibitemOpen
  \bibfield  {author} {\bibinfo {author} {\bibfnamefont {A.}~\bibnamefont
  {Leis}}, \bibinfo {author} {\bibfnamefont {M.}~\bibnamefont {Schleenvoigt}},
  \bibinfo {author} {\bibfnamefont {V.}~\bibnamefont {Cherepanov}}, \bibinfo
  {author} {\bibfnamefont {F.}~\bibnamefont {Lüpke}}, \bibinfo {author}
  {\bibfnamefont {P.}~\bibnamefont {Schüffelgen}}, \bibinfo {author}
  {\bibfnamefont {G.}~\bibnamefont {Mussler}}, \bibinfo {author} {\bibfnamefont
  {D.}~\bibnamefont {Grützmacher}}, \bibinfo {author} {\bibfnamefont
  {B.}~\bibnamefont {Voigtländer}},\ and\ \bibinfo {author} {\bibfnamefont
  {F.~S.}\ \bibnamefont {Tautz}},\ }\href
  {https://doi.org/https://doi.org/10.1002/qute.202100083} {\bibfield
  {journal} {\bibinfo  {journal} {Advanced Quantum Technologies}\ }\textbf
  {\bibinfo {volume} {4}},\ \bibinfo {pages} {2100083} (\bibinfo {year}
  {2021})},\ \Eprint
  {https://arxiv.org/abs/https://onlinelibrary.wiley.com/doi/pdf/10.1002/qute.202100083}
  {https://onlinelibrary.wiley.com/doi/pdf/10.1002/qute.202100083} \BibitemShut
  {NoStop}%
\bibitem [{\citenamefont {Zhang}\ \emph {et~al.}(2010)\citenamefont {Zhang},
  \citenamefont {He}, \citenamefont {Chang}, \citenamefont {Song},
  \citenamefont {Wang}, \citenamefont {Chen}, \citenamefont {Jia},
  \citenamefont {Fang}, \citenamefont {Dai}, \citenamefont {Shan},
  \citenamefont {Shen}, \citenamefont {Niu}, \citenamefont {Qi}, \citenamefont
  {Zhang}, \citenamefont {Ma},\ and\ \citenamefont {Xue}}]{TI-crossover}%
  \BibitemOpen
  \bibfield  {author} {\bibinfo {author} {\bibfnamefont {Y.}~\bibnamefont
  {Zhang}}, \bibinfo {author} {\bibfnamefont {K.}~\bibnamefont {He}}, \bibinfo
  {author} {\bibfnamefont {C.-Z.}\ \bibnamefont {Chang}}, \bibinfo {author}
  {\bibfnamefont {C.-L.}\ \bibnamefont {Song}}, \bibinfo {author}
  {\bibfnamefont {L.-L.}\ \bibnamefont {Wang}}, \bibinfo {author}
  {\bibfnamefont {X.}~\bibnamefont {Chen}}, \bibinfo {author} {\bibfnamefont
  {J.-F.}\ \bibnamefont {Jia}}, \bibinfo {author} {\bibfnamefont
  {Z.}~\bibnamefont {Fang}}, \bibinfo {author} {\bibfnamefont {X.}~\bibnamefont
  {Dai}}, \bibinfo {author} {\bibfnamefont {W.-Y.}\ \bibnamefont {Shan}},
  \bibinfo {author} {\bibfnamefont {S.-Q.}\ \bibnamefont {Shen}}, \bibinfo
  {author} {\bibfnamefont {Q.}~\bibnamefont {Niu}}, \bibinfo {author}
  {\bibfnamefont {X.-L.}\ \bibnamefont {Qi}}, \bibinfo {author} {\bibfnamefont
  {S.-C.}\ \bibnamefont {Zhang}}, \bibinfo {author} {\bibfnamefont {X.-C.}\
  \bibnamefont {Ma}},\ and\ \bibinfo {author} {\bibfnamefont {Q.-K.}\
  \bibnamefont {Xue}},\ }\href {https://doi.org/10.1038/nphys1689} {\bibfield
  {journal} {\bibinfo  {journal} {Nature Phys}\ }\textbf {\bibinfo {volume}
  {6}},\ \bibinfo {pages} {584} (\bibinfo {year} {2010})}\BibitemShut {NoStop}%
\bibitem [{\citenamefont {Sakamoto}\ \emph {et~al.}(2010)\citenamefont
  {Sakamoto}, \citenamefont {Hirahara}, \citenamefont {Miyazaki}, \citenamefont
  {Kimura},\ and\ \citenamefont {Hasegawa}}]{TI-thin-phase-trans}%
  \BibitemOpen
  \bibfield  {author} {\bibinfo {author} {\bibfnamefont {Y.}~\bibnamefont
  {Sakamoto}}, \bibinfo {author} {\bibfnamefont {T.}~\bibnamefont {Hirahara}},
  \bibinfo {author} {\bibfnamefont {H.}~\bibnamefont {Miyazaki}}, \bibinfo
  {author} {\bibfnamefont {S.-i.}\ \bibnamefont {Kimura}},\ and\ \bibinfo
  {author} {\bibfnamefont {S.}~\bibnamefont {Hasegawa}},\ }\href
  {https://doi.org/10.1103/PhysRevB.81.165432} {\bibfield  {journal} {\bibinfo
  {journal} {Phys. Rev. B}\ }\textbf {\bibinfo {volume} {81}},\ \bibinfo
  {pages} {165432} (\bibinfo {year} {2010})}\BibitemShut {NoStop}%
\bibitem [{\citenamefont {Geim}\ and\ \citenamefont
  {Grigorieva}(2013)}]{vanwaals-hetero}%
  \BibitemOpen
  \bibfield  {author} {\bibinfo {author} {\bibfnamefont {A.~K.}\ \bibnamefont
  {Geim}}\ and\ \bibinfo {author} {\bibfnamefont {I.~V.}\ \bibnamefont
  {Grigorieva}},\ }\href {https://doi.org/10.1038/nature12385} {\bibfield
  {journal} {\bibinfo  {journal} {Nature}\ }\textbf {\bibinfo {volume} {499}},\
  \bibinfo {pages} {419} (\bibinfo {year} {2013})}\BibitemShut {NoStop}%
\bibitem [{\citenamefont {Hu}\ \emph {et~al.}(2020)\citenamefont {Hu},
  \citenamefont {Gordon}, \citenamefont {Liu}, \citenamefont {Liu},
  \citenamefont {Zhou}, \citenamefont {Hao}, \citenamefont {Narayan},
  \citenamefont {Emmanouilidou}, \citenamefont {Sun}, \citenamefont {Liu},
  \citenamefont {Brawer}, \citenamefont {Ramirez}, \citenamefont {Ding},
  \citenamefont {Cao}, \citenamefont {Liu}, \citenamefont {Dessau},\ and\
  \citenamefont {Ni}}]{Nature-hetero-antiferr}%
  \BibitemOpen
  \bibfield  {author} {\bibinfo {author} {\bibfnamefont {C.}~\bibnamefont
  {Hu}}, \bibinfo {author} {\bibfnamefont {K.~N.}\ \bibnamefont {Gordon}},
  \bibinfo {author} {\bibfnamefont {P.}~\bibnamefont {Liu}}, \bibinfo {author}
  {\bibfnamefont {J.}~\bibnamefont {Liu}}, \bibinfo {author} {\bibfnamefont
  {X.}~\bibnamefont {Zhou}}, \bibinfo {author} {\bibfnamefont {P.}~\bibnamefont
  {Hao}}, \bibinfo {author} {\bibfnamefont {D.}~\bibnamefont {Narayan}},
  \bibinfo {author} {\bibfnamefont {E.}~\bibnamefont {Emmanouilidou}}, \bibinfo
  {author} {\bibfnamefont {H.}~\bibnamefont {Sun}}, \bibinfo {author}
  {\bibfnamefont {Y.}~\bibnamefont {Liu}}, \bibinfo {author} {\bibfnamefont
  {H.}~\bibnamefont {Brawer}}, \bibinfo {author} {\bibfnamefont {A.~P.}\
  \bibnamefont {Ramirez}}, \bibinfo {author} {\bibfnamefont {L.}~\bibnamefont
  {Ding}}, \bibinfo {author} {\bibfnamefont {H.}~\bibnamefont {Cao}}, \bibinfo
  {author} {\bibfnamefont {Q.}~\bibnamefont {Liu}}, \bibinfo {author}
  {\bibfnamefont {D.}~\bibnamefont {Dessau}},\ and\ \bibinfo {author}
  {\bibfnamefont {N.}~\bibnamefont {Ni}},\ }\href
  {https://doi.org/10.1038/s41467-019-13814-x} {\bibfield  {journal} {\bibinfo
  {journal} {Nature Communications}\ }\textbf {\bibinfo {volume} {11}},\
  \bibinfo {pages} {97} (\bibinfo {year} {2020})}\BibitemShut {NoStop}%
\bibitem [{\citenamefont {Chong}\ \emph {et~al.}(2018)\citenamefont {Chong},
  \citenamefont {Han}, \citenamefont {Nagaoka}, \citenamefont {Tsuchikawa},
  \citenamefont {Liu}, \citenamefont {Liu}, \citenamefont {Vardeny},
  \citenamefont {Pesin}, \citenamefont {Lee}, \citenamefont {Sparks},\ and\
  \citenamefont {Deshpande}}]{TI-hetero}%
  \BibitemOpen
  \bibfield  {author} {\bibinfo {author} {\bibfnamefont {S.~K.}\ \bibnamefont
  {Chong}}, \bibinfo {author} {\bibfnamefont {K.~B.}\ \bibnamefont {Han}},
  \bibinfo {author} {\bibfnamefont {A.}~\bibnamefont {Nagaoka}}, \bibinfo
  {author} {\bibfnamefont {R.}~\bibnamefont {Tsuchikawa}}, \bibinfo {author}
  {\bibfnamefont {R.}~\bibnamefont {Liu}}, \bibinfo {author} {\bibfnamefont
  {H.}~\bibnamefont {Liu}}, \bibinfo {author} {\bibfnamefont {Z.~V.}\
  \bibnamefont {Vardeny}}, \bibinfo {author} {\bibfnamefont {D.~A.}\
  \bibnamefont {Pesin}}, \bibinfo {author} {\bibfnamefont {C.}~\bibnamefont
  {Lee}}, \bibinfo {author} {\bibfnamefont {T.~D.}\ \bibnamefont {Sparks}},\
  and\ \bibinfo {author} {\bibfnamefont {V.~V.}\ \bibnamefont {Deshpande}},\
  }\href {https://doi.org/10.1021/acs.nanolett.8b04291} {\bibfield  {journal}
  {\bibinfo  {journal} {Nano Lett.}\ }\textbf {\bibinfo {volume} {18}},\
  \bibinfo {pages} {8047} (\bibinfo {year} {2018})},\ \bibinfo {note} {pMID:
  30406664},\ \Eprint
  {https://arxiv.org/abs/https://doi.org/10.1021/acs.nanolett.8b04291}
  {https://doi.org/10.1021/acs.nanolett.8b04291} \BibitemShut {NoStop}%
\bibitem [{\citenamefont {Kou}\ \emph {et~al.}(2014)\citenamefont {Kou},
  \citenamefont {Wu}, \citenamefont {Felser}, \citenamefont {Frauenheim},
  \citenamefont {Chen},\ and\ \citenamefont {Yan}}]{robust-2dhetero}%
  \BibitemOpen
  \bibfield  {author} {\bibinfo {author} {\bibfnamefont {L.}~\bibnamefont
  {Kou}}, \bibinfo {author} {\bibfnamefont {S.-C.}\ \bibnamefont {Wu}},
  \bibinfo {author} {\bibfnamefont {C.}~\bibnamefont {Felser}}, \bibinfo
  {author} {\bibfnamefont {T.}~\bibnamefont {Frauenheim}}, \bibinfo {author}
  {\bibfnamefont {C.}~\bibnamefont {Chen}},\ and\ \bibinfo {author}
  {\bibfnamefont {B.}~\bibnamefont {Yan}},\ }\href
  {https://doi.org/10.1021/nn503789v} {\bibfield  {journal} {\bibinfo
  {journal} {ACS Nano}\ }\textbf {\bibinfo {volume} {8}},\ \bibinfo {pages}
  {10448} (\bibinfo {year} {2014})},\ \bibinfo {note} {pMID: 25226453},\
  \Eprint {https://arxiv.org/abs/https://doi.org/10.1021/nn503789v}
  {https://doi.org/10.1021/nn503789v} \BibitemShut {NoStop}%
\bibitem [{\citenamefont {Husain}\ \emph {et~al.}(2020)\citenamefont {Husain},
  \citenamefont {Gupta}, \citenamefont {Kumar}, \citenamefont {Kumar},
  \citenamefont {Behera}, \citenamefont {Brucas}, \citenamefont {Chaudhary},\
  and\ \citenamefont {Svedlindh}}]{TMD-1}%
  \BibitemOpen
  \bibfield  {author} {\bibinfo {author} {\bibfnamefont {S.}~\bibnamefont
  {Husain}}, \bibinfo {author} {\bibfnamefont {R.}~\bibnamefont {Gupta}},
  \bibinfo {author} {\bibfnamefont {A.}~\bibnamefont {Kumar}}, \bibinfo
  {author} {\bibfnamefont {P.}~\bibnamefont {Kumar}}, \bibinfo {author}
  {\bibfnamefont {N.}~\bibnamefont {Behera}}, \bibinfo {author} {\bibfnamefont
  {R.}~\bibnamefont {Brucas}}, \bibinfo {author} {\bibfnamefont
  {S.}~\bibnamefont {Chaudhary}},\ and\ \bibinfo {author} {\bibfnamefont
  {P.}~\bibnamefont {Svedlindh}},\ }\href {https://doi.org/10.1063/5.0025318}
  {\bibfield  {journal} {\bibinfo  {journal} {Applied Physics Reviews}\
  }\textbf {\bibinfo {volume} {7}},\ \bibinfo {pages} {041312} (\bibinfo {year}
  {2020})},\ \Eprint {https://arxiv.org/abs/https://doi.org/10.1063/5.0025318}
  {https://doi.org/10.1063/5.0025318} \BibitemShut {NoStop}%
\bibitem [{\citenamefont {Varsano}\ \emph {et~al.}(2020)\citenamefont
  {Varsano}, \citenamefont {Palummo}, \citenamefont {Molinari},\ and\
  \citenamefont {Rontani}}]{TMD-topo}%
  \BibitemOpen
  \bibfield  {author} {\bibinfo {author} {\bibfnamefont {D.}~\bibnamefont
  {Varsano}}, \bibinfo {author} {\bibfnamefont {M.}~\bibnamefont {Palummo}},
  \bibinfo {author} {\bibfnamefont {E.}~\bibnamefont {Molinari}},\ and\
  \bibinfo {author} {\bibfnamefont {M.}~\bibnamefont {Rontani}},\ }\href
  {https://doi.org/10.1038/s41565-020-0650-4} {\bibfield  {journal} {\bibinfo
  {journal} {Nat. Nanotechnol.}\ }\textbf {\bibinfo {volume} {15}},\ \bibinfo
  {pages} {367} (\bibinfo {year} {2020})}\BibitemShut {NoStop}%
\bibitem [{\citenamefont {Ryu}\ \emph {et~al.}(2010)\citenamefont {Ryu},
  \citenamefont {Schnyder}, \citenamefont {Furusaki},\ and\ \citenamefont
  {Ludwig}}]{Ryu_2010}%
  \BibitemOpen
  \bibfield  {author} {\bibinfo {author} {\bibfnamefont {S.}~\bibnamefont
  {Ryu}}, \bibinfo {author} {\bibfnamefont {A.~P.}\ \bibnamefont {Schnyder}},
  \bibinfo {author} {\bibfnamefont {A.}~\bibnamefont {Furusaki}},\ and\
  \bibinfo {author} {\bibfnamefont {A.~W.~W.}\ \bibnamefont {Ludwig}},\ }\href
  {https://doi.org/10.1088/1367-2630/12/6/065010} {\bibfield  {journal}
  {\bibinfo  {journal} {New Journal of Physics}\ }\textbf {\bibinfo {volume}
  {12}},\ \bibinfo {pages} {065010} (\bibinfo {year} {2010})}\BibitemShut
  {NoStop}%
\bibitem [{\citenamefont {Schnyder}\ \emph {et~al.}(2008)\citenamefont
  {Schnyder}, \citenamefont {Ryu}, \citenamefont {Furusaki},\ and\
  \citenamefont {Ludwig}}]{Schnyder2008}%
  \BibitemOpen
  \bibfield  {author} {\bibinfo {author} {\bibfnamefont {A.~P.}\ \bibnamefont
  {Schnyder}}, \bibinfo {author} {\bibfnamefont {S.}~\bibnamefont {Ryu}},
  \bibinfo {author} {\bibfnamefont {A.}~\bibnamefont {Furusaki}},\ and\
  \bibinfo {author} {\bibfnamefont {A.~W.~W.}\ \bibnamefont {Ludwig}},\ }\href
  {https://doi.org/10.1103/PhysRevB.78.195125} {\bibfield  {journal} {\bibinfo
  {journal} {Phys. Rev. B}\ }\textbf {\bibinfo {volume} {78}},\ \bibinfo
  {pages} {195125} (\bibinfo {year} {2008})}\BibitemShut {NoStop}%
\bibitem [{\citenamefont {Kitaev}(2009)}]{kitaev2009}%
  \BibitemOpen
  \bibfield  {author} {\bibinfo {author} {\bibfnamefont {A.}~\bibnamefont
  {Kitaev}},\ }\href {https://doi.org/10.1063/1.3149495} {\bibfield  {journal}
  {\bibinfo  {journal} {AIP Conference Proceedings}\ }\textbf {\bibinfo
  {volume} {1134}},\ \bibinfo {pages} {22} (\bibinfo {year} {2009})},\ \Eprint
  {https://arxiv.org/abs/https://pubs.aip.org/aip/acp/article-pdf/1134/1/22/11584243/22\_1\_online.pdf}
  {https://pubs.aip.org/aip/acp/article-pdf/1134/1/22/11584243/22\_1\_online.pdf}
  \BibitemShut {NoStop}%
\bibitem [{\citenamefont {Cook}(2024)}]{qskhe2024}%
  \BibitemOpen
  \bibfield  {author} {\bibinfo {author} {\bibfnamefont {A.~M.}\ \bibnamefont
  {Cook}},\ }\href {https://doi.org/10.1103/PhysRevB.109.155123} {\bibfield
  {journal} {\bibinfo  {journal} {Phys. Rev. B}\ }\textbf {\bibinfo {volume}
  {109}},\ \bibinfo {pages} {155123} (\bibinfo {year} {2024})}\BibitemShut
  {NoStop}%
\bibitem [{\citenamefont {Sun}\ \emph {et~al.}(2015)\citenamefont {Sun},
  \citenamefont {Wu}, \citenamefont {Ali}, \citenamefont {Felser},\ and\
  \citenamefont {Yan}}]{yan2015prediction}%
  \BibitemOpen
  \bibfield  {author} {\bibinfo {author} {\bibfnamefont {Y.}~\bibnamefont
  {Sun}}, \bibinfo {author} {\bibfnamefont {S.-C.}\ \bibnamefont {Wu}},
  \bibinfo {author} {\bibfnamefont {M.~N.}\ \bibnamefont {Ali}}, \bibinfo
  {author} {\bibfnamefont {C.}~\bibnamefont {Felser}},\ and\ \bibinfo {author}
  {\bibfnamefont {B.}~\bibnamefont {Yan}},\ }\href
  {https://doi.org/10.1103/PhysRevB.92.161107} {\bibfield  {journal} {\bibinfo
  {journal} {Phys. Rev. B}\ }\textbf {\bibinfo {volume} {92}},\ \bibinfo
  {pages} {161107} (\bibinfo {year} {2015})}\BibitemShut {NoStop}%
\bibitem [{\citenamefont {Song}\ \emph {et~al.}(2018)\citenamefont {Song},
  \citenamefont {Hsu}, \citenamefont {Zhao}, \citenamefont {Zhao},
  \citenamefont {Chang}, \citenamefont {Teng}, \citenamefont {Lin},\ and\
  \citenamefont {Loh}}]{Song_2018}%
  \BibitemOpen
  \bibfield  {author} {\bibinfo {author} {\bibfnamefont {P.}~\bibnamefont
  {Song}}, \bibinfo {author} {\bibfnamefont {C.}~\bibnamefont {Hsu}}, \bibinfo
  {author} {\bibfnamefont {M.}~\bibnamefont {Zhao}}, \bibinfo {author}
  {\bibfnamefont {X.}~\bibnamefont {Zhao}}, \bibinfo {author} {\bibfnamefont
  {T.-R.}\ \bibnamefont {Chang}}, \bibinfo {author} {\bibfnamefont
  {J.}~\bibnamefont {Teng}}, \bibinfo {author} {\bibfnamefont {H.}~\bibnamefont
  {Lin}},\ and\ \bibinfo {author} {\bibfnamefont {K.~P.}\ \bibnamefont {Loh}},\
  }\href {https://doi.org/10.1088/2053-1583/aac78d} {\bibfield  {journal}
  {\bibinfo  {journal} {2D Materials}\ }\textbf {\bibinfo {volume} {5}},\
  \bibinfo {pages} {031010} (\bibinfo {year} {2018})}\BibitemShut {NoStop}%
\bibitem [{\citenamefont {Qian}\ \emph {et~al.}(2014)\citenamefont {Qian},
  \citenamefont {Liu}, \citenamefont {Fu},\ and\ \citenamefont
  {Li}}]{qian_2014}%
  \BibitemOpen
  \bibfield  {author} {\bibinfo {author} {\bibfnamefont {X.}~\bibnamefont
  {Qian}}, \bibinfo {author} {\bibfnamefont {J.}~\bibnamefont {Liu}}, \bibinfo
  {author} {\bibfnamefont {L.}~\bibnamefont {Fu}},\ and\ \bibinfo {author}
  {\bibfnamefont {J.}~\bibnamefont {Li}},\ }\href
  {https://doi.org/10.1126/science.1256815} {\bibfield  {journal} {\bibinfo
  {journal} {Science}\ }\textbf {\bibinfo {volume} {346}},\ \bibinfo {pages}
  {1344} (\bibinfo {year} {2014})},\ \Eprint
  {https://arxiv.org/abs/https://www.science.org/doi/pdf/10.1126/science.1256815}
  {https://www.science.org/doi/pdf/10.1126/science.1256815} \BibitemShut
  {NoStop}%
\bibitem [{\citenamefont {Qi}\ \emph {et~al.}(2006)\citenamefont {Qi},
  \citenamefont {Wu},\ and\ \citenamefont {Zhang}}]{qi2006topological}%
  \BibitemOpen
  \bibfield  {author} {\bibinfo {author} {\bibfnamefont {X.-L.}\ \bibnamefont
  {Qi}}, \bibinfo {author} {\bibfnamefont {Y.-S.}\ \bibnamefont {Wu}},\ and\
  \bibinfo {author} {\bibfnamefont {S.-C.}\ \bibnamefont {Zhang}},\ }\href
  {https://doi.org/10.1103/PhysRevB.74.085308} {\bibfield  {journal} {\bibinfo
  {journal} {Phys. Rev. B}\ }\textbf {\bibinfo {volume} {74}},\ \bibinfo
  {pages} {085308} (\bibinfo {year} {2006})}\BibitemShut {NoStop}%
\bibitem [{\citenamefont {Hosur}\ and\ \citenamefont
  {Qi}(2013)}]{hosur_wsmtransport_2013}%
  \BibitemOpen
  \bibfield  {author} {\bibinfo {author} {\bibfnamefont {P.}~\bibnamefont
  {Hosur}}\ and\ \bibinfo {author} {\bibfnamefont {X.}~\bibnamefont {Qi}},\
  }\href {https://doi.org/https://doi.org/10.1016/j.crhy.2013.10.010}
  {\bibfield  {journal} {\bibinfo  {journal} {Comptes Rendus Physique}\
  }\textbf {\bibinfo {volume} {14}},\ \bibinfo {pages} {857} (\bibinfo {year}
  {2013})},\ \bibinfo {note} {topological insulators / Isolants
  topologiques}\BibitemShut {NoStop}%
\bibitem [{\citenamefont {Nguyen}\ \emph {et~al.}(2021)\citenamefont {Nguyen},
  \citenamefont {Kobayashi}, \citenamefont {Wichmann},\ and\ \citenamefont
  {Nomura}}]{nguyen_chiralqhe_2021}%
  \BibitemOpen
  \bibfield  {author} {\bibinfo {author} {\bibfnamefont {D.-H.-M.}\
  \bibnamefont {Nguyen}}, \bibinfo {author} {\bibfnamefont {K.}~\bibnamefont
  {Kobayashi}}, \bibinfo {author} {\bibfnamefont {J.-E.~R.}\ \bibnamefont
  {Wichmann}},\ and\ \bibinfo {author} {\bibfnamefont {K.}~\bibnamefont
  {Nomura}},\ }\href {https://doi.org/10.1103/PhysRevB.104.045302} {\bibfield
  {journal} {\bibinfo  {journal} {Phys. Rev. B}\ }\textbf {\bibinfo {volume}
  {104}},\ \bibinfo {pages} {045302} (\bibinfo {year} {2021})}\BibitemShut
  {NoStop}%
\bibitem [{\citenamefont {Pal}\ \emph {et~al.}(2024)\citenamefont {Pal},
  \citenamefont {Winter},\ and\ \citenamefont {Cook}}]{pal2024mkc}%
  \BibitemOpen
  \bibfield  {author} {\bibinfo {author} {\bibfnamefont {A.}~\bibnamefont
  {Pal}}, \bibinfo {author} {\bibfnamefont {J.~H.}\ \bibnamefont {Winter}},\
  and\ \bibinfo {author} {\bibfnamefont {A.~M.}\ \bibnamefont {Cook}},\ }\href
  {https://doi.org/10.1103/PhysRevB.109.014516} {\bibfield  {journal} {\bibinfo
   {journal} {Phys. Rev. B}\ }\textbf {\bibinfo {volume} {109}},\ \bibinfo
  {pages} {014516} (\bibinfo {year} {2024})}\BibitemShut {NoStop}%
\bibitem [{Sup(2024)}]{SuppMat}%
  \BibitemOpen
  \href@noop {} {} (\bibinfo {year} {2024}),\ \bibinfo {note} {see
  Supplementary materials for the details on calculation for the edge state
  functional form and standing wave condition, wannier spectra for FST based
  non-trivial topology as a function of the Weyl node separation, Hall bar
  geometry used for numerical simulation of magnetoconductivity with KWANT,
  chiral anomaly based notes for the FST WSM and FST WI.}\BibitemShut {Stop}%
\bibitem [{\citenamefont {Grushin}(2012)}]{grushin_lorentz_2012}%
  \BibitemOpen
  \bibfield  {author} {\bibinfo {author} {\bibfnamefont {A.~G.}\ \bibnamefont
  {Grushin}},\ }\href {https://doi.org/10.1103/PhysRevD.86.045001} {\bibfield
  {journal} {\bibinfo  {journal} {Phys. Rev. D}\ }\textbf {\bibinfo {volume}
  {86}},\ \bibinfo {pages} {045001} (\bibinfo {year} {2012})}\BibitemShut
  {NoStop}%
\bibitem [{\citenamefont {Abdulla}(2024)}]{abdulla2024pairwise}%
  \BibitemOpen
  \bibfield  {author} {\bibinfo {author} {\bibfnamefont {F.}~\bibnamefont
  {Abdulla}},\ }\href {https://doi.org/10.48550/arXiv.2312.02463} {\bibinfo
  {title} {Pairwise annihilation of weyl nodes induced by magnetic fields in
  the hofstadter regime}} (\bibinfo {year} {2024}),\ \Eprint
  {https://arxiv.org/abs/2312.02463} {arXiv:2312.02463 [cond-mat.mes-hall]}
  \BibitemShut {NoStop}%
\bibitem [{\citenamefont {Devakul}\ \emph {et~al.}(2021)\citenamefont
  {Devakul}, \citenamefont {Kwan}, \citenamefont {Sondhi},\ and\ \citenamefont
  {Parameswaran}}]{devakul2021serpentine}%
  \BibitemOpen
  \bibfield  {author} {\bibinfo {author} {\bibfnamefont {T.}~\bibnamefont
  {Devakul}}, \bibinfo {author} {\bibfnamefont {Y.~H.}\ \bibnamefont {Kwan}},
  \bibinfo {author} {\bibfnamefont {S.~L.}\ \bibnamefont {Sondhi}},\ and\
  \bibinfo {author} {\bibfnamefont {S.~A.}\ \bibnamefont {Parameswaran}},\
  }\href {https://doi.org/10.1103/PhysRevLett.127.116602} {\bibfield  {journal}
  {\bibinfo  {journal} {Phys. Rev. Lett.}\ }\textbf {\bibinfo {volume} {127}},\
  \bibinfo {pages} {116602} (\bibinfo {year} {2021})}\BibitemShut {NoStop}%
\bibitem [{\citenamefont {Potter}\ \emph
  {et~al.}(2014{\natexlab{b}})\citenamefont {Potter}, \citenamefont {Kimchi},\
  and\ \citenamefont {Vishwanath}}]{potter_qosc_2014}%
  \BibitemOpen
  \bibfield  {author} {\bibinfo {author} {\bibfnamefont {A.~C.}\ \bibnamefont
  {Potter}}, \bibinfo {author} {\bibfnamefont {I.}~\bibnamefont {Kimchi}},\
  and\ \bibinfo {author} {\bibfnamefont {A.}~\bibnamefont {Vishwanath}},\
  }\href {https://doi.org/10.1038/ncomms6161} {\bibfield  {journal} {\bibinfo
  {journal} {Nature Commun}\ }\textbf {\bibinfo {volume} {5}},\ \bibinfo
  {pages} {1} (\bibinfo {year} {2014}{\natexlab{b}})}\BibitemShut {NoStop}%
\bibitem [{\citenamefont {Zhang}\ \emph
  {et~al.}(2016{\natexlab{b}})\citenamefont {Zhang}, \citenamefont {Bulmash},
  \citenamefont {Hosur}, \citenamefont {Potter},\ and\ \citenamefont
  {Vishwanath}}]{zhang_qosc_2016}%
  \BibitemOpen
  \bibfield  {author} {\bibinfo {author} {\bibfnamefont {Y.}~\bibnamefont
  {Zhang}}, \bibinfo {author} {\bibfnamefont {D.}~\bibnamefont {Bulmash}},
  \bibinfo {author} {\bibfnamefont {P.}~\bibnamefont {Hosur}}, \bibinfo
  {author} {\bibfnamefont {A.~C.}\ \bibnamefont {Potter}},\ and\ \bibinfo
  {author} {\bibfnamefont {A.}~\bibnamefont {Vishwanath}},\ }\href
  {https://doi.org/10.1038/srep23741} {\bibfield  {journal} {\bibinfo
  {journal} {Sci. Rep.}\ }\textbf {\bibinfo {volume} {6}},\ \bibinfo {pages}
  {23741} (\bibinfo {year} {2016}{\natexlab{b}})}\BibitemShut {NoStop}%
\bibitem [{\citenamefont {Wang}\ \emph {et~al.}(2017)\citenamefont {Wang},
  \citenamefont {Sun}, \citenamefont {Lu},\ and\ \citenamefont
  {Xie}}]{wang20173dqhe}%
  \BibitemOpen
  \bibfield  {author} {\bibinfo {author} {\bibfnamefont {C.~M.}\ \bibnamefont
  {Wang}}, \bibinfo {author} {\bibfnamefont {H.-P.}\ \bibnamefont {Sun}},
  \bibinfo {author} {\bibfnamefont {H.-Z.}\ \bibnamefont {Lu}},\ and\ \bibinfo
  {author} {\bibfnamefont {X.~C.}\ \bibnamefont {Xie}},\ }\href
  {https://doi.org/10.1103/PhysRevLett.119.136806} {\bibfield  {journal}
  {\bibinfo  {journal} {Phys. Rev. Lett.}\ }\textbf {\bibinfo {volume} {119}},\
  \bibinfo {pages} {136806} (\bibinfo {year} {2017})}\BibitemShut {NoStop}%
\bibitem [{\citenamefont {{Groth}}\ \emph {et~al.}(2014)\citenamefont
  {{Groth}}, \citenamefont {{Wimmer}}, \citenamefont {{Akhmerov}},\ and\
  \citenamefont {{Waintal}}}]{kwantpaper}%
  \BibitemOpen
  \bibfield  {author} {\bibinfo {author} {\bibfnamefont {C.~W.}\ \bibnamefont
  {{Groth}}}, \bibinfo {author} {\bibfnamefont {M.}~\bibnamefont {{Wimmer}}},
  \bibinfo {author} {\bibfnamefont {A.~R.}\ \bibnamefont {{Akhmerov}}},\ and\
  \bibinfo {author} {\bibfnamefont {X.}~\bibnamefont {{Waintal}}},\ }\href
  {https://doi.org/10.1088/1367-2630/16/6/063065} {\bibfield  {journal}
  {\bibinfo  {journal} {New Journal of Physics}\ }\textbf {\bibinfo {volume}
  {16}},\ \bibinfo {eid} {063065} (\bibinfo {year} {2014})}\BibitemShut
  {NoStop}%
\bibitem [{\citenamefont {Belopolski}\ \emph {et~al.}(2017)\citenamefont
  {Belopolski}, \citenamefont {Yu}, \citenamefont {Sanchez}, \citenamefont
  {Ishida}, \citenamefont {Chang}, \citenamefont {Zhang}, \citenamefont {Xu},
  \citenamefont {Zheng}, \citenamefont {Chang}, \citenamefont {Bian},
  \citenamefont {Jeng}, \citenamefont {Kondo}, \citenamefont {Lin},
  \citenamefont {Liu}, \citenamefont {Shin},\ and\ \citenamefont
  {Hasan}}]{belopolski2017signature}%
  \BibitemOpen
  \bibfield  {author} {\bibinfo {author} {\bibfnamefont {I.}~\bibnamefont
  {Belopolski}}, \bibinfo {author} {\bibfnamefont {P.}~\bibnamefont {Yu}},
  \bibinfo {author} {\bibfnamefont {D.~S.}\ \bibnamefont {Sanchez}}, \bibinfo
  {author} {\bibfnamefont {Y.}~\bibnamefont {Ishida}}, \bibinfo {author}
  {\bibfnamefont {T.-R.}\ \bibnamefont {Chang}}, \bibinfo {author}
  {\bibfnamefont {S.~S.}\ \bibnamefont {Zhang}}, \bibinfo {author}
  {\bibfnamefont {S.-Y.}\ \bibnamefont {Xu}}, \bibinfo {author} {\bibfnamefont
  {H.}~\bibnamefont {Zheng}}, \bibinfo {author} {\bibfnamefont
  {G.}~\bibnamefont {Chang}}, \bibinfo {author} {\bibfnamefont
  {G.}~\bibnamefont {Bian}}, \bibinfo {author} {\bibfnamefont {H.-T.}\
  \bibnamefont {Jeng}}, \bibinfo {author} {\bibfnamefont {T.}~\bibnamefont
  {Kondo}}, \bibinfo {author} {\bibfnamefont {H.}~\bibnamefont {Lin}}, \bibinfo
  {author} {\bibfnamefont {Z.}~\bibnamefont {Liu}}, \bibinfo {author}
  {\bibfnamefont {S.}~\bibnamefont {Shin}},\ and\ \bibinfo {author}
  {\bibfnamefont {M.~Z.}\ \bibnamefont {Hasan}},\ }\href
  {https://doi.org/10.1038/s41467-017-00938-1} {\bibfield  {journal} {\bibinfo
  {journal} {Nature Commun}\ }\textbf {\bibinfo {volume} {8}},\ \bibinfo
  {pages} {942} (\bibinfo {year} {2017})}\BibitemShut {NoStop}%
\bibitem [{\citenamefont {Liu}\ \emph {et~al.}(2017)\citenamefont {Liu},
  \citenamefont {Wang}, \citenamefont {Fang}, \citenamefont {Fu},\ and\
  \citenamefont {Qian}}]{liu_2017_vdW}%
  \BibitemOpen
  \bibfield  {author} {\bibinfo {author} {\bibfnamefont {J.}~\bibnamefont
  {Liu}}, \bibinfo {author} {\bibfnamefont {H.}~\bibnamefont {Wang}}, \bibinfo
  {author} {\bibfnamefont {C.}~\bibnamefont {Fang}}, \bibinfo {author}
  {\bibfnamefont {L.}~\bibnamefont {Fu}},\ and\ \bibinfo {author}
  {\bibfnamefont {X.}~\bibnamefont {Qian}},\ }\href
  {https://doi.org/10.1021/acs.nanolett.6b04487} {\bibfield  {journal}
  {\bibinfo  {journal} {Nano Lett.}\ }\textbf {\bibinfo {volume} {17}},\
  \bibinfo {pages} {467} (\bibinfo {year} {2017})},\ \bibinfo {note} {pMID:
  27935725},\ \Eprint
  {https://arxiv.org/abs/https://doi.org/10.1021/acs.nanolett.6b04487}
  {https://doi.org/10.1021/acs.nanolett.6b04487} \BibitemShut {NoStop}%
\bibitem [{\citenamefont {Polatkan}\ \emph {et~al.}(2020)\citenamefont
  {Polatkan}, \citenamefont {Goerbig}, \citenamefont {Wyzula}, \citenamefont
  {Kemmler}, \citenamefont {Maulana}, \citenamefont {Piot}, \citenamefont
  {Crassee}, \citenamefont {Akrap}, \citenamefont {Shekhar}, \citenamefont
  {Felser}, \citenamefont {Dressel}, \citenamefont {Pronin},\ and\
  \citenamefont {Orlita}}]{polatkan2020magopt}%
  \BibitemOpen
  \bibfield  {author} {\bibinfo {author} {\bibfnamefont {S.}~\bibnamefont
  {Polatkan}}, \bibinfo {author} {\bibfnamefont {M.~O.}\ \bibnamefont
  {Goerbig}}, \bibinfo {author} {\bibfnamefont {J.}~\bibnamefont {Wyzula}},
  \bibinfo {author} {\bibfnamefont {R.}~\bibnamefont {Kemmler}}, \bibinfo
  {author} {\bibfnamefont {L.~Z.}\ \bibnamefont {Maulana}}, \bibinfo {author}
  {\bibfnamefont {B.~A.}\ \bibnamefont {Piot}}, \bibinfo {author}
  {\bibfnamefont {I.}~\bibnamefont {Crassee}}, \bibinfo {author} {\bibfnamefont
  {A.}~\bibnamefont {Akrap}}, \bibinfo {author} {\bibfnamefont
  {C.}~\bibnamefont {Shekhar}}, \bibinfo {author} {\bibfnamefont
  {C.}~\bibnamefont {Felser}}, \bibinfo {author} {\bibfnamefont
  {M.}~\bibnamefont {Dressel}}, \bibinfo {author} {\bibfnamefont {A.~V.}\
  \bibnamefont {Pronin}},\ and\ \bibinfo {author} {\bibfnamefont
  {M.}~\bibnamefont {Orlita}},\ }\href
  {https://doi.org/10.1103/PhysRevLett.124.176402} {\bibfield  {journal}
  {\bibinfo  {journal} {Phys. Rev. Lett.}\ }\textbf {\bibinfo {volume} {124}},\
  \bibinfo {pages} {176402} (\bibinfo {year} {2020})}\BibitemShut {NoStop}%
\bibitem [{\citenamefont {Xie}\ \emph {et~al.}(2021)\citenamefont {Xie},
  \citenamefont {Wu}, \citenamefont {Jin},\ and\ \citenamefont
  {Song}}]{xie2021trsymm}%
  \BibitemOpen
  \bibfield  {author} {\bibinfo {author} {\bibfnamefont {L.~C.}\ \bibnamefont
  {Xie}}, \bibinfo {author} {\bibfnamefont {H.~C.}\ \bibnamefont {Wu}},
  \bibinfo {author} {\bibfnamefont {L.}~\bibnamefont {Jin}},\ and\ \bibinfo
  {author} {\bibfnamefont {Z.}~\bibnamefont {Song}},\ }\href
  {https://doi.org/10.1103/PhysRevB.104.165422} {\bibfield  {journal} {\bibinfo
   {journal} {Phys. Rev. B}\ }\textbf {\bibinfo {volume} {104}},\ \bibinfo
  {pages} {165422} (\bibinfo {year} {2021})}\BibitemShut {NoStop}%
\bibitem [{\citenamefont {Bosnar}\ \emph {et~al.}(2023)\citenamefont {Bosnar},
  \citenamefont {Vyazovskaya}, \citenamefont {Petrov}, \citenamefont
  {Chulkov},\ and\ \citenamefont {Otrokov}}]{bosnar2023chernvdw}%
  \BibitemOpen
  \bibfield  {author} {\bibinfo {author} {\bibfnamefont {M.}~\bibnamefont
  {Bosnar}}, \bibinfo {author} {\bibfnamefont {A.~Y.}\ \bibnamefont
  {Vyazovskaya}}, \bibinfo {author} {\bibfnamefont {E.~K.}\ \bibnamefont
  {Petrov}}, \bibinfo {author} {\bibfnamefont {E.~V.}\ \bibnamefont
  {Chulkov}},\ and\ \bibinfo {author} {\bibfnamefont {M.~M.}\ \bibnamefont
  {Otrokov}},\ }\href {https://www.nature.com/articles/s41699-023-00396-y}
  {\bibfield  {journal} {\bibinfo  {journal} {npj 2D Materials and
  Applications}\ }\textbf {\bibinfo {volume} {7}},\ \bibinfo {pages} {33}
  (\bibinfo {year} {2023})}\BibitemShut {NoStop}%
\bibitem [{\citenamefont {Koshino}\ and\ \citenamefont
  {McCann}(2009)}]{koshino2009graphene}%
  \BibitemOpen
  \bibfield  {author} {\bibinfo {author} {\bibfnamefont {M.}~\bibnamefont
  {Koshino}}\ and\ \bibinfo {author} {\bibfnamefont {E.}~\bibnamefont
  {McCann}},\ }\href {https://doi.org/10.1103/PhysRevB.80.165409} {\bibfield
  {journal} {\bibinfo  {journal} {Phys. Rev. B}\ }\textbf {\bibinfo {volume}
  {80}},\ \bibinfo {pages} {165409} (\bibinfo {year} {2009})}\BibitemShut
  {NoStop}%
\bibitem [{\citenamefont {Zhang}\ \emph {et~al.}(2011)\citenamefont {Zhang},
  \citenamefont {Jung}, \citenamefont {Fiete}, \citenamefont {Niu},\ and\
  \citenamefont {MacDonald}}]{zhang2011layergraphene}%
  \BibitemOpen
  \bibfield  {author} {\bibinfo {author} {\bibfnamefont {F.}~\bibnamefont
  {Zhang}}, \bibinfo {author} {\bibfnamefont {J.}~\bibnamefont {Jung}},
  \bibinfo {author} {\bibfnamefont {G.~A.}\ \bibnamefont {Fiete}}, \bibinfo
  {author} {\bibfnamefont {Q.}~\bibnamefont {Niu}},\ and\ \bibinfo {author}
  {\bibfnamefont {A.~H.}\ \bibnamefont {MacDonald}},\ }\href
  {https://doi.org/10.1103/PhysRevLett.106.156801} {\bibfield  {journal}
  {\bibinfo  {journal} {Phys. Rev. Lett.}\ }\textbf {\bibinfo {volume} {106}},\
  \bibinfo {pages} {156801} (\bibinfo {year} {2011})}\BibitemShut {NoStop}%
\bibitem [{\citenamefont {Wang}\ \emph {et~al.}(2024)\citenamefont {Wang},
  \citenamefont {Zhang}, \citenamefont {Li}, \citenamefont {Sanborn},
  \citenamefont {Zhao}, \citenamefont {Wang}, \citenamefont {Watanabe},
  \citenamefont {Taniguchi}, \citenamefont {Crommie}, \citenamefont {Chen}
  \emph {et~al.}}]{wang2024chern}%
  \BibitemOpen
  \bibfield  {author} {\bibinfo {author} {\bibfnamefont {S.}~\bibnamefont
  {Wang}}, \bibinfo {author} {\bibfnamefont {Z.}~\bibnamefont {Zhang}},
  \bibinfo {author} {\bibfnamefont {H.}~\bibnamefont {Li}}, \bibinfo {author}
  {\bibfnamefont {C.}~\bibnamefont {Sanborn}}, \bibinfo {author} {\bibfnamefont
  {W.}~\bibnamefont {Zhao}}, \bibinfo {author} {\bibfnamefont {S.}~\bibnamefont
  {Wang}}, \bibinfo {author} {\bibfnamefont {K.}~\bibnamefont {Watanabe}},
  \bibinfo {author} {\bibfnamefont {T.}~\bibnamefont {Taniguchi}}, \bibinfo
  {author} {\bibfnamefont {M.~F.}\ \bibnamefont {Crommie}}, \bibinfo {author}
  {\bibfnamefont {G.}~\bibnamefont {Chen}}, \emph {et~al.},\ }\href
  {https://pubs.acs.org/doi/10.1021/acs.nanolett.3c05145} {\bibfield  {journal}
  {\bibinfo  {journal} {Nano Letters}\ } (\bibinfo {year} {2024})}\BibitemShut
  {NoStop}%
\end{thebibliography}%

\pagebreak

\appendix

\clearpage

\makeatletter
\renewcommand{\theequation}{S\arabic{equation}}
\renewcommand{\thefigure}{S\arabic{figure}}
\renewcommand{\thesection}{S\arabic{section}}
\setcounter{equation}{0}
\setcounter{section}{0}
\onecolumngrid


\begin{center}
  \textbf{\large Supplemental material for ``Finite-size topological phases from semimetals''}\\[.2cm]
  Adipta Pal,$^{1,2}$ and Ashley M. Cook$^{1,2,*}$\\[.1cm]
  {\itshape ${}^1$Max Planck Institute for Chemical Physics of Solids, Nöthnitzer Strasse 40, 01187 Dresden, Germany\\
  ${}^2$Max Planck Institute for the Physics of Complex Systems, Nöthnitzer Strasse 38, 01187 Dresden, Germany\\}
  ${}^*$Electronic address: cooka@pks.mpg.de\\
(Dated: \today)\\[1cm]
\end{center}

\section{Calculation of edge states}
Let us open the boundary along the x-axis for the WSM we considered in Eqn.~\ref{bulk3d}. This must create edge states flowing along y at $x=1$ and $x=L_x$ boundaries. We must hence choose an ansatz for our edge states which decays into the bulk but also vanishes at $x=0$ and $x=L_x+1$. Let us focus on the first boundary condition at $x=0$ and choose the ansatz,
\begin{equation}
\Psi(x,k_y,k_z) = (e^{-\lambda_1x}-e^{-\lambda_2x})\phi(k_y,k_z)\chi.
\end{equation}
where $\chi$ is a two-component spinor. Here the wavefunction vanishes at $x=0$ and $\lambda_1$ and $\lambda_2$ are not necessarily real but they must be distinct. The Hamiltonian is modified as $k_x\rightarrow i\lambda$ for a decaying solution,
\begin{equation}\label{Hlambda}
H(i\lambda,k_y,k_z) = (M-2t\cosh\lambda-2t\cos k_y-2t\cos k_z)\sigma^z+2i\Delta\sinh \lambda\sigma^x+2\Delta\sin k_y\sigma^y,
\end{equation}
so that the ansatz must be an eigenstate of the Hamiltonian at two distinct values of $\lambda$ i.e. $H(\lambda_1)\Psi =H(i\lambda_2)\Psi =E\Psi$. Therefore, the determinant, $\text{det}(H(i\lambda_1)-H(i\lambda_2))$ must be zero, and we look for the spinor in the nullspace of $H(i\lambda_1)-H(i\lambda_2)$. From the determinant, we have the condition, for distinct $\lambda_1$ and $\lambda_2$,
\begin{equation}\label{edgecond}
2t\sinh(\lambda_+)=\pm 2\Delta\cosh(\lambda_+),\quad \lambda_+=\frac{\lambda_1+\lambda_2}{2},
\end{equation}
which leads us to the spinor which satisfies, $2\Delta\cosh(\lambda_+)(\sigma^z\mp i\sigma^x)\chi =0$. The spinor, $\chi$ is therefore the ladder operator for eigenvectors of $\sigma^y$. Choosing the positive sign in Eqn.~\ref{edgecond}, we have $\chi_L\sim(1,-i)^T$. Next we solve for energy $E$ and $\lambda$ in the eigenvalue equation, $H(i\lambda,k_y,k_z)\chi_L=E\chi_L$, which we expand in terms of $\sigma^y$ and its ladder operators, $\sigma^\pm_y=\frac{1}{2}(\sigma^z\pm i\sigma^x)$,
\begin{equation}
\begin{split}
((m-2t\cosh\lambda + 2\Delta\sinh\lambda)\sigma^+_y+(m-2t\cosh\lambda - 2\Delta\sinh\lambda)\sigma^-_y+2\Delta\sin k_y\sigma^y)\chi_L=E\chi_L
\end{split}
\end{equation}
where $m=M-2t\cos k_y-2t\cos k_z$. Then, $\chi_L$ must naturally lead to the eigenvalue, $E=-2\Delta\sin k_y$, however, subject to the condition,$m-2t\cosh \lambda + 2\Delta\sinh\lambda = 0$. From this condition, we obtain a quadratic equation for $e^{-\lambda}$, which we solve to get our two assumed values of $\lambda$,
\begin{equation}
\begin{split}
&(t+\Delta)e^{-2\lambda}-me^{-\lambda}+(t-\Delta) = 0,\\
\implies & e^{-\lambda} = \frac{m\pm\sqrt{m^2-4(t^2-\Delta^2)}}{2(t+\Delta)},\\
\implies & e^{-\lambda} = \frac{m\pm i\sqrt{4(t^2-\Delta^2)-m^2}}{2(t+\Delta)} = Re^{\pm i\theta}.
\end{split}
\end{equation}
Here $R$ and $\theta$ represent the absolute value and arguments of the complex $\lambda$s. The transition from the second line to the third line by changing the sign inside the square root is possible because we never assumed that $\lambda$ was real and we need complex $\lambda$ to fit our decaying wavefunction inside a finite size lattice based on our second boundary condition, $\Psi(x=L_x+1)=0$. Before moving forward, we comment on the choice of sign in Eqn.~\ref{edgecond}. We observe from our calculation that the two $\lambda$s are complex conjugates of each other, therefore, $\lambda_+$ is real and is proportional to the decay constant for our edge state which for the left edge must be positive. Choosing the positive sign in Eqn.~\ref{edgecond} therefore assumes that $\text{sgn}(t)=\text{sgn}{\Delta}$ and therefore in a self-consistent way, we get that the real part of our $\lambda$s are positive. On the other hand, if $t$ and $\Delta$ had different signs, the negative sign in Eqn.~\ref{edgecond} supplies us a decaying wavefunction at the left edge and the changes in the spinor and energy follows.\\
Now we consider our ansatz and substitute for $\lambda_1$ and $\lambda_2$. $\Psi(x,k_y,k_z)$ should now have the form, $\Psi \sim R^x\sin(\theta x)\phi(k_y,k_z)\chi_L$. Putting in boundary conditions at $x=L_x+1$, sets the general solution for $\theta$ as $\theta = \frac{n\pi}{L_x+1}$, $n=1,...,L_x$. The edge state can finally be written as,
\begin{equation}
\Psi(x,k_y,k_z) \sim \bigg{(}\frac{t-\Delta}{t+\Delta}\bigg{)}^{\frac{x}{2}}\sin\big{(}\frac{n\pi}{L_x+1}x\big{)}e^{ik_yy+ik_zz}
\begin{pmatrix}
1\\
-i
\end{pmatrix}
, \quad (n=1,...,L_x).
\end{equation}
We have assumed plane wave solution for $\phi(k_y,k_z)$. Physically this is equivalent to a lattice version of the particle in a box problem, and the solutions of $\theta$ are the standing wave wavenumbers which fit in the lattice. For each standing wave wavenumber, we have a constraint on the value of $m=M-2t\cos k_y-2t\cos k_z$, which is computed from the real part of the $\lambda$s,
\begin{equation}\label{mcond}
M-2t\cos k_y-2t\cos k_z = 2\sqrt{t^2-\Delta^2}\cos\big{(}\frac{n\pi}{L_x+1}\big{)}, \quad (n=1,...,L_x).
\end{equation}
These are exactly the edge states and constraint on $m$ we have presented in the main text.

\section{Density of states for (3-1)d edge modes in thin film systems}

\begin{figure}[htb!]
    \centering
    \includegraphics[width=0.95\textwidth]{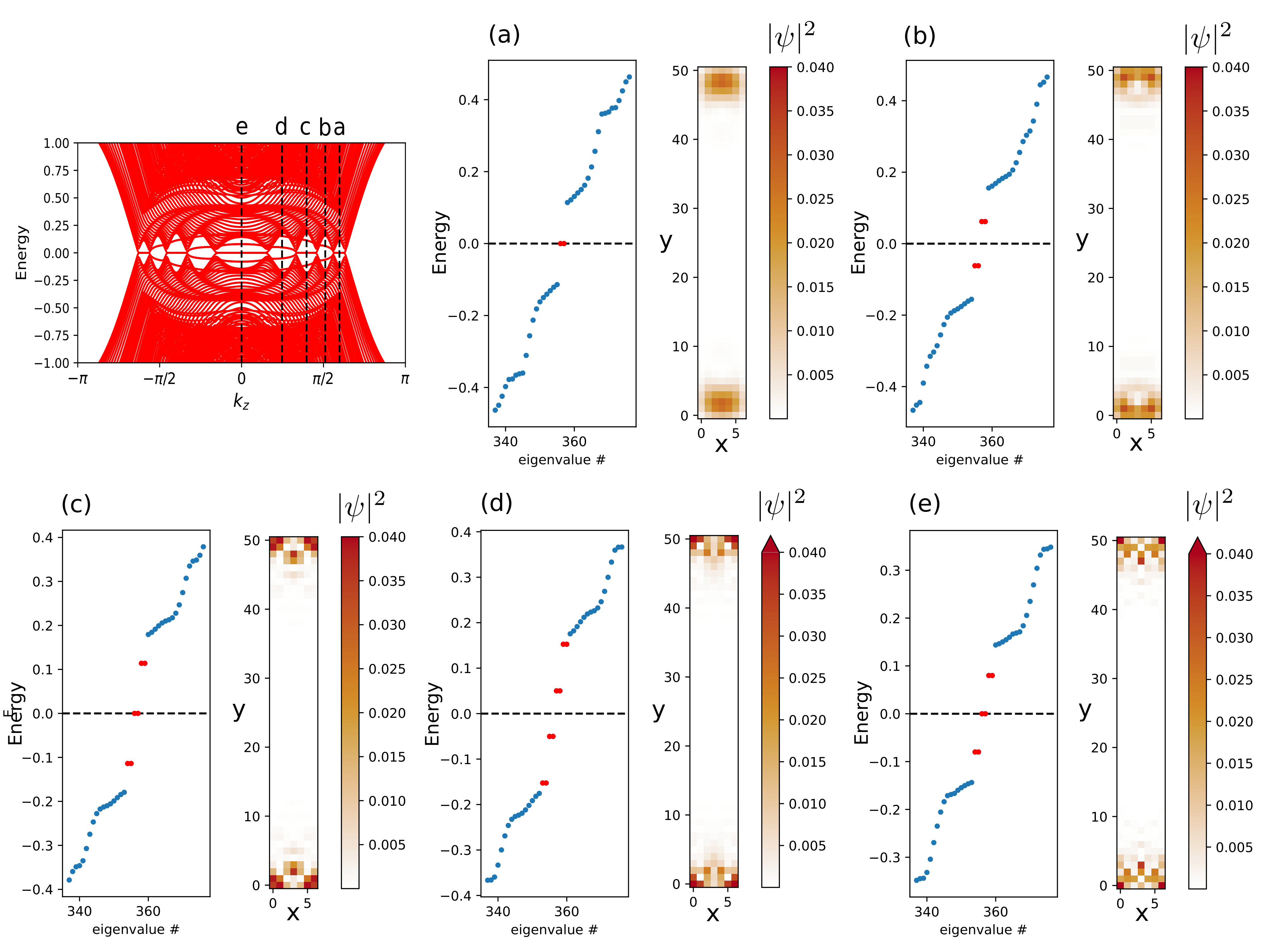}
    \caption{We show clearly the in-gap states present in the slab spectra for a FST-WSM with small system size along x, for different values of $k_z$ representing different odd-even regimes as mentioned in the main text. The probability density of the state at or closest to zero energy is plotted in real space for x and y OBC. The five plots (a), (b), (c), (d) and (e) refer to the five $k_z$ values shown in the plot in the top left corner.}
    \label{fig:probdenandeigE}
\end{figure}

\section{Spectral flow as a function of M and kz}
We have shown the spectral flow along $k_y$ for a thin film along x by computing the Wilson loop in the main text as a function of $k_z$ for a single value of $M$. We provide here a heat map of the number of flat bands in $k_z$ at zero energy as a function of $M$. It can be observed that the highest number of flat bands in a thin film along x with $L_x$ sites is $L_x$.
\begin{center}
\begin{figure}[htb!]
    \centering
    \includegraphics[width=0.95\textwidth]{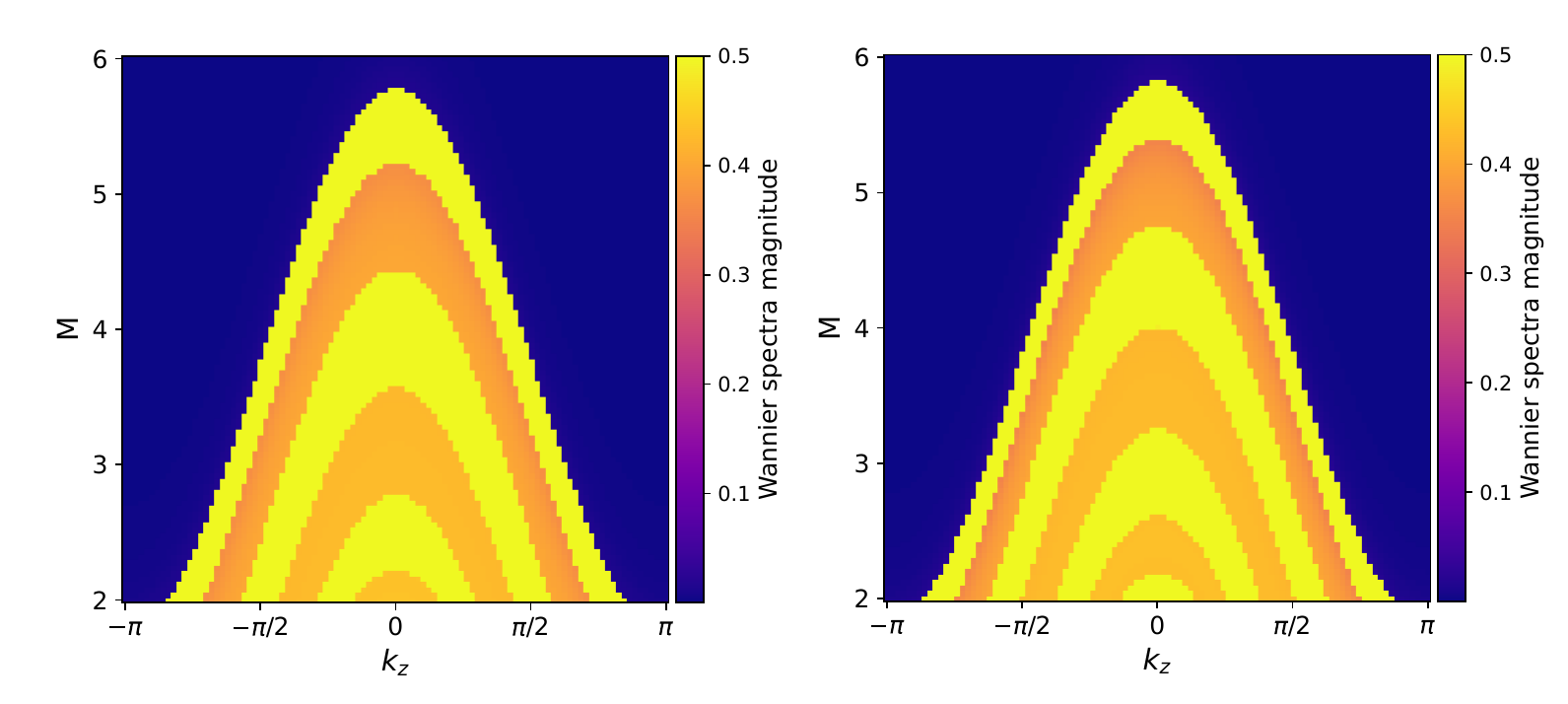}
    \caption{Wannier spectra for Weyl semimetal thin film along x with (left) $L=6$ and (right) $L=7$ vs. $k_z$ and $M$. Other parameters include $t=1$ and $\Delta=0.2$. We find that the maximum number of flat bands possible at any $M$ within the bulk Weyl semimetal parameter regime is equal to the number of sites, $L_x$ along the thin film.}
    \label{fig:WLvskzandM}
\end{figure}
\end{center}

\section{Numerical calculation of conductivity for the chiral anomaly}
\begin{figure}
    \centering
    \includegraphics[width=0.5\textwidth]{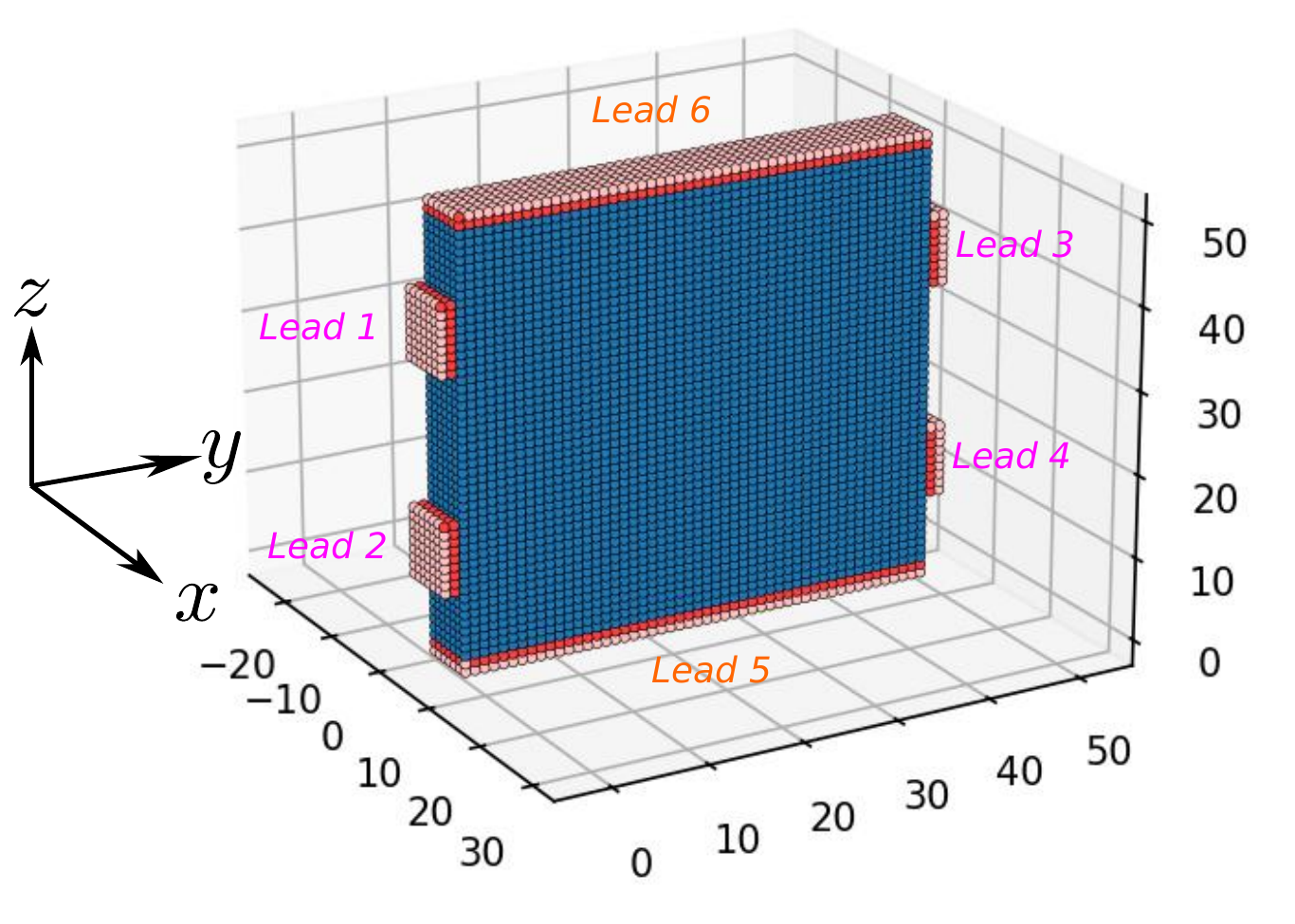}
    \caption{Six terminal Hall bar geometry set-up used in KWANT to calculate chiral anomaly behaviour in the main text. The voltage measuring leads 1-4 are highlighted in magenta and the current carrying leads 4-5 are highlighted in orange. The voltage across lead 1 and 2 and across 2 and 4 are measured for the longitudinal and transverse conductivity for unit current flowing between leads 4 and 5. The above setup has a small system size along x ($L_x=7$) and large system sizes along y and z ($L_y=L_z=51$).}
    \label{fig:hallbarsetup}
\end{figure}

We use the KWANT Python package to calculate the conductivity for a six terminal Hall bar geometry where the scattering region is given by the finite size TRB WSM. We consider open boundary conditions along all three axis and consider the z-axis to attach out current carrying leads which in the formalism in KWANT must be translation invariant systems of the same material as the scattering region. The other four voltage leads are metallic of the form, $\sum_{i=x,y,z}8t\cos k_i\sigma^0$. The resulting configuration is shown in Fig.~\ref{fig:hallbarsetup}. We will consider unit current flowing from the left through current lead 5 and calculate the voltage across leads 1 and 2 and voltage across leads 2 and 4 to get the longitudinal conductivity, $\sigma^{zz}$ and Hall conductivity $\sigma^{yz}$. This is done by evaluating the conductance matrix for the scattering region using one of the built-in functions in KWANT and solving the voltage for unit current through lead 5.

We simulations in the main text are carried out in presence of external magnetic field of the form, $\mathbf{B}=(B\sin\theta,0,B\cos\theta)$. We use the following Peierls substitution for the momenta, $k_x\rightarrow k_x-yB\cos\theta$ and $k_y\rightarrow k_y+zB\sin\theta$.

\section{Chiral anomaly in the FST WSM}
\begin{figure}[htb!]
    \centering
    \includegraphics[width=0.95\textwidth]{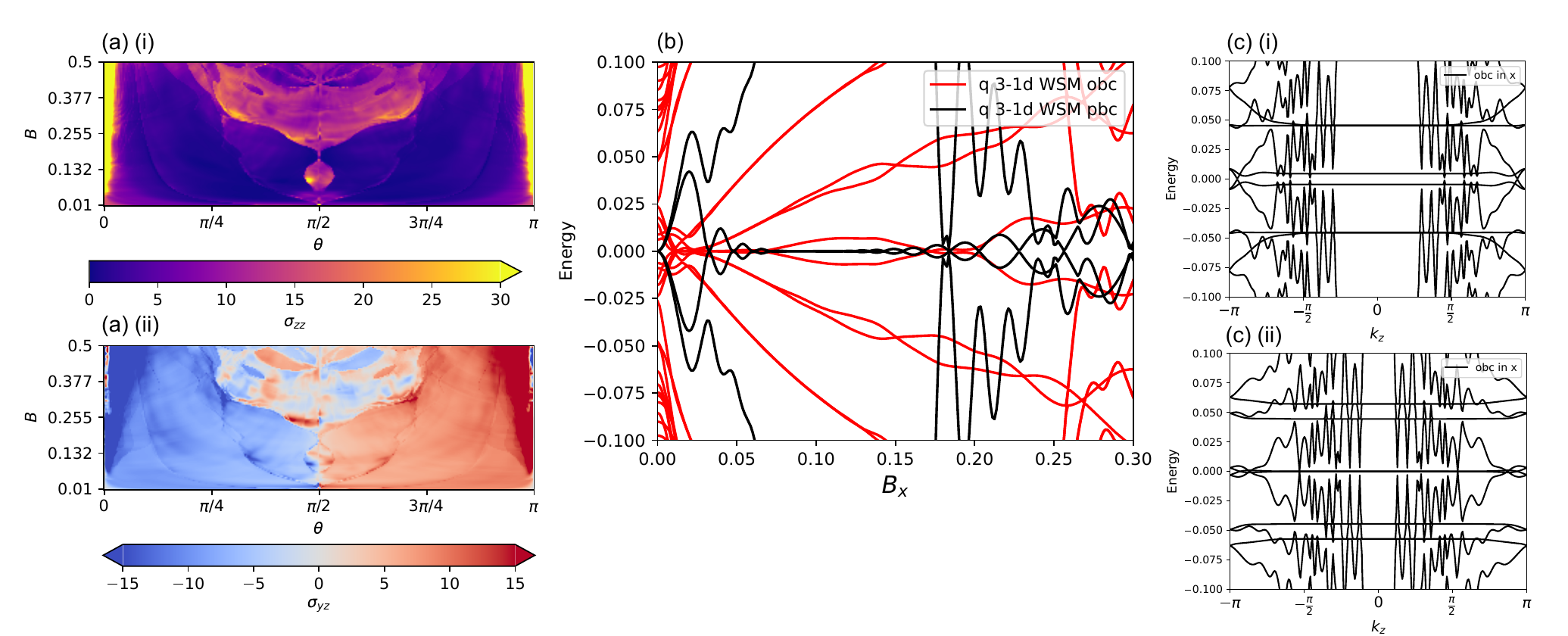}
    \caption{Magnetoconductivity in the FST WSM and the influence of serpentine LLs. The FST WSM has thin film along x ($L_x=7$) with $M=3$ and $\Delta= 0.2t$. Subfigure (a)(i) and (a)(ii) shows the longitudinal ($\sigma_{zz}$) and transverse magnetoconductivity ($\sigma_{yz}$) for the FST WSM as a function of the relative angle $\theta$ between the magnetic and electric field in the x-axis and the magnitude of the magnetic field, $B$ along the y-axis of the plot. Comparing with subfigure (b) for the magic magnetic field which is the exact point in (a)(i) and (a)(ii) where the conductivity vanishes. Subfigures (c)(i) and (c)(ii) show the slab spectra vs $k_z$ at $\theta=\frac{\pi}{2}$ for magnetic field magnitudes $B=0.1725$ and $B=0.15$ respectively. The former is approximately the magic field strength and the latter is away from the magic field.}
    \label{fig:chiralanomalySM}
\end{figure}

We provide a brief explanation about the nature of the longitudinal and transverse magnetoconductivity as a function of angle $\theta$ between the electric and magnetic field and the magnitude of the magnetic field. We refer to Fig.~\ref{fig:chiralanomalySM}. Here we observe bubbles in the heatmap which closes at certain points in (a). We show here that the closings directly correspond to the magic magnetic fields in the serpentine Landau levels derived from the SWNs in a FST WSM (thin film along x) in (b). We further show the slab spectra as a function of $k_z$ which shows at those magic magnetic fields there exists a two flat band Landau levels at exactly zero energy leading to zero conductivity at those magnetic fields in (c)(i). At other magnetic fields, there exists some dispersing cones which contribute to the conductivity as shown in (c)(ii). The two Landau levels near zero are gapped at those points.

\section{Chiral anomaly in FST insulator}
We start with the bulk Hamiltonian fo the TRB WSm in the main text. In the presence of a magnetic field along z, $B\hat{z}$, we expand the system around $k_x=k_y=0$ and change $k_y\rightarrow k_y^\prime=k_y+Bx$ (consider charge as 1),
\begin{equation}
\mathcal{H}(k_z)\approx (M-4t-2t\cos k_z+t(k_x^2+{k_y^{\prime}}^2))\sigma^z+2\Delta k_x\sigma^x+2\Delta k_y^\prime\sigma^y.
\end{equation}
The energy for the lowest Landau level (LLL) is given by the following expression,
\begin{equation}
E_{LLL}(k_z) = -(M-4t-2t\cos k_z+tB).
\end{equation}
Denote the system size along z as $L_z$. As we showed in the main text, opening boundary along the z axis block diagonalizes the system, and is equivalent to the substitution, $k_z\rightarrow \frac{n\pi}{L_z+1}$ $(n=1,...,L_z)$. Therefore the LLL expression changes to,
\begin{equation}
E_{LLL}(n) = -(M-4t-2t\cos \frac{n\pi}{L_z+1}+tB),\quad (n=1,...,L_z).
\end{equation}
We observe the LLL energy shows a linear dependence in magnetic field $B$ and does not contain any $\Delta$ dependent term. The $\Delta$ dependence only affects the non-chiral LLs.

\end{document}